\documentclass[prb, twocolumn, groupedaddress, nofootinbib, longbibliography, aps]{revtex4-2}
\usepackage{tikz}
\usepackage{physics}
\usepackage{bbold}
\usepackage{color}
\usepackage{graphicx}
\usepackage{float}
\usepackage{mathtools}
\usepackage{amsmath}
\usepackage{bm}
\usepackage[colorlinks, bookmarks=true, breaklinks=true,linkcolor=red, citecolor=blue, linktocpage=true, urlcolor=blue]{hyperref}

\newcommand{\tree}[5]{
\begin{tikzpicture}[xscale=.8, yscale=.8, nodes={execute at begin node=$, execute at end node=$}, baseline={(current bounding box.center)}]
\draw (-3,0)node[below]{#1}--(0,0)node[below]{#4};
\draw (-2,0)--(-2,1)node[above]{#2};
\draw (-1,0)--(-1,1)node[above]{#3};
\node[below] at (-1.5,0){#5};
\end{tikzpicture}
}

\newcommand{\treeb}[5]{
\begin{tikzpicture}[xscale=.8, yscale=.8, nodes={execute at begin node=$, execute at end node=$}, baseline={(current bounding box.center)}]
\draw (-2.4,0)node[below]{#1}--(0,0)node[below]{#4};
\draw (-1.2,0)--(-1.2,0.5)--(-2,1)node[above]{#2};
\draw (-1.2,0.5)--(-0.4,1)node[above]{#3};
\node[left] at (-1.3,0.25){#5};
\end{tikzpicture}
}

\begin{document}

\title{Symmetry breaking phases and transitions in an Ising fusion category lattice model}

\author{Soumil Roychowdhury}
\affiliation{Department of Physics and HK Institute of Quantum Science \& Technology, The University of Hong Kong, Pokfulam Road, Hong Kong, China}

\author{Chenjie Wang}
\email{cjwang@hku.hk}
\affiliation{Department of Physics and HK Institute of Quantum Science \& Technology, The University of Hong Kong, Pokfulam Road, Hong Kong, China}

\date{\today}

\begin{abstract}
An anyon-chain-like lattice model with symmetry described by the Ising fusion category is studied. Combining numerical and analytical studies, we uncover a rich phase diagram that contains three phases: a symmetric critical phase and two categorical symmetry breaking phases. The symmetric phase lies in the same universality class as the usual critical Ising model. The first symmetry-breaking phase, dubbed the \emph{categorical ferromagnetic} phase, has the Ising fusion category fully broken and exhibits a threefold ground-state degeneracy, as expected from the generalized Landau paradigm. The other symmetry-breaking phase is analogous to a conventional antiferromagnet: it breaks lattice translation and part of the Ising fusion category, and therefore is termed the \emph{categorical antiferromagnetic} phase. Unlike ordinary antiferromagnetic states associated with finite invertible symmetry breaking, this phase itself is critical, being described by a fourfold degenerate Ising conformal field theory. We argue more generally that antiferromagnetic states associated with broken non-invertible symmetries have a large low-energy manifold that grows exponentially in system size, due to the greater-than-one quantum dimension of domain walls. We also numerically study the transitions between the three phases. The transition between the symmetric and categorical ferromagnetic phase is described by the $c=7/10$ tricritical Ising CFT, while the transition between the symmetric and categorical antiferromagnetic phases is less understood. Our numerical data suggest that the latter transition is continuous and described by a conformal field theory with central charge $c=3/2$.
\end{abstract}

\maketitle

\section{Introduction}

Symmetry is a cornerstone of modern physics, and many of the most successful theories are founded on symmetry principles, most notably the Standard Model. Traditionally, symmetries are described by groups. In recent years, however, the notion of symmetry has been significantly broadened through the emphasis on the correspondence between  symmetries and topological defects\footnote{This perspective has a long history, likely tracing back to the studies of conformal field theory\cite{Verlinde1988}.} (see \cite{McGreevyRev2023, shao2023s,  schafer-nameki_ictp_2023, bhardwaj_lectures_2023, LuoWangWangRev2024} for reviews). This correspondence allows one to relax the requirement that symmetries should be invertible, leading to the notion of \emph{non-invertible symmetries}\cite{feiguin_interacting_2007,BhardwajTachikawa2018JHEP,Chang_etal_2019JHEP,JiWen2020PRR}, which are described by fusion categories\footnote{In this paper, we consider only finite symmetries. How symmetries of continuous groups can be generalized to categorical symmetries with appropriate mathematical structures remains largely open; see, for example, a recent work \cite{YiNanWangarXiv2025}.}\cite{Etingof2015tensor} rather than groups. The same perspective  also allows one to consider symmetries generated by topological defects of different dimensions, giving rise to \emph{higher-form symmetries}\cite{gaiotto_generalized_2015}. In addition, this view naturally incorporates 't Hooft anomaly\cite{tHooft1979} into the very definition of symmetry. More broadly, \emph{generalized symmetries}  can be unified within the mathematical framework of higher fusion categories\cite{Douglas2018,KongWen2014,KLWZZ_higherAlg_2020,Freed2022}. 

Interestingly, many familiar physical phenomena can be reinterpreted as instances of spontaneous breaking of generalized symmetries. For example, the order-disorder coexistence Ising transition line can be understood as a result of the spontaneous breaking of the Kramers-Wannier duality, one of the best known non-invertible symmetries \cite{BlumeEmeryPRB1971, OBrienFendleyPRL2018, seiberg_non-invertible_2024}. For another example, two-dimensional topological orders\cite{Wee_Zoo_paper_2017} --- long regarded as examples beyond the conventional Landau paradigm --- can now be viewed as symmetry-breaking phases of 1-form symmetries\cite{Wen2019PRB, McGreevyRev2023}. Consequently, considerable effort has been devoted to exploring  new physics within this \emph{generalized Landau paradigm} \cite{Lake2018,ThorngrenWang1, choi2022noninvertible, bhardwaj2024categorical, Bhardwaj_generalizedLandau2, bhardwaj_illustrating_2024,Apoorv2023SciPost, chatterjee2024quantum,  Cordova2024PRX, ZhaoWan2025arXiv}.  

Generalized symmetries also provide new tools for constraining the IR physics of field theories and lattice models.  A prominent modern example is the family of anyon chain models\cite{feiguin_interacting_2007, Gils2013PRB, trebst_short_2008, AMF1,AMF2,Vanhove2018PRL}, in which non-invertible symmetries are often---though not always---found to pin a model at quantum criticality. The underlying mechanism can be understood as a generalization of anomaly matching to generalized symmetries. Just as ordinary group-like symmetries with a 't Hooft anomaly forbid a trivially gapped ground state---as in edge theories of symmetry-protected topological phases\cite{Senthil_sptReview_2015} and in Lieb-Schultz-Mattis-type constraints\cite{LSM-review}---generalized symmetries naturally incorporate 't Hooft anomalies in the definition and can thereby place strong constraints on the low-energy physics of a theory\cite{Chang_etal_2019JHEP, ThorngrenWang1}. More recently, generalized symmetries have been used extensively to construct quantum critical models, although mostly for 1D systems\cite{Chatterjee2023PRB,chatterjee2024quantum,ning_building_2023, Hung2025arXiv,Haagerup1,Haagerup2} (also see a series of earlier works \cite{Fuchs1,Fuchs3,Fuchs2}).  Extensions to higher dimensions are certainly possible\cite{Inamura2024arXiv}, yet progress is limited by the lack of reliable methods for identifying the low-energy theory.

In this work, we study the phase diagram of a specific 1D lattice model whose symmetry is described by the Ising fusion category $\mathcal{C}_{\rm Ising}$ (the fusion category of Verlinde lines of the critical Ising theory; see Sec.~\ref{sec:model} for the definitions of $\mathcal{C}_{\rm Ising}$ and the model). This model belongs to a family of models introduced in Ref.~\cite{ning_building_2023} as an extension of the anyon chains for realizing edge theories of two-dimensional symmetry-protected and symmetry-enriched topological phases\cite{ChengGuLiuWen-SPT, SET-StationQ}. Several members of the family were shown in Ref.~\cite{ning_building_2023} to host critical phases. In the present work, we focus on the member with $\mathcal{C}_{\rm Ising}$ symmetry and perform  a detailed analytical and numerical study on the phases realized in the model, including both symmetry-breaking and critical phases. 

\begin{figure}
    \centering
    \includegraphics[width=1\linewidth]{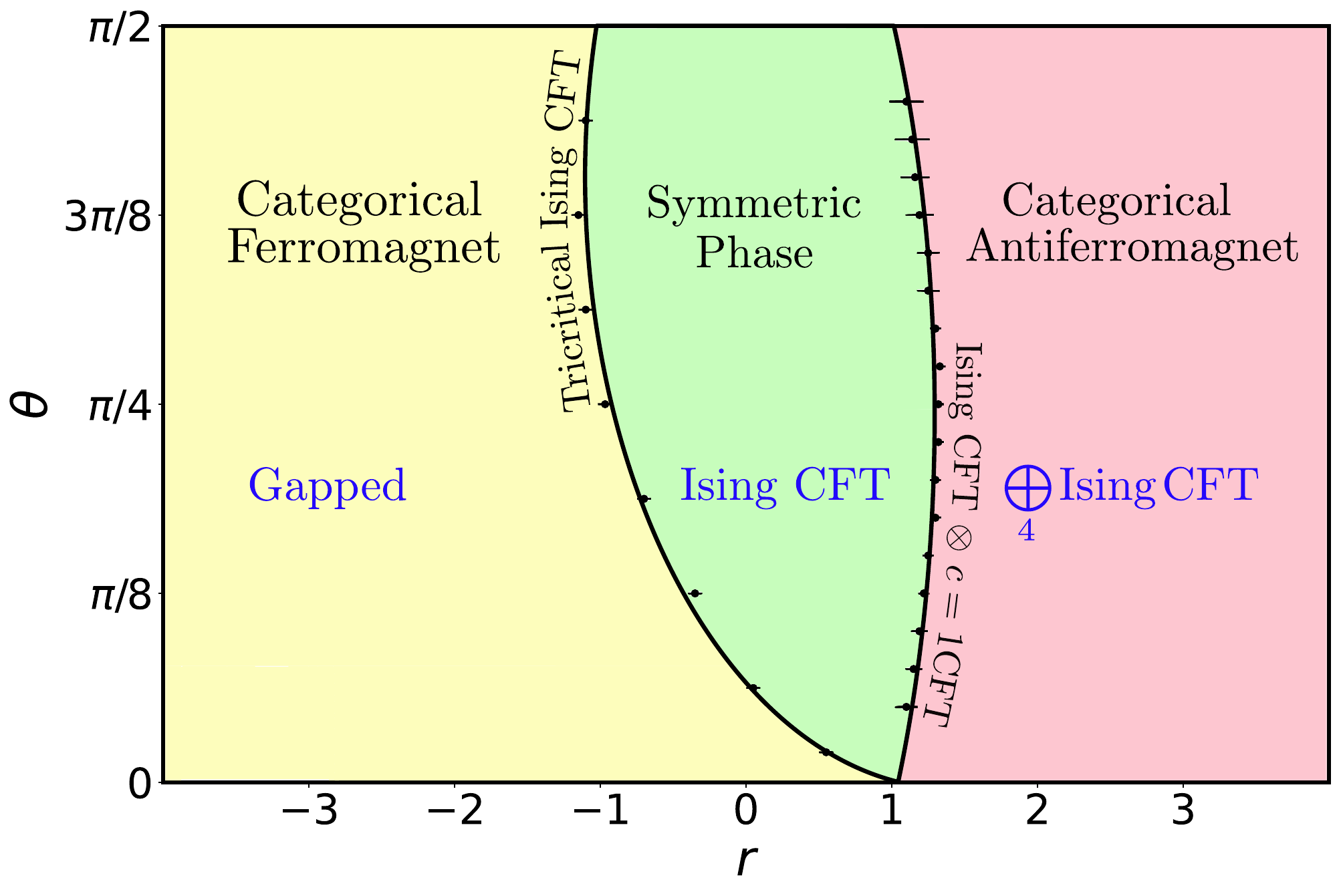}
    \caption{Phase diagram of the Ising fusion category lattice model. We identify three phases: a gapless symmetric phase (green) whose low-energy physics is described by the Ising CFT; a gapped phase with $\mathcal{C}_{\rm Ising}$ completely broken and a threefold ground state degeneracy, termed as the \emph{categorical ferromagnetic} (CatFM) phase (yellow); and a \emph{categorical anti-ferromagnetic} (CatAFM) phase (red), where lattice translation and the non-invertible element of $\mathcal{C}_{\rm Ising}$ are broken and the ground state is fourfold degenerate.  The categorical anti-ferromagnetic phase is gapless and its low-energy physics is described by the direct sum (not direct product) of four Ising CFTs.  The transition between the CatFM  and symmetric phases is characterized by the tricritical Ising CFT. The transition between the symmetric and CatAFM phases is less understood, and we conjecture that it is an Ising CFT stacked with a $c=1$ CFT (i.e., Luttinger liquid). The data points on the CatFM-Symmetric transition line are from the loop-TNR numerical calculations\cite{yang_loop_2017}, and those data points on the symmetric-CatAFM transition line are obtained from DMRG calculations\cite{fishman_itensor_2022}.}
    \label{fig:phasediagram}
\end{figure}

By tuning the two parameters $r$ and $\theta$ in the  model, we identify three phases in the phase diagram (see Fig.~\ref{fig:phasediagram}): (i) a symmetric gapless phase whose low-energy physics is the Ising conformal field theory (CFT); (ii) a threefold degenerate gapped $\mathcal{C}_{\rm Ising}$-breaking phase, which we dub the \emph{categorical ferromagnetic} (CatFM) phase; and (iii) a fourfold degenerate gapless symmetry-breaking phase, which we dub the \emph{categorical antiferromagnetic} (CatAFM) phase, where both lattice translation and the non-invertible element of $\mathcal{C}_{\rm Ising}$ are broken. Among these phases, the most interesting is the CatAFM phase, which highlights an important difference between conventional and categorical antiferromagnets. Antiferromagnetic states are associated with simultaneous breaking of lattice and internal symmetries. They may be viewed as arrays of ferromagnetic domains separated by domain walls. When the broken internal symmetry is non-invertible, the associated domain walls have quantum dimension greater than 1, giving rise to an extensive Hilbert space whose dimension grows exponentially with the domain wall number. Appearance of this infinitely large Hilbert space (in the thermodynamic limit) is a unique hallmark of the simultaneous breaking both non-invertible and lattice translation symmetries. In our model, the antiferromagnetic states finally stabilize into a gapless phase described by a direct sum (rather than a direct product) of four Ising CFTs.  

Transitions between the three phases are also studied numerically. The transition between the symmetric and CatFM phase is described by the $c=7/10$ tricritical Ising CFT. The transition between the symmetric and CatAFM phase is particularly interesting, which occurs in the presence of a background Ising CFT. Our numerical results suggest that it is continuous and described by a  conformal field theory with central charge $c=3/2$, which is a stack of the Ising CFT and the $c=1$ Luttinger liquid.

The rest of the paper is organized as follows. In Sec.~\ref{sec:model}, we introduce the model and review the basics of the $\mathcal{C}_{\rm Ising}$ symmetry. In Sec.~\ref{sec:phase}, we present detailed analytical and numerical results for the three phases of the Ising fusion category lattice model. Numerical results on the phase transitions are given in Sec.~\ref{sec:transition}. We conclude in Sec.~\ref{sec:conclusion}. The appendices contain several technical discussions. In particular, Appendix~\ref{app:perturbation} presents a perturbative study of the CatFM and CatAFM phases, while Appendix~\ref{app:entropy_gap2} discusses an interesting entanglement entropy gap that appears in our study.

\section{Model}
\label{sec:model}

In this section, we present the model, including the structure of the Hilbert space and its symmetry. This model is a member of the family of models introduced in Ref.~\cite{ning_building_2023}, which generalizes the anyon chain models\cite{feiguin_interacting_2007}, using the so-called $G$-graded fusion category. An earlier version of this model was proposed in Ref.~\cite{jones2021majorana-spt} as a lattice edge theory of 2D Majorana fermion symmetry-protected topological phases.

\subsection{Hilbert space}
\label{sec:Hilbert}

The Hilbert space of the model is built from the fusion spaces of the Ising fusion category, which we denote as $\mathcal{C}_{\rm Ising}$. Unlike conventional spin models, such a Hilbert space does not have a tensor product structure.

To begin, we introduce the Ising fusion category $\mathcal{C}_{\rm Ising}$. A fusion category contains three pieces of data: (i) a finite set of simple objects $\{\mathbb{1},a,b,\dots\}$, where $\mathbb{1}$ is the vacuum anyon; (ii) fusion rules between the simple objects, $a\times b = \sum_c N_{ab}^c c$, where  $N_{ab}^c$ is called the fusion multiplicity and is an integer; and (iii) $F$-symbols that characterize the associative properties of the category.  $\mathcal{C}_{\rm Ising}$ contains three simple objects: the vacuum anyon $\mathbb{1}$, a fermion $\psi$, and an anyon $\sigma$ of quantum dimension $d_\sigma = \sqrt{2}$.\footnote{While the braiding data is irrelevant to this work, we still occasionally refer to the simple objects $\psi$ and $\sigma$ as ``anyons''. Indeed, braiding data can be well defined in $\mathcal{C}_{\rm Ising}$.} The fusion rules are
\begin{align}
\mathbb{1} \times x & = x\times \mathbb{1} = x, \quad \forall x; \nonumber \\
\psi\times\psi & = \mathbb{1},\quad \psi\times\sigma  = \sigma\times\psi = \sigma; \nonumber \\
\sigma\times\sigma & = \mathbb{1} + \psi.
\label{eq:1}
\end{align}
A fusion process is represented by a trivalent vertex, with two of the legs labeled by the fusing objects and the third leg by the fusion outcome. Alternatively, a trivalent vertex may also represent a splitting process, with an incoming object splitting into two outcomes. This is why in general fusion categories, the legs carry an orientation (see, e.g., Ref. \cite{simon2023topological}, for an introduction to graphical calculus of fusion categories). However, since all simple objects in $\mathcal{C}_{\rm Ising}$ are self-dual, no orientation is needed on the legs.  For example, the graph
\begin{equation*}
\tree{\sigma}{\sigma}{\sigma}{\sigma}{x}
\end{equation*}
is a fusion diagram of four external legs and one internal leg. With the fusion rules in \eqref{eq:1}, either $x=\mathbb{1}$ or $\psi$ is valid. Each valid fusion diagram represents a physical state. The states associated with all valid fusion diagrams span a Hilbert space, called the fusion space. 

The $F$-symbols (or matrices) are associated with basis transformations of a fusion space. Graphically, it is defined by
\begin{align}
\tree{a}{b}{c}{d}{e} = \sum_f\left(F^{abc}_d\right)_e^f\treeb{a}{b}{c}{d}{f}
\end{align}
The matrix $F^{abc}_d$ is unitary. Specifically, for $\mathcal{C}_{\rm Ising}$, $F$-matrices can be chosen to be real. The non-trivial $F$-matrices are
\begin{align}
    \left(F^{\psi\sigma\psi}_{\sigma}\right)^\sigma_\sigma & = \left(F_{\psi}^{\sigma\psi\sigma}\right)_\sigma^\sigma = -1\nonumber\\
    F^{\sigma\sigma\sigma}_{\sigma} & = \frac{\kappa}{\sqrt{2}}
    \left(
    \begin{matrix}
    1 & 1 \\
    1 & -1 
    \end{matrix}
    \right)
    \label{eq:F-symbols}
\end{align}
All other $F$-matrices are the identity matrix. Note that there are two inequivalent Ising fusion categories associated with $\kappa = +1$ and $-1$, respectively. In this work, we only consider $\kappa =1$, and the notation $\mathcal{C}_{\rm Ising}$ always refers to the Ising fusion category with $\kappa =1$. The model associated with $\kappa = -1$  is numerically more challenging, and we leave it for future work.

Next, we build the Hilbert space of the model.  Consider the fusion diagram
\begin{equation}
\begin{tikzpicture}[baseline={(current bounding box.center)}]
    \draw (0,0)--(6,0);
    \foreach \x in {1,2,3,4,5}:
        \draw (\x,0)--(\x,1);  
    \node at (1,1.2){$a_{i-2}$};
    \node at (2,1.2){$a_{i-1}$};
    \node at (3,1.2){$a_{i}$};
    \node at (4,1.2){$a_{i+1}$};
    \node at (5,1.2){$a_{i+2}$};
    \node at (0.5,-0.2){$x_{i-3}$};
    \node at (1.5,-0.2){$x_{i-2}$};
    \node at (2.5,-0.2){$x_{i-1}$};
    \node at (3.5,-0.2){$x_{i}$};
    \node at (4.5,-0.2){$x_{i+1}$};
    \node at (5.5,-0.2){$x_{i+2}$};
    \node at (0,0.5){$\cdots$};
    \node at (6,0.5){$\cdots$};
\end{tikzpicture}
\label{eq:lattice}
\end{equation}
with periodic boundary conditions imposed in the horizontal direction. The labels $a_i$ and $x_i$ run through all simple objects in $\mathcal{C}_{\rm Ising} = \{\mathbb{1}, \psi,\sigma\}$, with the fusion rules satisfied at every vertex. The index $i$ takes values $i=1,2,\dots, L$. We view this fusion diagram as a one-dimensional lattice of length $L$. Every valid configuration, labeled by $\{a_i\}$ and $\{x_i\}$, is a basis vector in the Hilbert space.

However, in addition to the constraint from fusion rules, we will impose another constraint on the choice of $\{a_i\}$ and $\{x_i\}$. To describe this, we need the $\mathbb{Z}_2$ grading structure of $\mathcal{C}_{\rm Ising}$. Consider the decomposition 
\begin{equation}
    \mathcal{C}_{\rm Ising} = \mathcal{C}_+ \oplus \mathcal{C}_-
    \label{eq:decomposition}
\end{equation}
where $\mathcal{C}_+ = \{1,\psi\}$ and $\mathcal{C}_- = \{\sigma\}$. We notice that the fusion rules respect the decomposition: for $a \in \mathcal{C}_\mu$ and $b\in\mathcal{C}_\nu$, the fusion outcomes in
$a\times b$ must belong to  $\mathcal{C}_{\mu\nu}$,  where $\mu\nu$ is the product of $\mu$ and $\nu$ by viewing ``$\pm$'' as ``$\pm1$''. A decomposition \eqref{eq:decomposition} that respects the fusion rules is called a \emph{$\mathbb{Z}_2$ grading structure}. Let
\begin{align}
    \phi(x) = \left\{
    \begin{array}{cc}
    +,  & \text{if } x\in\mathcal{C}_+ \\[3pt]
    -,  & \text{if } x\in\mathcal{C}_- 
    \end{array}
    \right.
\end{align}
Then, at every vertex in the lattice \eqref{eq:lattice}, the fusion rules give rise to
\begin{align}
\left\{
\begin{array}{ll}
 a_i\in \mathcal{C}_+, & \quad \text{if }\phi(x_i)=\phi(x_{i+1}),\\[3pt] a_i\in\mathcal{C}_-, & \quad \text{if }\phi(x_i)\neq \phi(x_{i+1}).
\end{array}
\right.
\end{align}
Now we impose the constraint
\begin{align}
    a_i=\mathbb{1},\quad \text{if } a_i\in \mathcal{C}_+\label{eq:constraint}
\end{align}
That is, we impose $a_i\neq \psi$ at any $i$. Accordingly, $a_i$ is constrained to take values in the subset  $\{\mathbb{1}, \sigma\}$. 

With the constraint \eqref{eq:constraint},  $\{a_i\}$ are completely determined by $\{x_i\}$. This leads to the simplified notation for the states in the Hilbert space:
\begin{equation}
\label{eq:3}
\Bigg|\tree{x_{i-1}}{a_i}{a_{i+1}}{x_{i+1}}{x_i} \Bigg\rangle \equiv |x_{i-1}x_ix_{i+1}\rangle
\end{equation}
It is easy to see that any state in the Hilbert space must be of the form
\begin{align}
    |\cdots \sigma\sigma \mu\mu\mu \sigma\sigma\sigma \mu'\mu'\sigma\sigma\cdots\rangle
    \label{eq:basis}
\end{align}
where $\mu,\mu' = \mathbb{1}$ or $\psi$. Note that all $\mu$'s must be \emph{the same} between two $\sigma$ segments, which is precisely due to the condition $a_i\neq \psi$. Since $\mu\in\mathcal{C}_+$ and $\sigma\in \mathcal{C}_-$, we view a $\mu$-string segment  as a ``$+$'' domain and a $\sigma$-string segment as a ``$-$'' domain. Accordingly, the state in \eqref{eq:basis} is associated with the following domain picture:
\begin{align}
\label{eq:domain}
        |\cdots --+++---++--\cdots\rangle
\end{align}
While  a ``$-$'' domain always means a $\sigma$-string, a ``$+$'' domain can be either a string ``$\cdots\mathbb{1}\mathbb{1}\mathbb{1}\cdots$'' or a string ``$\cdots\psi\psi\psi\cdots$''. The domain picture resembles that of the conventional spin-$1/2$ chain. However, the Hilbert space of our model generalizes the spin-$1/2$ chain by further splitting a ``$+$'' domain into two states, ``$\cdots\mathbb{1}\mathbb{1}\mathbb{1}\cdots$'' and ``$\cdots\psi\psi\psi\cdots$''.

\subsection{Hamiltonian}
\label{sec:hamiltonian}
The Hamiltonian of the model is of the form 
\begin{align}
H = -\sum_i H_i,
\label{eq:model}
\end{align}
where $H_i$ is a local operator that acts on three lattice sites. The local interaction $H_i$ consists of two pieces 
\begin{align}
    H_i=H_i^{\rm dw} + H_i^{\rm flip}
\label{eq:Hi}
\end{align}
with
\begin{align}\label{eq:Hr}
    H_i^{\rm dw}\ket{\mu\mu\mu} &= 0 \nonumber\\
    H_i^{\rm dw}\ket{\mu\mu\sigma} &= r\ket{\mu\mu\sigma}\nonumber\\
    H_i^{\rm dw}\ket{\mu\sigma\nu} &=0 \nonumber\\
    H_i^{\rm dw}\ket{\sigma\mu\mu} &= r\ket{\sigma\mu\mu}\nonumber\\
    H_i^{\rm dw}\ket{\mu\sigma\sigma} &= r\ket{\mu\sigma\sigma} \nonumber\\
    H_i^{\rm dw}\ket{\sigma\mu\sigma} &= 0\nonumber\\
    H_i^{\rm dw}\ket{\sigma\sigma\mu} &= r\ket{\sigma\sigma\mu} \nonumber\\
    H_i^{\rm dw}\ket{\sigma\sigma\sigma}&=0
\end{align}
and
\begin{align}\label{eq:Hth}
    H_i^{\rm flip}\ket{\mu\mu\mu} &= \cos\theta\ket{\mu\sigma\mu} \nonumber\\
    H_i^{\rm flip}\ket{\mu\mu\sigma} &= \sin\theta\ket{\mu\sigma\sigma}\nonumber\\
    H_i^{\rm flip}\ket{\mu\sigma\nu} &= \delta_{\mu\nu}\cos\theta\ket{\mu\mu\mu}\nonumber\\
    H_i^{\rm flip}\ket{\sigma\mu\mu} &= \sin\theta\ket{\sigma\sigma\mu}\nonumber\\
    H_i^{\rm flip}\ket{\mu\sigma\sigma} &=  \sin\theta\ket{\mu\mu\sigma}\nonumber\\
    H_i^{\rm flip}\ket{\sigma\mu\sigma} &= \frac{\cos\theta}{\sqrt{2}}\ket{\sigma\sigma\sigma}\nonumber\\
    H_i^{\rm flip}\ket{\sigma\sigma\mu} &=  \sin\theta\ket{\sigma\mu\mu}\nonumber\\
    H_i^{\rm flip}\ket{\sigma\sigma\sigma}&= \frac{\cos\theta}{\sqrt{2}}(\ket{\sigma\mathbb{1}\sigma} + \ket{\sigma\psi\sigma})
\end{align}
where $\mu=\mathbb{1}$ or $\psi$. Since $H_i$ only acts on three sites, the values of $x_i$ outside the relevant three sites are not shown. The three-site configurations listed above exhaust all valid possibilities. The term $H_i^{\rm dw}$ describes an interaction between ``domain walls'' on the two sides of the $i$-th domain: the energy is $-r$, if the domain walls on the two sides are different; the energy is zero, otherwise. This interaction favors magnetically ordered states, either ferromagnetism or antiferromagnetism, depending on the sign of $r$. We will elaborate further on this in Sec.~\ref{sec:phase}. The term $H_i^{\rm flip}$ corresponds to an interaction that flips domains, thereby disordering the domains. In the special case $r=0$, our model reduces to the model of Ref.~\cite{jones2021majorana-spt}.  We do not discuss the construction of the model here, but note that it was designed to respect the categorical symmetry $\mathcal{C}_{\rm Ising}$ (symmetry of the model is discussed below). Interested readers may consult Ref.~\onlinecite{ning_building_2023} for details. 

The ranges of the parameters are $r \in (-\infty, \infty)$ and $\theta\in[0,2\pi]$. However, in Appendix \ref{app:para_range}, we show that it is sufficient to consider $\theta\in [0,\pi/2]$. For other values of $\theta$, the spectrum of $H$ is related to that of $[0,\pi/2]$ by appropriate transformations.

\subsection{$\mathcal{C}_{\rm Ising}$ symmetry}
\label{sec:symmetry}

The category $\mathcal{C}_{\rm Ising}$ is extensively used above to construct the Hilbert space of the model. On the other hand, $\mathcal{C}_{\rm Ising}$ also characterizes the symmetry of the model in \eqref{eq:model}-\eqref{eq:Hth}. Each simple object in $\mathcal{C_{\rm Ising}}=\{1,\psi,\sigma\}$ is associated with a symmetry operator that commutes with the Hamiltonian. That is,
\begin{align}
    U(y)H = H U(y),
\end{align}
where $y=\mathbb{1}, \psi$ and $\sigma$. The identity anyon is associated with the identity operator, i.e., $U(\mathbb{1})=I$. The symmetry operator $U(y)$ is obtained by fusing a line of the anyon $y$ onto a fusion state. Graphically, 
\begin{widetext}
\begin{align}
&  U(y) \Bigg|\tree{x_{i-1}}{a_i}{a_{i+1}}{x_{i+1}}{x_i}\Bigg\rangle  = \Bigg|\begin{tikzpicture}[xscale=.8, yscale=.8, nodes={execute at begin node=$, execute at end node=$}, baseline={(current bounding box.center)}]
\draw (-3,0)node[above]{x_{i-1}}--(0,0)node[above]{x_{i+1}};
\draw (-2,0)--(-2,1)node[above]{a_{i}};
\draw (-1,0)--(-1,1)node[above]{a_{i+1}};
\node[above] at (-1.5,0){x_i};
\draw [thick](-3,-0.3)--(0,-0.3);
\node[below] at (-1.5,-0.5){y};
\end{tikzpicture}
\Bigg\rangle  =\sum_{\{x_i'\}} \prod_{i=1}^L \left(F_{x'_{i+1}}^{y,x_i,a_{i+1}}\right)^{ x_i'}_{x_{i+1}}\Bigg|\tree{x'_{i-1}}{a_i}{a_{i+1}}{x'_{i+1}}{x'_i}\Bigg\rangle
\label{eq:symmetry}
\end{align}
\end{widetext}
The $F$-symbols in the last expression come from the diagrammatic manipulations of absorbing the line $y$ into the fusion tree. One may refer to Ref.~\onlinecite{ning_building_2023} for details of this calculation.

Specifically, the action of $U(\psi)$ is to flip between the $\mathbb{1}$ and $\psi$ anyons. In the short-hand notation \eqref{eq:basis}, we have
\begin{equation}\label{eq:10}
U(\psi)\ket{\cdots\sigma\mu\mu\sigma\sigma\mu'\mu'\cdots} = \ket{\cdots\sigma\bar{\mu}\bar{\mu}\sigma\sigma\bar{\mu'}\bar{\mu'}\cdots}
\end{equation}
where $\mu,\mu'=\mathbb{1},\psi$. The bar means $\bar{\mathbb{1}}=\psi$ and $\bar{\psi} = \mathbb{1}$, denoting that $U(\psi)$ flips between $\mathbb{1}$ and $\psi$. Note that the nontrivial $F$-symbols in \eqref{eq:F-symbols} do not enter this expression. 

The action of $U(\sigma)$ is slightly more complicated, and the non-trivial $F$-symbols do enter the expression of $U(\sigma)$. As discussed in Sec.~\ref{sec:Hilbert}, a state consists of segments of $\sigma$-strings and $\mu$-strings.  Periodic boundary conditions enforce the number of $\sigma$-segments and $\mu$-segments to be equal (the lengths of the individual segments may vary), except the special cases of the uniform $\sigma$-string state and the uniform $\mu$-string state which will be considered separately. Let this number be $n$. With the fusion rules of $\mathcal{C}_{\rm Ising}$, we have
\begin{align}
U(\sigma)\ket{\cdots\sigma\sigma\mu_k\mu_k\sigma\mu_{k+1}\cdots} \rightarrow \ket{\cdots\mu_k'\mu_k'\sigma\sigma\mu_{k+1}'\sigma\cdots}
\label{eq:sigma_action}
\end{align}
where $\mu_k,\mu_k'=\mathbb{1},\psi$, and the right side is understood as a linear superposition of states with different values of $\{\mu_k'\}$. We see that the $\sigma$ and $\mu$ segments are exchanged under $U(\sigma)$. Inserting the $F$-symbols from \eqref{eq:F-symbols} into expression \eqref{eq:symmetry},  we find the matrix element of the symmetry operator $U(\sigma)$ 
\begin{equation}\label{eq:11}
    \bra{\{\mu_k'\}}U(\sigma)\ket{\{\mu_k\}} = \left(\frac{1}{\sqrt{2}}\right)^{n}\prod_{k=1}^n(-1)^{(\mu_k + \mu_{k-1})\mu_k'}
\end{equation}
where $\ket{\{\mu_k\}}$ and $\ket{\{\mu_k'\}}$ are short-hand notations for the states in \eqref{eq:sigma_action}. For the special uniform states, one may directly check that
\begin{align}
    U(\sigma)|\mathbb{1}\rangle & = |\sigma\rangle \nonumber\\
    U(\sigma)|\mathbb{\psi}\rangle & = |\sigma\rangle \nonumber\\
    U(\sigma)|\mathbb{\sigma}\rangle & = |\mathbb{1}\rangle + |\psi\rangle
    \label{eq:FM}
\end{align}
Note that these expressions are simply \eqref{eq:11} with $n=0$.

The symmetry operators $I$, $U(\psi)$ and $U(\sigma)$ satisfy the following algebra:
\begin{subequations}
\label{eq:9}
\begin{align}
         &U(\psi)^2 = I \label{eq:9.1}\\
         &U(\psi) U(\sigma) = U(\sigma) U(\psi) = U(\sigma) \label{eq:9.2}\\
         &U(\sigma)^2 = I + U(\psi)\label{eq:9.3}
\end{align}
\end{subequations}
Also, all operators are real, $U(y)^\dag = U(y)$\footnote{This property is due to the particular gauge choice of $F$ symbols in \eqref{eq:F-symbols}. }. One may explicitly verify this algebra from the expressions of $U(\psi)$ and $U(\sigma)$. However, there is an easier way to see it. Consider the product of two symmetry operators, $U(y_1)U(y_2)$. Graphically, it means fusing the lines of $y_2$ and $y_1$ successively into the fusion tree. Since fusion is associative, we can also fuse the two lines of $y_1$ and $y_2$ first. By doing so, the graphical calculus\cite{simon2023topological} would immediately give
\begin{align}
    U(y_1)U(y_2) = \sum_{y} N^y_{y_1y_2} U(y)
\end{align}
Applying the fusion rules of $\mathcal{C}_{\rm Ising}$ to this relation, we obtain the equations in \eqref{eq:9}.  

The symmetry algebra \eqref{eq:9} is commutative. It is easy to find the irreducible representations of the algebra. There are three irreducible representations, all of which are one-dimensional. Let the eigenvalues of $I$, $U(\psi)$ and $U(\sigma)$ be $\lambda_\mathbb{1}$, $\lambda_{\psi}$ and $\lambda_\sigma$, respectively. Then,  the three irreducible representations correspond to
\begin{align}
   \Vec{\lambda}\equiv (\lambda_{\mathbb{1}}, \lambda_\psi, \lambda_\sigma)  = 
   \left\{
   \begin{array}{l}
    (1, 1, \sqrt{2}),\\[3pt]
    (1,1,-\sqrt{2}), \\[3pt]
    (1, -1, 0).
   \end{array}
   \right.
\end{align}
We will refer to $\lambda_y$ the ``charge''  as of the symmetry $U(y)$. It is obvious that the symmetry $U(\sigma)$ is \emph{non-invertible} as it has a zero eigenvalue.\footnote{Having a zero eigenvalue is not a necessary condition for a symmetry to be called ``non-invertible''. For example, the non-invertible symmetry $Y_\tau$ in the Fibonacci golden chain is invertible as a matrix.\cite{buican_anyonic_2017, feiguin_interacting_2007}} The representation corresponding to $\vec{\lambda} = (1, 1, \sqrt{2})$ is understood as the identity representation. For the other two representations, either $U(\psi)$ or $U(\sigma)$ is associated with a minus sign. 

We remark that the algebra \eqref{eq:9} being commutative is due to the fact that $\mathcal{C}_{\rm Ising}$ is actually braided. The fusion rules of any braided fusion category are commutative. In this case, the symmetry algebra (identical to the fusion algebra) is called the Verlinde algebra\cite{VERLINDE1988360}.  Irreducible representations of the Verlinde algebra are all one-dimensional, and the total number is equal to the number of anyons in the braided fusion category. The eigenvalues $\{\lambda_a\}$, called \emph{fusion characters}, are the common eigenvalues of the fusion matrices $\{N_a\}$, where $(N_a)_{bc} = N_{ab}^c$ (see, e.g., Appendix E of Ref.~\onlinecite{kitaev_anyons_2006} for a review). In this sense, as symmetries, braided fusion categories generalize finite Abelian groups.

\section{Phases}
\begin{figure*}[t]
    \centering
    \includegraphics[width=1.0\linewidth]{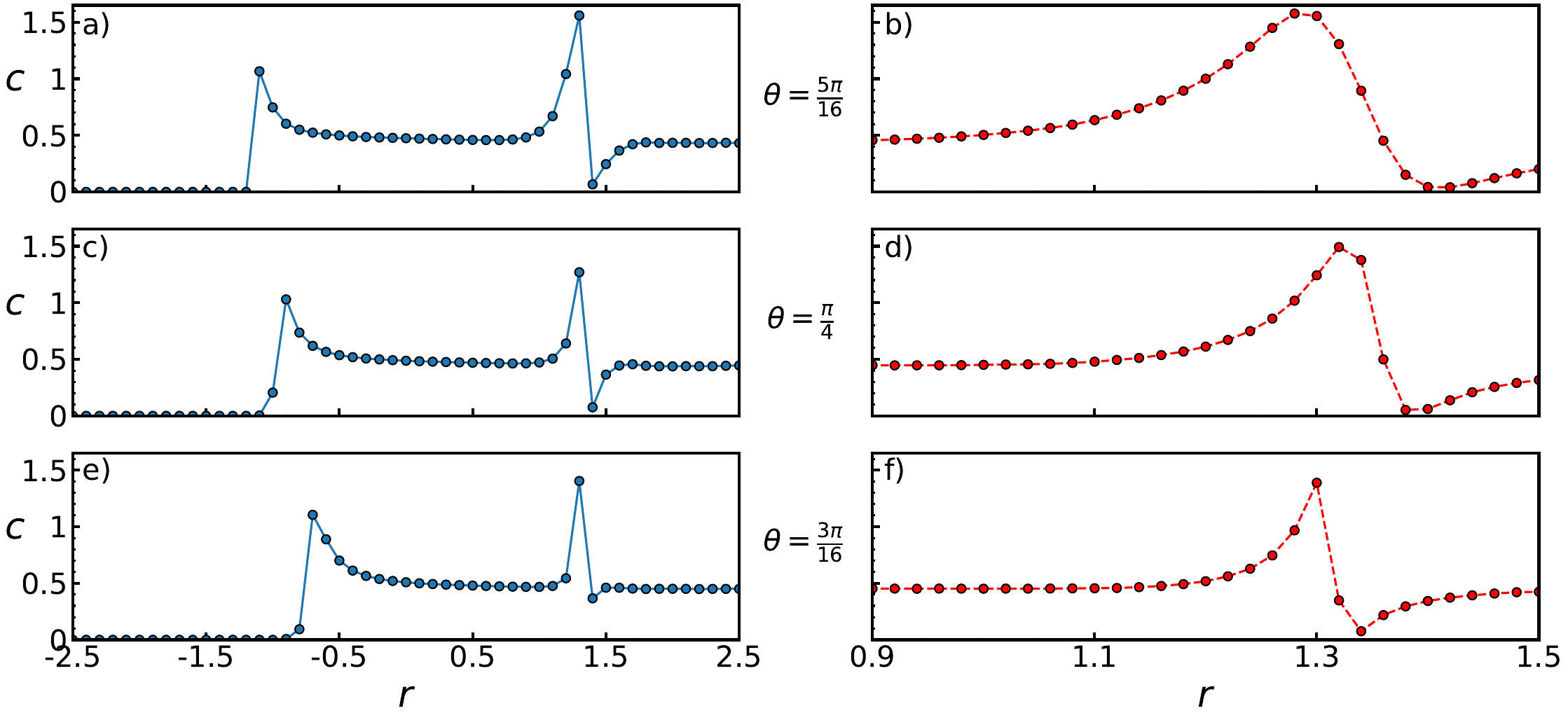}
    \caption{Central charge $c$ extracted from DMRG calculations of entanglement entropy, plotted against $r$ along three horizontal cuts of the phase diagram at $\theta=5\pi/16, \pi/4$ and $3\pi/16$. Panels (a), (c) and (e) reveal three distinct regimes with $c = 0$, $c\approx 0.5$ and $c\approx 0.5$, corresponding to the gapped CatFM phase, gapless symmetric phase, and gapless CatAFM phase, respectively. These regimes are separated by spikes in the central charge, which mark phase transitions. We note, however, that central charge $c$ near the transition points are not quantitatively reliable due to strong finite-size effects. The calculations were done for a chain of length $L = 400$ with a maximum bond dimension $\chi = 250$. Panels (b), (d) and (f) show zoomed-in views of the central charge near the symmetric-CatAFM transition.}
    \label{fig:extended_dmrg}
\end{figure*}
\label{sec:phase}

\begin{figure*}[t]
    \centering
    \includegraphics[width = 1.0\linewidth]{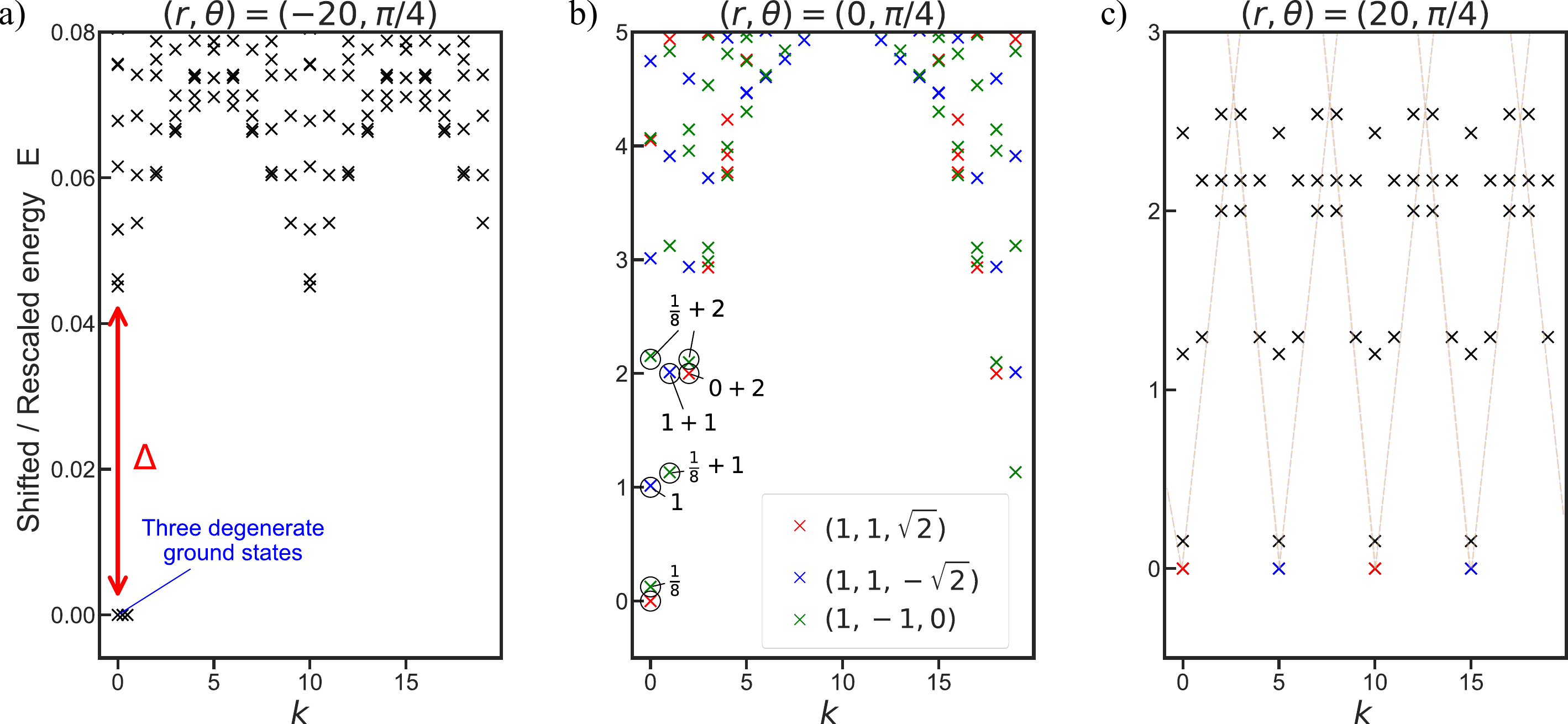}
    \caption{Low energy spectra obtained via exact diagonalization in the different phases. All calculations were performed for a chain of length $L = 20$. (a) Spectrum at $(r,\theta)=(-20,\pi/4)$ in the CatFM phase. A gapped spectrum with a threefold ground-state degeneracy is observed (the ground states have been slightly shifted along the momentum axis for clarity). Note that $\Delta$ does not represent the true gap in the thermodynamic limit, as it is strongly affected by strong finite-size effects (see discussions in the main text). (b) Spectrum at $(r,\theta)=(0,\pi/4)$ in the symmetric critical phase. The conformal tower of the Ising CFT  is observed (the energies are rescaled such that the lowest energy state with $k=2$ ,corresponding to the stress-energy tensor, has energy 2).  States are colored according to their $\mathcal{C}_{\rm Ising}$ charges (see legend). (c) Spectrum at $(r,\theta)=(20,\pi/4)$ in the CatAFM phase. Four degenerate conformal towers of the Ising CFT are observed, highlighted with orange dashed lines (again energies are appropriately rescaled). The ground states are colored according to their  $\mathcal{C}_{\rm Ising}$ symmetry charges. Specifically, the ground states carry charges $\lambda=(1,1,\pm \sqrt{2})$, indicating the spontaneous breaking of $U(\sigma)$, whereas $U(\psi)$ remains unbroken. }
    \label{fig:ED}
\end{figure*}

In this section, we study the phases appearing in the model \eqref{eq:model}.  We will discuss the phase transitions in Sec.~\ref{sec:transition}.  We combine a variety of numerical methods (exact diagonalization, density matrix renormalization group, and tensor renormalization group methods) and analytical methods (perturbation analysis) for the study.

\subsection{Phase diagram}

Figure \ref{fig:phasediagram} shows the phase diagram of the model in the $(r, \theta)$ plane, mapped out through our numerical calculations. We find three distinct phases:
\begin{enumerate}
    \item[(i)] A symmetric phase  that is gapless. The low-energy physics is described by the critical Ising conformal field theory.
    
    \item[(ii)] A gapped phase in which $\mathcal{C}_{\rm Ising}$ is spontaneously broken. We call this phase a \emph{Categorical Ferromagnet} (CatFM). It has a three-fold ground state degeneracy.

    \item[(iii)] A gapless phase in which  lattice translation and the non-invertible element $U(\sigma)$ of $\mathcal{C}_{\rm Ising}$ are spontaneously broken. We call this phase a \emph{Categorical Antiferromagnet} (CatAFM). The low-energy physics is characterized by a direct sum of four Ising CFTs.
\end{enumerate}
Note that no gapped symmetric phase exists. Indeed, $\mathcal{C}_{\rm Ising}$ is an anomalous symmetry in the sense that it does not allow a symmetric gapped ground state \cite{ThorngrenWang1}. We will discuss each phase in detail in the following subsections.

The transition lines between these phases are observed to be continuous. We find that the Symmetric-CatFM transition is described by the tricritical Ising CFT with the central charge $c=0.7$, while the Symmetric-CatAFM transition displays a central charge $c \approx 1.5$. We conjecture that the latter is a stack of the Ising CFT and a $c=1$ compactified free boson theory (Luttinger Liquid). The special lines of $\theta=0$ and $\theta=\pi/2$ are discussed in Sec.~\ref{sec:special_lines}.

\subsection{Gapless symmetric phase}

The gapless symmetric phase is studied by combining exact diagonalization (ED) and density matrix renormalization group (DMRG)\cite{White1992PRL,xiang2023density,schollwock_density-matrix_2005,schollwock_density-matrix_2011} (implemented using the ITensor library\footnote{In order to enforce the constrained Hilbert space, we embed the Hamiltonian into a larger spin-1 local Hilbert space and then enforce the fusion rules by adding a penalty term to the Hamiltonian.}\cite{fishman_itensor_2022}). Figure \ref{fig:extended_dmrg} plots the central charge along the lines $\theta=\frac{3\pi}{16}$, $\frac{\pi}{4}$ and $\frac{5\pi}{16}$, respectively, obtained from the DMRG ground states. A central charge $c=1/2$ is observed throughout the symmetric phase, in agreement with that of the Ising CFT. Figure \ref{fig:ED} shows the low-energy spectrum at $(r,\theta) = (0,\pi/4)$ from an ED calculation,
after appropriate shifting and rescaling of the data. According to conformal field theory\cite{cardy1986operator}, the energy spectrum can be organized into the formula
\begin{equation}\label{eq:14}
    E = E_1L + \frac{2\pi v}{L}\left(-\frac{c}{12} + h + \bar{h}\right)
\end{equation}
where $h$ and $\bar{h}$ are scaling dimensions, $L$ is the length of the lattice under periodic boundary conditions, and $v$ is the velocity. Every eigenstate is associated with a pair of scaling dimensions $(h,\bar{h})$.  For the Ising CFT, there are three primary states corresponding to $(h, \bar{h})=(0,0)$, $(\frac{1}{16},\frac{1}{16})$ and $(\frac{1}{2}, \frac{1}{2})$, respectively. All other states---descendants of the primaries---have scaling dimensions that differ by some integers. We find that the ED spectrum of our model agrees well with Eq.~\ref{eq:14}, up to a reasonable finite size effect (see the caption of Fig.~\ref{fig:ED}). 

Notably, exact diagonalization allows us to investigate the symmetry properties of the eigenstates as well. We find that the ground state --- the $(0,0)$ primary state --- lies in the $\vec{\lambda}=(1, 1, \sqrt{2})$ symmetry sector, i.e., the trivial representation of $\mathcal{C}_{\rm Ising}$.  For the primary states  $(\frac{1}{16},\frac{1}{16})$ and $(\frac{1}{2}, \frac{1}{2})$ (and their descendants),  we find that they lie in the $\vec{\lambda}=(1, -1,0)$ and $\vec{\lambda}=(1,1, -\sqrt{2})$ sectors, respectively, i.e., nontrivial representations of $\mathcal{C}_{\rm Ising}$. These symmetry properties agree with the CFT expectation. A particularly useful result in CFT is the state-operator correspondence: there is a one-to-one correspondence between eigenstates and local operators.  With this correspondence, we conclude that there are no relevant operators (those with $h+\bar{h}<2$) that are symmetric under both $\mathcal{C}_{\rm Ising}$ and translation. This explains the perturbative stability of the symmetric gapless phase, i.e., it extends to a finite region in the phase diagram. This behavior has been widely observed in lattice models with categorical symmetry \cite{feiguin_interacting_2007} and is also in agreement with the arguments of \cite{buican_anyonic_2017} regarding the symmetry protection of criticality in anyon chains. 

\subsection{Categorical ferromagnetic phase}

The yellow part of the phase diagram (Fig.~\ref{fig:phasediagram}) is a gapped phase, which is well established by the vanishing central charge from our DMRG calculations, as shown in Fig.~\ref{fig:extended_dmrg}. It is associated with a three-fold ground-state degeneracy (see Fig.~\ref{fig:ED}a), corresponding to a full spontaneous breaking of the category $\mathcal{C}_{\rm Ising}$. Our ED calculation confirms that the three ground states are associated with the three irreducible representations of $\mathcal{C}_{\rm Ising}$, respectively. In analogy to the ferromagnetic phase in the transverse-field Ising model, we refer to this phase as a \emph{Categorical Ferromagnet} (CatFM).

To have a better understanding of the CatFM phase, we perform a perturbative analysis in the regime $r \ll -1$. We separate the Hamiltonian \eqref{eq:model} into two pieces
\begin{align}
    H = H^0 + H^1
    \label{eq:pert1}
\end{align}
where 
\begin{align}
    H^0 = -\sum_i H_i^{\rm dw}, \quad H^1 = -\sum_i H_i^{\rm flip}
    \label{eq:pert2}
\end{align}
The terms $H_i^{\rm dw}$ and $H_i^{\rm flip}$ are given in Eq. \eqref{eq:Hr} and \eqref{eq:Hth}, respectively. We take $H^0$ to be the unperturbed Hamiltonian and $H^1$ to be the perturbation. Note that $H^0$ depends only on the parameter $r$, and $H^1$ depends only on the parameter $\theta$. Below we describe the essential steps and results. The detailed perturbative calculations can be found in Appendix \ref{app:perturbation}.

Since $H^0$ is diagonal in the basis, the unperturbed ground states can be easily identified. However, the structure of the unperturbed ground states depends on whether $L$ is even or odd. For odd $L$, we find three FM-type ground states
$\ket{\cdots\mathbb{1}\mathbb{1}\mathbb{1}\cdots}$,  $\ket{\cdots\psi\psi\psi\cdots}$, $\ket{\cdots\sigma\sigma\sigma\cdots}$. The ground state energy is 0.  For the convenience of later discussions, we denote the three ground states as
\begin{align}
    \ket{\mathbb{1}}& \equiv \ket{\cdots\mathbb{1}\mathbb{1}\mathbb{1}\cdots}, \nonumber\\
    \ket{\psi} & \equiv \ket{\cdots\psi\psi\psi\cdots},\nonumber\\
    \ket{\sigma} & \equiv \ket{\cdots\sigma\sigma\sigma\cdots}.
    \label{eq:16}
\end{align}
All excited states have energy $E_{\rm ex}\ge |2r|$.

For even $L$, there are two kinds of unperturbed ground states: (i) three FM-type states in \eqref{eq:16}; (ii) many AFM-type states of the form $\ket{\cdots \sigma \mu_i\sigma\mu_{i+1} \sigma\mu_{i+2} \cdots}$, where $\mu_i=\mathbb{1}$ or $\psi$. The AFM-type states can be separated into {two sectors} that differ by translating a lattice constant. All these states have energy $E_{\rm 0} = 0$. Hence, the total degeneracy of the unperturbed ground-state manifold is 
\begin{equation}
    \mathrm{G.S.D.} = 3+ 2^{L/2+1}
\end{equation}
Note that the AFM-type states are not ground states for odd $L$. When $L$ is odd, it is unavoidable to have a segment of either $\mu\sigma\sigma\mu'$ or $\sigma\mu\mu\sigma$ embedded in the AFM states, such that the energy  becomes $|2r|$. 

We observe an apparent discrepancy between the odd and even $L$ cases. This discrepancy can be resolved by including corrections from the perturbation $H^1$. We show in Appendix \ref{app:perturbation} that the AFM ground states for even $L$ are gapped out by the second-order perturbative correction from $H^1$ (note that the first-order correction vanishes), leaving the true ground states to be the three FM states in \eqref{eq:16}. To be precise,  consider the  degenerate space of one of the two AFM sectors. It can be mapped to a spin-$1/2$ chain with length $L/2$:
\begin{align*}
    \ket{\cdots\mu_i\sigma\mu_{i+1}\sigma\mu_{i+2}\cdots} \Rightarrow \ket{\cdots\mu_i\mu_{i+1}\mu_{i+2}\cdots}
\end{align*}
It turns out that the second order correction from $H^1$ is identical to the transverse-field Ising model at the critical point. The effective Hamiltonian is 
\begin{align}
    H_{\rm eff} = \bigoplus_2 H_{\rm Ising}
\end{align}
where
\begin{align}
    H_{\rm Ising} = \frac{\cos^2\theta}{4r}\left[L + \sum_j\sigma_j^z\sigma_{j+1}^z + \sum_j\sigma_j^x\right]
\end{align}
We have two copies of $H_{\rm Ising}$ because, in second order perturbation theory, the two AFM sectors do not mix (and they do not mix with the FM-type states either). The degeneracy of the AFM-type states is then lifted, and the energy of the lowest state can be obtained from the exact solution to the transverse-field Ising model. After considering the second-order correction to the FM-type states, we find an energy gap between the AFM- and FM-type states:
\begin{equation}\label{eq:17}
    \Delta = \frac{L \cos^2\theta}{4|r|}\left(1-\frac{2}{\pi}\right)
\end{equation}
The details of the calculations can be found in Appendix \ref{app:perturbation}. Indeed, the ED spectrum in Fig.~\ref{fig:ED}a shows two conformal towers above the three degenerate ground states. However, $\Delta$ in \eqref{eq:17} is not the true energy gap in the thermodynamic limit. Since $\Delta \propto L$, the AFM conformal towers will be pushed to infinity as $L\rightarrow \infty$. It is easy to show that the thermodynamic energy gap is approximately $|2r|$ for $r\ll -1$.

Having established the three FM-type ground states, we now look at their symmetry properties. From the general expression Eq.~\eqref{eq:11}, we find that
\begin{align}\label{eq:18}
    \ket{\mathbb{1}} &\xleftrightarrow{U(\psi)} \ket{\psi}\nonumber\\
    \ket{\mathbb{1}} + \ket{\psi} & \xleftrightarrow{U(\sigma)}  \ket{\sigma} \nonumber\\
    \ket{\mathbb{1}} - \ket{\psi} & \xrightarrow{U(\sigma)}  0
\end{align}
The presence of three degenerate ground states and their transformations under the symmetry operators  justify that it is indeed a phase with $\mathcal{C}_{\rm Ising}$ spontaneously broken. It is a case of the \textit{Categorical Landau paradigm} as discussed in \cite{bhardwaj_illustrating_2024, bhardwaj2024categorical}.

\subsection{Categorical antiferromagnetic phase}

To study the CatAFM phase, we first perform a perturbative analysis in the regime $r \gg 1$. Again, we separate the Hamiltonian into $H^0+H^1$, with $H^0$ and $H^1$ given in Eq.~\ref{eq:pert2}. For $r\gg 1$, the unperturbed ground states take the form
$$
\ket{\cdots\sigma\sigma\mu_j\mu_j\sigma\sigma\mu_{j+1}\mu_{j+1}\cdots}
$$
where $\mu_i=\mathbb{1}$ or $\psi$.  They are AFM-type states with a unit cell consisting of four lattice sites, meaning the lattice translation group $\mathbb{Z}^{\rm trans}_L$ has been spontaneously broken down to $\mathbb{Z}^{\rm trans}_{L/4}$.  The above AFM states are divided into four orthogonal sectors  that are related by translation. The degeneracy of the unperturbed ground states is 
\begin{align}
    \text{G.S.D.} = 4\cdot 2^{L/4}
\end{align}
where, for simplicity, we have assumed that the length $L$ is a multiple of four. When $L$ is not a multiple of four, one may think of the model as being defined on a lattice with a translation symmetry defect. Properties of translation defects are not important for our discussions, so we will take $L$ to be a multiple of four below. 

\begin{figure}[t]
    \centering
    \includegraphics[width=1.0\linewidth]{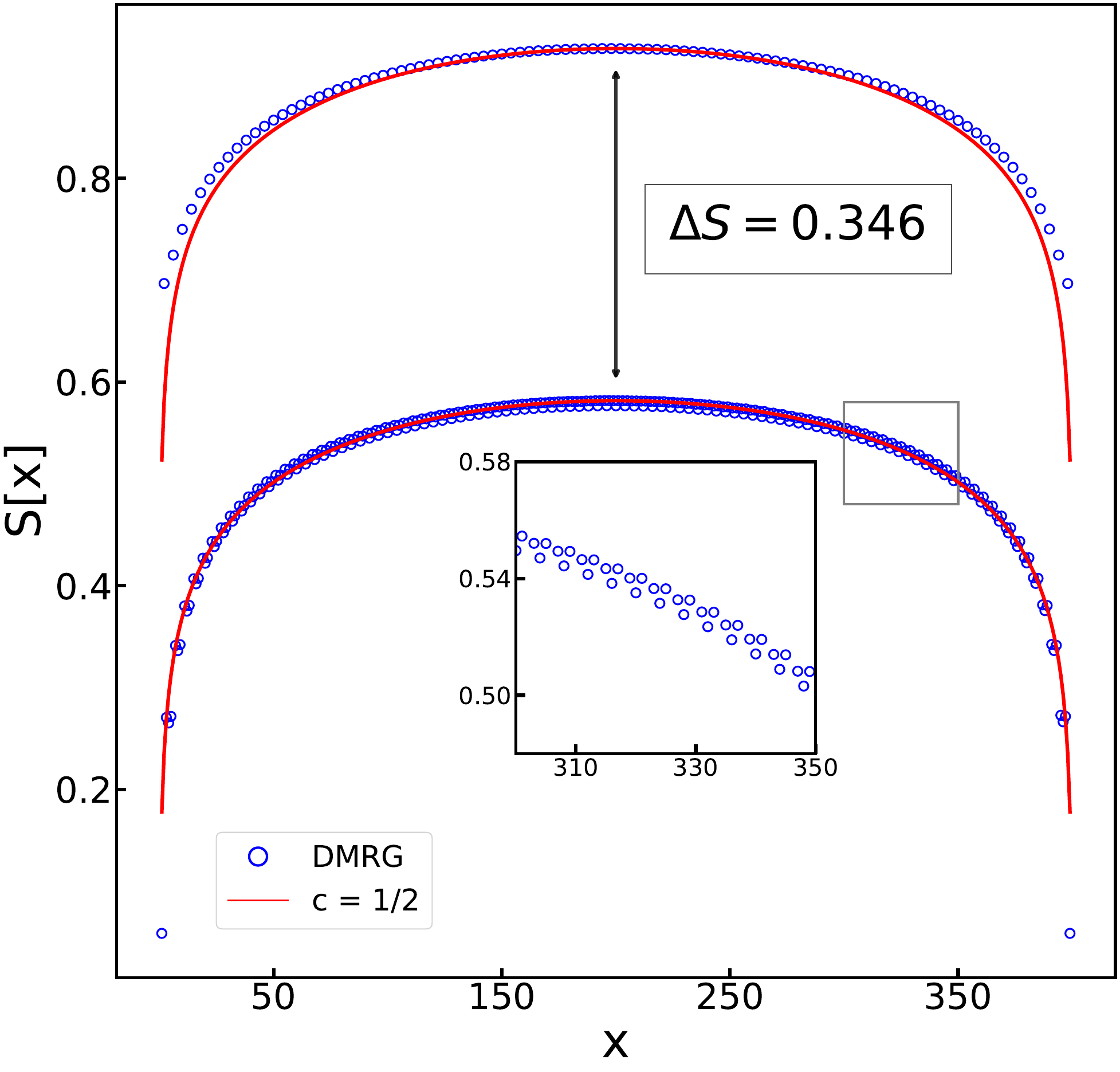}
    \caption{Entanglement entropy $S$ of a ground state deep in the CatAFM phase, plotted against the subsystem size $x$. The $S[x]$ curve splits into four branches, distinguished by the value of $x$ modulo 4. The three lower branches nearly overlap (see inset), while the upper one is separated  with a gap $\Delta S \approx 0.346$. Each branch is well described by the formula \ref{eq:13}, although a small discrepancy is visible for the upper branch near the edges. The calculations were performed at $(r,\theta) = (5,\pi/4)$ for  a chain  of length $L = 400$.} 
    \label{fig:entropy}
\end{figure}

To lift the degeneracy and obtain nontrivial corrections from $H^1$, we have to perform a fourth-order perturbative calculation (corrections to the lower orders are trivial; details are given in Appendix \ref{app:perturbation}). Similar to the CatFM case, we map the AFM-type states to a spin-1/2 chain as follows
\begin{align}
    |\cdots \sigma\sigma\mu_j\mu_j\sigma\sigma\mu_{j+1}\mu_{j+1}\cdots\rangle \ \Rightarrow \ |\cdots \mu_j\mu_{j+1}\cdots\rangle
\end{align}
With this mapping, we find that the fourth-order perturbative effective theory is  given by four copies of the critical transverse-field Ising model 
\begin{equation}\label{:eq 19}
    H_{\text{eff}} = \bigoplus_4 H_{\rm Ising} 
\end{equation}
with
\begin{align}
\label{eq:AFM_Ising_eff}
    H_{\rm Ising} = \frac{\sin^2\theta\cos^2\theta}{8r^3}\left(\sum_j\sigma_j^z\sigma_{i+1}^z + \sum_i\sigma_j^x\right) 
\end{align}
where an unimportant constant energy has been omitted. The four copies of $H_{\rm Ising}$ come from the four sectors of AFM states. This means that there are four degenerate ground states, above each of which resides a gapless spectrum that is identical to that of the critical Ising model.

The perturbative results are verified by our exact diagonalization study at a point deep in the CatAFM phase, as shown in Fig.~\ref{fig:ED}c. We are able to observe the four degenerate ground states at different momenta, verifying the spontaneous breaking of lattice translation. The Ising conformal towers in each sector can be roughly identified. However, the scaling dimensions are obscured by strong finite size effects. Note that a chain of length $L = 20$ in our model corresponds to an effective spin-$1/2$ Ising chain of length $L_{\text{Ising}} = L/4= 5$. 

We further confirm the perturbative picture through DMRG calculations. In particular, we calculate the  entanglement entropy on open boundary conditions\footnote{\label{footnote5}To be specific, the boundary conditions are set as $x_1=x_{L} = \sigma$. It is equivalent to set $H_1|x_1\rangle=U\delta_{x_1,\sigma} |x_1\rangle$ and $H_{L}|x_{L}\rangle=U\delta_{x_{L},\sigma} |x_{L}\rangle$ in the Hamiltonian \eqref{eq:model}, with $U\rightarrow \infty$. It corresponds to the free boundary conditions for the effective Ising chain $H_{\rm Ising}$.} and compare the numerical results with the celebrated Cardy-Calabrese formula \cite{calabrese2004entanglement}
\begin{equation}\label{eq:13}
    S[x] = \frac{c}{6}\log(\frac{2L}{\pi}\sin(\frac{\pi x}{L})) + \text{const.}
\end{equation}
where $x$ is the subsystem size, $L$ is the chain length, and $c$ is the central charge. Our results are shown in Fig.~\ref{fig:entropy}. A rather unique behavior is observed: Instead of a single smooth curve, the entanglement entropy $S[x]$ splits into four branches according to the location of the entanglement cut modulo 4. Among the four branches, three are nearly degenerate, while the last one has a gap above the others, with the gap given by 
\begin{align}
    \Delta S \approx 0.346\approx \frac{1}{2}\ln 2.
\end{align}
The gap $\Delta S$ varies only slightly across the chain (likely due to finite size effects), and the value 0.346 is measured  at the center of the chain.  Each of the four branches appears to follow the behavior predicted by Eq.~\ref{eq:13} independently. We fit the data into this formula and obtain a central charge from each of the curves, finding that they all have approximately $c \approx 0.5$. This verifies that the low-energy physics is described by the Ising CFT.

How to understand the observed behavior of $S[x]$? The degeneracy of the lower three branches in Fig.~\ref{fig:entropy} is easy to understand, which is given below. The entanglement entropy gap $\Delta S$ is discussed in Appendix \ref{app:entropy_gap2}, where we relate $\Delta S$ to  a measurement-induced drop of the entanglement entropy in the critical Ising chain. While not having an analytical derivation of $\Delta S$, we strongly believe that the gap $\Delta S$ has an exact value $\frac{1}{2}\ln{2}$ in the limit $x, L \gg1$. 

Below we argue for the degeneracy of the  lower three curves of entanglement entropy in Fig.~\ref{fig:entropy}. Assume that we are deep in the CatAFM phase so that the perturbative analysis applies. Then, the ground state is given by
\begin{align}
    |\Psi_{\rm GS}\rangle = \sum_{\{\mu_j\}} & \psi_{\rm Ising} (\cdots\mu_j\mu_{j+1}\cdots)\nonumber\\
&\times |\sigma\cdots\mu_j\mu_j\sigma\sigma\mu_{j+1}\mu_{j+1}\cdots \sigma\rangle
\label{eq:catAFM_gs}
\end{align}
where $\psi_{\rm Ising}(\cdots\mu_j\mu_{j+1}\cdots)$ is the ground-state wave function of the critical Ising chain of length $L/4$ (i.e., $j=1,2,\dots, L/4$) under free boundary conditions in the $\sigma^z$ basis.  Due to our choice of boundary conditions (see footnote \ref{footnote5}), the open-chain ground state $|\Psi_{\rm GS}\rangle$ is unique. Given the particular translational symmetry breaking pattern in the ground state,  entanglement cuts naturally decompose into four distinct types:
\begin{equation}\label{eq:20}
    \begin{tikzpicture}[xscale=.7, yscale=.5, nodes={execute at begin node=$, execute at end node=$}]
    \node at (-0.15,0) {|\cdots};
    \node at (0.50,0){\mu_j};
    \draw[red, thick] (0.77,-0.8) -- (0.77, 0.8)node[above]{A};
    \node at (1.1,0) {\mu_j};
    \draw[red, thick] (1.4,-0.8) -- (1.4, 0.8)node[above]{B};
    \node at (1.65,0) {\sigma};
    \draw[red, thick] (1.85,-0.8) -- (1.85, 0.8)node[above]{C};
     \node at (2.1,0) {\sigma};
     \draw[red, thick] (2.3,-0.8) -- (2.3, 0.8)node[above]{D};
     \node at (2.9,0){\mu_{j+1}};
     \draw[red, thick] (3.4,-0.8) -- (3.4, 0.8)node[above]{A};
     \node at (4.7,0){\mu_{j+1}\sigma\sigma\cdots\rangle};
     \node[red] at (4,1.2){\cdots};
    \end{tikzpicture}
\end{equation}
which we call type-A, B, C, D, respectively. The key observation is that the type-B, C and D cuts are all mapped to the same entanglement cut in the corresponding Ising chain:
\begin{equation}\label{eq:21}
    \begin{tikzpicture}[xscale=.7, yscale=.5, nodes={execute at begin node=$, execute at end node=$}]
    \node at (-0.1,0) {|\cdots};
    \node at (0.6,0){\mu_j};
    \node at (1.1,0) {\mu_j};
    \draw[red, thick] (1.4,-0.8) -- (1.4, 0.8)node[above]{B};
    \node at (1.65,0) {\sigma};
    \draw[red, thick] (1.85,-0.8) -- (1.85, 0.8)node[above]{C};
     \node at (2.1,0) {\sigma};
     \draw[red, thick] (2.3,-0.8) -- (2.3, 0.8)node[above]{D};
     \node at (3.35,0){\mu_{j+1}\cdots\rangle};
     \draw [->] (4.4,0) -- (5.5,0);
     \node at (6.2,0) {|\cdots\mu_j};
     \draw[red, thick] (6.95,-0.8) -- (6.95, 0.8);
     \node at (8,0){\mu_{j+1}\cdots\rangle};
    \end{tikzpicture}
\end{equation}
This, in turn, means that the reduced density matrices for the entanglement cuts B, C, D are identical to the one for the corresponding cut in the Ising chain. Therefore, the entanglement entropy for three consecutive sites corresponding to the type-B, C, D cuts is identical to that for the Ising chain. This explains the three overlapping branches observed in Fig.~\ref{fig:entropy}, and also the excellent agreement in the central charge with that of the Ising CFT. On the other hand, there is no obvious correspondence between type-A entanglement cuts and any cut in the Ising chain. To explain the fourth branch in Fig.~\ref{fig:entropy}, which corresponds to type-A entanglement cuts, we must rely on alternative methods. We discuss the type-A cuts and the entanglement entropy gap $\Delta S$ in Appendix \ref{app:entropy_gap2}.

Finally, we turn to the symmetry properties of the four CatAFM ground states. For this purpose, we impose periodic boundary conditions so that the model is symmetric under lattice translations. The full symmetry of the model is $\mathcal{C}_{\rm Ising}\times \mathbb{Z}^{\rm trans}_L$, where $\mathbb{Z}^{\rm trans}_L$ is the cyclic group of translations. We label the degenerate ground states of the four sectors as follows:
\begin{align}\label{eq:22}
&\mu\mu\sigma\sigma: \quad |1\rangle \nonumber \\
&\sigma\mu\mu\sigma:\quad \ket{2} \nonumber \\
&\sigma\sigma\mu\mu:\quad \ket{3} \nonumber \\ &\mu\sigma\sigma\mu: \quad \ket{4} 
\end{align}
Each state $|n\rangle$ behaves as a critical Ising ground state.

Let us first check the transformations of the ground states under translation. Let $T$ be the operator of a unit translation. Obviously, we have 
\begin{align}
    T|n\rangle = |n+1\rangle,
\end{align}
where the addition $n+1$ is taken modulo 4. This nontrivial action of $T$ on the ground states is a manifestation of the spontaneous translation breaking: $\mathbb{Z}_{L}^{\rm trans}\rightarrow \mathbb{Z}_{L/4}^{\rm trans}$. It is convenient to define $T_{\rm Ising}\equiv T^4$, the four-site translation. Meanwhile, $T_{\rm Ising}$ corresponds to a \emph{single-site} translation of the effective Ising chain $H_{\rm Ising}$ in \eqref{eq:AFM_Ising_eff}. We notice that $T_{\rm Ising}$ acts as the identity operator on the ground states, a consequence of the fact that the ground state of the Ising chain is translation invariant. Since the states $\{|n\rangle\}$ are not eigenstates of $T$, we define a basis transformation
\begin{align}
    |a\rangle = \sum_{n=1}^4 \frac{e^{-ian\pi /2}}{2}|n\rangle
\end{align}
where $a=0,1,2,3$. Then, we have
\begin{align}
    T|a\rangle = i^{a}|a\rangle
\end{align}
The four ED ground states in Fig.~\ref{fig:ED}(c) are the translation eigenstates $\{|a\rangle \}$, with the lattice momentum being $a\pi/2$, respectively.

For transformations under $\mathcal{C}_{\rm Ising}$, it is clear that the ground states are invariant under $U(\psi)$.   Under the action of $U(\sigma)$, we have
\begin{equation}\label{eq:26}
   U(\sigma) \ket{n} = \sqrt{2}|n+2\rangle
\end{equation}
The coefficient $\sqrt{2}$ is inferred from the symmetry algebra \eqref{eq:9} and the fact that $U(\psi)|n\rangle = U(\mathbb{1})|n\rangle = |n\rangle$. Then, for the translation eigenstates, we have
\begin{align}
    U(\sigma)|a\rangle = (-1)^a\sqrt{2}|a\rangle
\end{align}
Therefore, the symmetry $U(\sigma)$ is also broken. The category symmetry $\mathcal{C}_{\rm Ising}$ is broken down to $\mathbb{Z}_{2}^\psi=\{U(\mathbb{1}), U(\psi)\}$.

However, the remaining unbroken symmetry is larger than $\mathbb{Z}^\psi_{2}\times \mathbb{Z}_{L/4}^{\rm trans}$.  Define the operator 
\begin{align}
    \tilde{U}(\sigma) = U(\sigma)T^{-2}.
\end{align}
We find that the ground states are mapped to themselves under $\tilde{U}(\sigma)$, namely
\begin{align}
    \tilde{U}(\sigma)|a\rangle = \sqrt{2}|a\rangle.
\end{align}
Accordingly, $\tilde{U}(\sigma)$ is not broken in the ground states. 
Noting that $T^{-4} = T_\text{Ising}^{-1}$ and using the original symmetry algebra \eqref{eq:9}, we obtain the following algebra of the unbroken symmetries:
    \begin{align}
    &U(\psi)\cdot\tilde{U}(\sigma) = \tilde{U}(\sigma)\cdot U(\psi) = \tilde{U}(\sigma)\nonumber\\
    &\tilde{U}(\sigma)^2 = \left(I + U(\psi)\right)T_\text{Ising}^{-1}
\end{align}
This is exactly the symmetry algebra of the usual  critical transverse-field Ising model (see e.g. Ref.~ \cite{seiberg_non-invertible_2024} for a detailed discussion). The \emph{non-invertible lattice translation} $\tilde{U}(\sigma)$ corresponds to the conventional Kramers-Wannier duality and pins the low-energy effective theory in the CatAFM phase at criticality.

\subsection{Special lines}
\label{sec:special_lines}

In the phase diagram (Fig.~\ref{fig:phasediagram}), the two horizontal lines corresponding to $\theta = \pi/2$ and $\theta = 0$ require special attention. Along these lines, the model often, though not always, exhibits a large ground-state degeneracy that grows exponentially with system size, indicating  the instability of the low-energy physics.

\subsubsection{$\theta = \frac{\pi}{2}$}

For $\theta=\pi/2$, the Hamiltonian $H=-\sum H_i$ acts on the basis states as follows 
\begin{align}
& H_i\ket{\mu\mu\mu} = 0 \nonumber\\
&H_i\ket{\mu\mu\sigma} = r\ket{\mu\mu\sigma} + \ket{\mu\sigma\sigma}\nonumber\\
& H_i\ket{\mu\sigma\nu} = 0 \nonumber\\
&H_i\ket{\sigma\mu\mu} = r\ket{\sigma\mu\mu} + \ket{\sigma\sigma\mu}\nonumber\\
&H_i\ket{\mu\sigma\sigma} = r\ket{\mu\sigma\sigma} + \ket{\mu\mu\sigma}\nonumber\\
& H_i\ket{\sigma\mu\sigma} = 0 \nonumber\\
&H_i\ket{\sigma\sigma\mu} = r\ket{\sigma\sigma\mu} + \ket{\sigma\mu\mu} \nonumber\\
& H_i\ket{\sigma\sigma\sigma} = 0 
\label{eq:theta_pi_2}
\end{align}
To analyze the model, we make use of a domain wall picture as presented below. Consider that the domain-wall state of the following anyon state is
\begin{align*}
    |\cdots \sigma\sigma \mu\mu\mu \sigma\sigma \mu'\sigma\sigma\cdots\rangle\rightarrow |\cdots \uparrow\downarrow\uparrow\uparrow\downarrow\uparrow\downarrow\downarrow\uparrow \cdots\rangle
\end{align*}
where ``$\uparrow$'' represents the absence of a domain wall and  ``$\downarrow$'' represents the presence of a domain wall. Note that the mapping from the anyon states to the domain-wall configurations is \emph{many-to-one}: typically, the number of anyon states corresponding to a given domain-wall configuration grows exponentially with system size due to the choices of $\mu_i=\mathbb{1},\psi$. However, if we ignore the underlying anyon degrees of freedom and  focus only on the domain-wall configuration, the Hamiltonian action reduces to
\begin{align}
    & H_i |\uparrow\uparrow\rangle = 0 \nonumber\\
    & H_i |\downarrow\downarrow\rangle = 0 \nonumber\\
    &H_i\ket{\uparrow\downarrow} = r\ket{\uparrow\downarrow} + \ket{\downarrow\uparrow}\nonumber\\
    &H_i\ket{\downarrow\uparrow} = r\ket{\downarrow\uparrow} + \ket{\uparrow\downarrow} 
    \label{eq:H_domain-wall-picture}
\end{align}
Such a reduction is consistent only for $\theta=\pi/2$; for other values of $\theta$, the corresponding eight equations in \eqref{eq:theta_pi_2} cannot be consistently reduced to the four equations in \eqref{eq:H_domain-wall-picture}.  Using the Pauli matrices, the Hamiltonian in \eqref{eq:H_domain-wall-picture} can be written as 
\begin{equation}
    H_i \rightarrow \frac{1}{2}\left(r + \sigma_i^x\sigma_{i+1}^x + \sigma_i^y\sigma_{i+1}^y - r\sigma_i^z\sigma_{i+1}^z\right)
    \label{eq:XXZ}
\end{equation}
It is just the XXZ model where the anisotropy parameter $\Delta = -r$, where $r<0$ corresponds to the ferromagnetic exchange interaction and $r>0$ corresponds to the anti-ferromagnetic interaction.

It must be stressed that the XXZ model in Eq.~\ref{eq:XXZ} \emph{does not exactly} recover the original model due to the many-to-one nature of the mapping to the domain-wall configurations. $H_i$ has a nontrivial action on the anyon states that correspond to the same domain-wall configuration. Take the state $|\sigma\sigma\mu\mu\sigma\sigma\mu'\mu'\rangle$, with $L=8$ on periodic boundary conditions, for illustration. One can check that
\begin{align}
     H_8H_6H_4H_2& H_7H_5H_3H_1 |\sigma\sigma\mu\mu\sigma\sigma\mu'\mu'\rangle\nonumber\\
     & \rightarrow |\mu'\mu'\sigma\sigma\mu\mu\sigma\sigma\rangle \label{eq:shifting}
\end{align}
where ``$\rightarrow$'' means that the final state after a sequence of actions by $H_i$ has a nonzero component in the basis state $|\mu'\mu'\sigma\sigma\mu\mu\sigma\sigma\rangle$. The initial and final states in \eqref{eq:shifting} are different, but they are mapped to the same domain-wall state $|\!\uparrow\downarrow\uparrow\downarrow\uparrow\downarrow\uparrow\downarrow\rangle$, indicating that the Hamiltonian remains nontrivial even after the domain-wall state is fixed.

Nevertheless, the shifting effect in \eqref{eq:shifting} is a high-order effect, of the $L$-th order, and hence should be negligible in the thermodynamic limit if the XXZ model is gapped.  Indeed, the XXZ model is gapped for $|r|>1$. Hence, we believe the XXZ model captures the key physics on the $\theta=\pi/2$ line when $|r|>1$. The energy spectrum at $|r|>1$ should be that of the XXZ model, with each energy associated with a large degeneracy that grows exponentially with system size. We also assert that  the two  transition lines in the phase diagram Fig.~\ref{fig:phasediagram} should terminate at the points $r=\pm 1$ on the $\theta=\pi/2$ line. 

We have performed an ED study that confirms the large degeneracy for $|r|>1$. The behaviors of the energy-momentum spectra of finite size systems are slightly different in the ferromagnetic and anti-ferromagnetic regimes: for $r<-1$, the degeneracy is quite exact even at very small sizes (e.g. $L=8$); for $r>1$, the degeneracy is approximate, and it becomes more and more accurate as $L$ increases (this degeneracy actually arises from the collapse of the four conformal towers in the Cat-AFM phase as $\theta$ is gradually tuned towards $\pi/2$).  We have also performed DMRG calculations of the entanglement entropy for $\theta\lesssim \pi/2$, confirming that the two transition lines terminate near $r=-1$ and $r=1$, respectively (see the data points in Fig.~\ref{fig:phasediagram}). 

For $|r|<1$, the XXZ model is a Luttinger liquid, i.e., a $c=1$ conformal field theory. Since it is gapless, the shifting effect in \eqref{eq:shifting} may not be negligible. Indeed, our small-size ED calculations ($L\le 16$) suggest that the low-energy dispersion is non-linear, indicating a nontrivial consequence of the effect \eqref{eq:shifting}. Nevertheless, our DMRG calculations do not give conclusive results on the central charge. A further analysis is needed to understand the low-energy physics in the regime $|r|<1$.

\subsubsection{$\theta = 0$}

For $\theta=0$, the mapping to the domain-wall states as above is not possible. Hence, we are unable to make any analytical deductions and rely completely on numerical calculations. From our ED and DMRG calculations, we find that both transition lines appear to meet the  $\theta = 0$ line at $r = 1$. In addition, both ED and DMRG results suggest that the  model is gapped with a threefold ground state degeneracy for $r<1$ (i.e., in the CatFM phase). For $r>1$, we again find a large ground-state degeneracy, similar to the $|r|>1$ regime on the $\theta=\pi/2$ line.

\section{Phase Transitions}
\label{sec:transition}
\begin{figure}[t]
    \centering
    \includegraphics[width=0.95\linewidth]{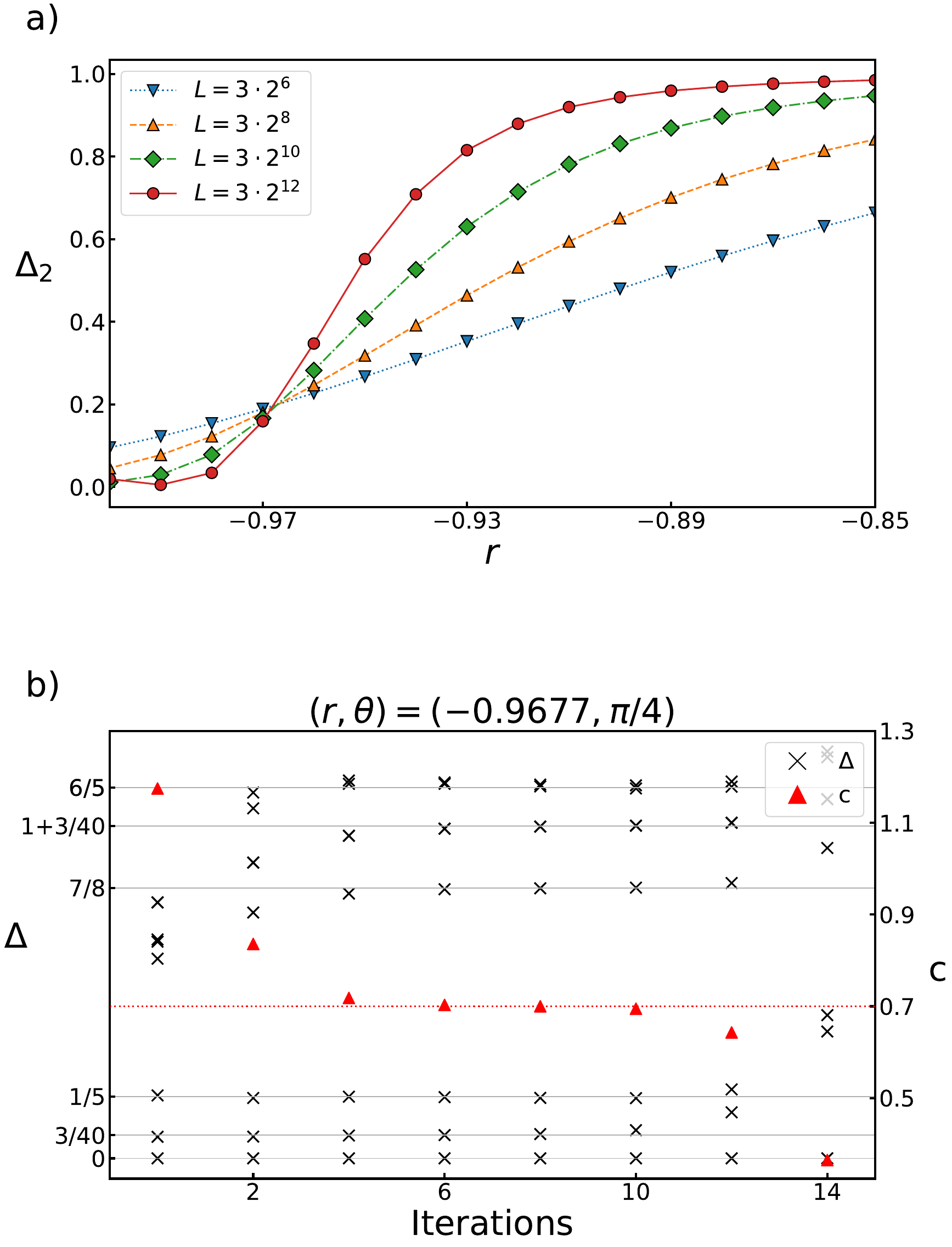}
    \caption{Loop-TNR results for the CatFM-symmetric transition along the $\theta=\pi/4$ line. a) The second scaling dimension $\Delta_2$  is plotted as a function of $r$ for several system sizes. The intersection point of the curves is identified as the transition point, yielding $r_c = -0.9677(7)$ and $\Delta_2\approx 0.2$. On the left side of the transition, $\Delta_2\rightarrow 0$ as $L$ increases, whereas on the right side $\Delta_2\to 1$, as expected for the Ising CFT. The calculations were done with a bond dimension $\chi = 34$. b) Conformal data calculated at the transition point, $(r,\theta) = (-0.9677,\pi/4)$, in agreement with the tricritical Ising CFT. The bond dimension used here is $\chi = 40$.}
    \label{fig:Loop-TNR}
\end{figure}

In this section, we discuss the transitions between the phases described above. Our understanding of the transitions is primarily obtained from numerical methods, including tensor network renormalization and DMRG methods.

\subsection{CatFM - Symmetric}
The main goal of our numerical calculations is to identify the low-energy theory along the CatFM-Symmetric transition line. We find that the whole line is a continuous transition characterized by the tricritical Ising CFT whose central charge $c=7/10$. First of all, one can see in Fig.~\ref{fig:extended_dmrg} that DMRG calculations (for $L=400$) on the central charge show a strong finite size effect near the transitions. This leads us to turn to the class of numerical algorithms based on tensor network renormalization group (TRG) \cite{LevinNavePRL2007,evenbly2015tensor} to obtain more accurate conformal data. In particular, we have implemented the loop-optimized tensor network renormalization group method (Loop-TNR) \cite{yang_loop_2017} (see \cite{bao2019loop, li_tensor-network_2022, wei_tensor_2023} for further useful details on implementation). The primary advantage of this method is that it allows us to reach much larger system sizes and extract the conformal data to a high degree of accuracy.

We perform Loop-TNR calculations\footnote{As previously done in DMRG calculations, we enforce the fusion rules by implementing a penalty term in the Hamiltonian.} near the transition for a given $\theta$, a range of values of $r$, and several system sizes, using a relatively small bond dimension ($\chi = 34$). From the calculations, we obtain a few low-lying scaling dimensions and the central charge of the low-energy theory. Plotting the second scaling dimension $\Delta_2$ against $r$ for different system sizes, we locate the transition point (the data quality of $\Delta_2$ is a bit better than that of $\Delta_1$). Figure \ref{fig:Loop-TNR}a shows the $\Delta_2\sim r$ plots at $\theta=\pi/4$; calculations at other values of $\theta$ are also done, from which we map out the transition line in Fig.~\ref{fig:phasediagram}. Having located the transition point, we can use a larger bond dimension ($\chi = 40$) to get higher-quality results on the conformal data (shown in Fig.~\ref{fig:Loop-TNR}b). Comparing our numerical conformal data with the exact data of CFTs \cite{di_francesco_conformal_1997}, we confirm that the transition is described by the tricritical Ising CFT ($\mathcal{M}_{5,4}$ minimal model). It must be noted that the quality of data begins to degrade after as few as 12 iterations in our calculations, likely due to the uncertainty in the precise location of the transition point.

The transition between the CatFM  and symmetric Ising CFT phases was known long ago to be the tri-critical Ising transition\cite{BlumeEmeryPRB1971}, although it was not described in the language of categorical symmetry breaking. In particular, the CatFM phase was known as the phase coexistence of the paramagnetic and ferromagnetic states. More recent related studies can be found in Refs.~\cite{RahmaniPRL2015, RahmaniPRB2015, OBrienFendleyPRL2018, CameronSciPost2024,seiberg_non-invertible_2024, cortes2022duality}. It is worth noting some minor differences between our model and those studied in previous works: (1) our model does not have a tensor product structure, whereas the previous works are formulated in the context of spin chains, interacting Majorana fermions, or in continuum field theory; and (2) in these works, $\mathcal{C}_{\rm Ising}$  is realized as the usual Kramers-Wannier duality, whose square gives rise to lattice translation, while in our model $\mathcal{C}_{\rm Ising}$  is purely an internal symmetry. Despite these differences, the infrared universal properties at the transition should, and to our knowledge do, agree across all the works.

\subsection{Symmetric - CatAFM}

\begin{figure}[t]
    \centering
    \includegraphics[width=0.85\linewidth]{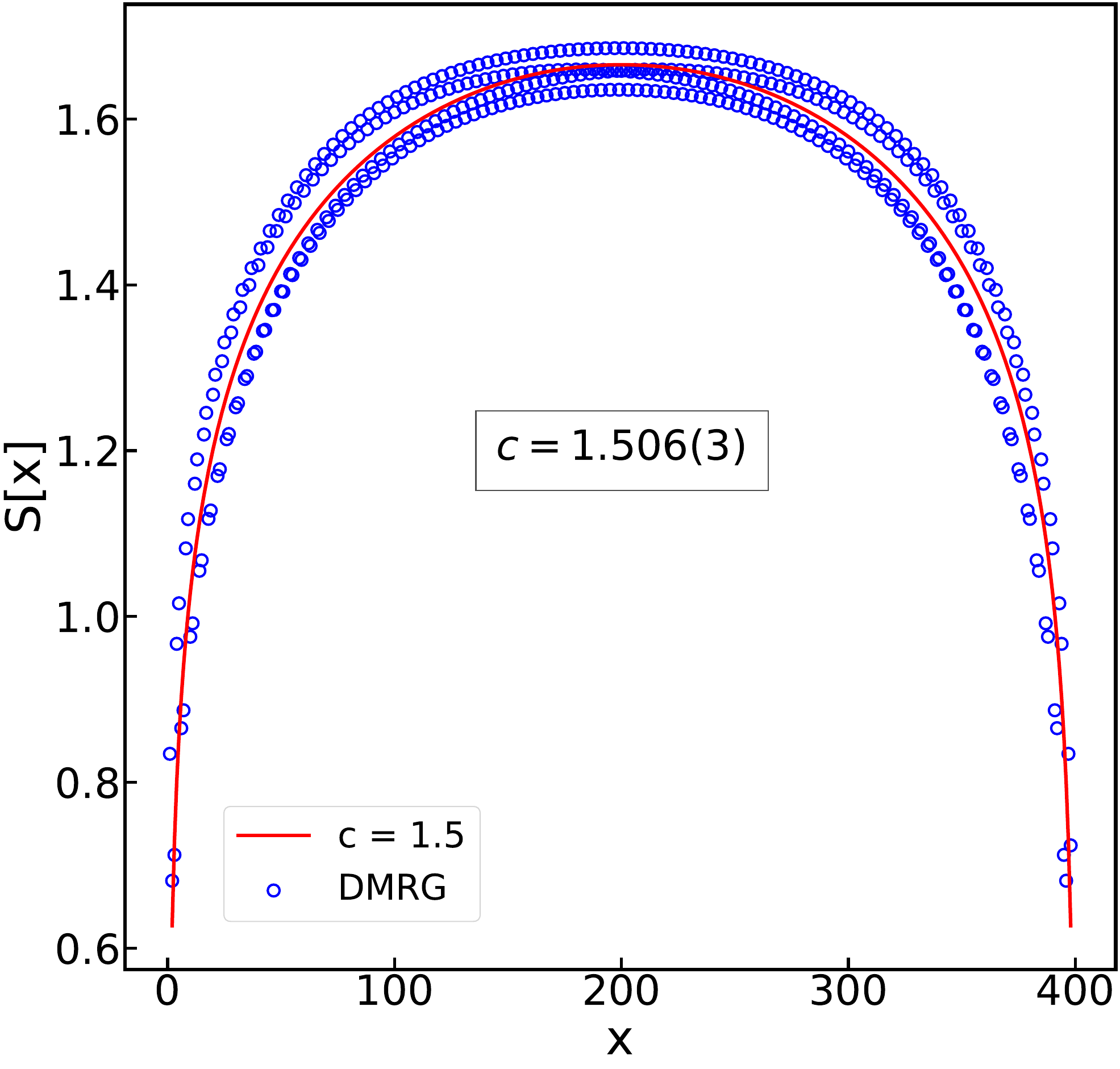}
    \caption{Entanglement entropy $S$ plotted against subsystem size $x$ at a symmetric-CatAFM transition point, $(r,\theta) = (1.33, \pi/4)$. The curve splits into four approximately degenerate branches, according to the value of $x$ modulo 4. Such a splitting may arise from the limited precision of the transition point and can be understood as a signature of the entry into the CatAFM phase.  Fitting the data to the formula \eqref{eq:13} yields $c = 1.506(3)$.}
    \label{fig:right_transition}
\end{figure}

It is unfortunate that when applying the Loop-TNR method to the Symmetric-CatAFM transition, our calculations do not converge well.\footnote{We observed that the speed of light failed to stabilize across iterations, consequently making it impossible to extract reliable conformal data.} Hence, the following discussion will mainly rely on DMRG calculations and will be qualitative due to the  relatively low quality of our data. 

We compute the entanglement entropy and extract the central charge  according to the formula \eqref{eq:13}. Knowing that the central charge should be $c = 1/2$ in both the symmetric and CatAFM phases, we identify the transition points with the peaks in the $c\sim r$ plots in the right panels in Fig.~\ref{fig:extended_dmrg} (it is not a particularly precise way to locate the transition but enough for our discussions). The central charge at the transition is then extracted from the entanglement entropy, as shown in Fig.~\ref{fig:right_transition} for $\theta=\pi/4$ (it is similar for other values of $\theta$). We observe that the entanglement entropy curve splits into four nearly degenerate branches. Such a splitting possibly stems from our imprecise determination of the location of the transition point or the finite size effect. Nevertheless, we still fit the curve into \eqref{eq:13} and  obtain $c = 1.506(3)$.

The Symmetric-CatAFM transition is especially interesting because it occurs in the presence of a background Ising CFT. The observed central charge $c\approx 1.5$ suggests that the infrared theory at the transition is possibly a $c=1$ Luttinger liquid together with the background Ising CFT, yielding a total central charge $c=1+1/2$. Many aspects of this transition remain poorly understood. For instance, if the conjecture of a $c=1+1/2$ CFT is correct, what is the Luttinger parameter along the transition line, and how does it vary? How does the categorical symmetry $\mathcal{C}_{\rm Ising}$ act on the low-energy states at the transition? We leave these questions for a more careful future study.

\section{Conclusion}
\label{sec:conclusion}

In summary, we have investigated a lattice model that respects the Ising fusion category symmetry, $\mathcal{C}_{\rm Ising}$, using a combination of numerical and analytical methods, including exact diagonalization, DMRG, tensor-network renormalization, and perturbative analysis. We have identified three phases of the model: a symmetric critical phase and two categorical symmetry breaking phases. The first symmetry breaking phase, analogous to the conventional ferromagnet, is denoted the CatFM phase, in which the full $\mathcal{C}_{\rm Ising}$ symmetry is broken. The other, denoted the CatAFM phase, breaks the non-invertible element of $\mathcal{C}_{\rm Ising}$ together with lattice translation. Unlike the conventional anti-ferromagnet, however, the CatAFM phase is gapless and critical. Transitions between these phases are also studied numerically. Our work is a concrete lattice investigation of categorical symmetry and its associated symmetry breaking phenomena.

Several questions remain open for future study. One particularly interesting direction is to study anti-ferromagnetic states associated with general non-invertible symmetry breaking. An anti-ferromagnetic state may be viewed as an array of ferromagnetic domains separated by domain walls.  When the broken symmetry is non-invertible, each domain wall is associated with a defect $a$ of quantum dimension $d_a>1$ (one may also consider domain walls associated with different types of defects). As a result, the low-energy subspace is extensive, in the sense that the dimension grows exponentially with the number $N$ of domain walls, $\sim d_a^N$, and $N$ should be proportional to the system size $L$.  A natural question is whether such a low-energy manifold will generally give rise to a critical phase, as in our model, or whether it is unstable and eventually gapped out. By a rough analogy, this situation resembles the appearance of Goldstone modes when a continuous symmetry is spontaneously broken, although in that case, the infinitely large low-energy manifold arises from the continuity of a Lie group and does not scale with system size.

Another important open problem is to clarify the nature of the phase transition between the symmetric gapless Ising CFT phase and the CatAFM phase. This transition involves the breaking of both non-invertible and lattice translation symmetries, and it occurs in the presence of a background CFT. It would be interesting to understand the coupling between the symmetry-breaking order parameter and the background CFT across the transition.

\begin{acknowledgments}
We are grateful to Zheng-Cheng Gu, Jiaqi Guo, Shuo Yang and Yueshui Zhang for the enlightening discussions, and especially to Shuo Yang for valuable suggestions on numerical techniques. We thank Information Technology Services at the University of Hong Kong and Beijing Paratera Tech
Corp., Ltd for providing computational services. The work was supported by Research Grants Council of Hong Kong (GRF 17311322, CRF C7012-21GF, CRF C7015-24GF, CRS HKU701/24, and AoE/P-604/25-R) and National Natural Science Foundation of China (Grant No. 12222416). 
\end{acknowledgments}

\appendix

\section{Restriction to $0\leq \theta\leq \pi/2$}
\label{app:para_range}

In this appendix, we show that it is sufficient to restrict $\theta$ to the range $[0,\pi/2]$. For other values of $\theta$, the model is mapped to this range via unitary transformations.

Consider a basis state $|\{x_i\}\rangle$ in the Hilbert space. Let us define the domain wall number operator $\hat N_i$ between the site $i$ and $i+1$:
\begin{align}
    \hat N_i | \dots x_ix_{i+1}\dots\rangle = n_i| \dots x_ix_{i+1}\dots\rangle
\end{align}
where $n_i = 0$ if $x_i =x_{i+1}$, and $n_i=1 $ if $x_i \neq x_{i+1}$. With the domain wall number operators, we define the following two operators
\begin{align}
    U = (-1)^{(\sum_k \hat N_k)/2}, \quad V  = (-1)^{\sum_k k \hat N_k}
\end{align}
Due to periodic boundary conditions, the total number of domain walls $\sum_{k}n_k$ must be even. Then, we have $U^2=V^2=I$.  Both $U$ and $V$ are unitary operators.

Now consider the Hamiltonian in \eqref{eq:model}. We find that
\begin{align}
    UH_iU\ket{\mu\mu\mu} &= -\cos\theta\ket{\mu\sigma\mu} \nonumber\\
    UH_iU\ket{\mu\mu\sigma} &= r\ket{\mu\mu\sigma} + \sin\theta\ket{\mu\sigma\sigma}\nonumber\\
    UH_iU\ket{\mu\sigma\nu} &= -\delta_{\mu\nu}\cos\theta\ket{\mu\mu\mu}\nonumber\\
   UH_iU\ket{\sigma\mu\mu} &= r\ket{\sigma\mu\mu} + \sin\theta\ket{\sigma\sigma\mu}\nonumber\\
    UH_iU\ket{\mu\sigma\sigma} &= r\ket{\mu\sigma\sigma} + \sin\theta\ket{\mu\mu\sigma}\nonumber\\
    UH_iU\ket{\sigma\mu\sigma} &= -\frac{\cos\theta}{\sqrt{2}}\ket{\sigma\sigma\sigma}\nonumber\\
    UH_iU\ket{\sigma\sigma\mu} &= r\ket{\sigma\sigma\mu} + \sin\theta\ket{\sigma\mu\mu}\nonumber\\
    UH_iU\ket{\sigma\sigma\sigma}&= -\frac{\cos\theta}{\sqrt{2}}(\ket{\sigma\mathbb{1}\sigma} + \ket{\sigma\psi\sigma})
\end{align}
and
\begin{align}
    VH_iV\ket{\mu\mu\mu} &= -\cos\theta\ket{\mu\sigma\mu} \nonumber\\
    VH_iV\ket{\mu\mu\sigma} &= r\ket{\mu\mu\sigma} -\sin\theta\ket{\mu\sigma\sigma}\nonumber\\
    VH_iV\ket{\mu\sigma\nu} &= -\delta_{\mu\nu}\cos\theta\ket{\mu\mu\mu}\nonumber\\
   VH_iV\ket{\sigma\mu\mu} &= r\ket{\sigma\mu\mu} -\sin\theta\ket{\sigma\sigma\mu}\nonumber\\
    VH_iV\ket{\mu\sigma\sigma} &= r\ket{\mu\sigma\sigma} -\sin\theta\ket{\mu\mu\sigma}\nonumber\\
    VH_iV\ket{\sigma\mu\sigma} &= -\frac{\cos\theta}{\sqrt{2}}\ket{\sigma\sigma\sigma}\nonumber\\
    VH_iV\ket{\sigma\sigma\mu} &= r\ket{\sigma\sigma\mu} -\sin\theta\ket{\sigma\mu\mu}\nonumber\\
    VH_iV\ket{\sigma\sigma\sigma}&= -\frac{\cos\theta}{\sqrt{2}}(\ket{\sigma\mathbb{1}\sigma} + \ket{\sigma\psi\sigma})
\end{align}
Denote the original Hamiltonian as $H(r,\theta)$. Then, the above results lead to
\begin{align}
    UH(r,\theta)U & = H(r,\pi -\theta)\nonumber\\
    VH(r,\theta)V &  = H(r,\pi +\theta)
\end{align}
Since $U$ and $V$ are unitary,  $ H(r,\pi -\theta)$ and $H(r,\pi +\theta)$ have the same energy spectrum as $H(r,\theta)$. Therefore, for identifying the ground-state phase diagram of the model, it is sufficient to restrict $\theta$ to the interval $[0, \pi/2]$.

\section{Perturbation analysis}
\label{app:perturbation}

In this appendix, we shall work through the perturbative calculations in detail. We separate the Hamiltonian into two pieces  $H=H^0+H^1$ 
with
\begin{align}
    H^0 = - \sum_i H_i^{\rm dw},\quad H^1 = -\sum_i H_i^{\rm flip}
\end{align}
where  $H_i^{\rm dw}$ is the nearest-neighbor domain-wall interaction, and $H_i^{\rm flip}$ is associated with flipping a single domain (see Eqs.~\ref{eq:Hr} and \ref{eq:Hth} in Sec.~\ref{sec:hamiltonian} for the explicit expressions). We take $H^0$ to be the unperturbed Hamiltonian and $H^1$ as the perturbation. $H^0$ depends only on the parameter $r$, and $H^1$ depends only on the parameter $\theta$. We consider two perturbative regimes, corresponding to $r\ll -1$ (CatFM) and $r\gg 1$ (CatAFM), respectively.  The following analysis relies on the standard formulation of degenerate perturbation theory (see e.g. Ref.~\cite{sakurai2020modern}).

\subsection{CatFM phase: $r\ll -1$}
\label{app:pert1}

For $r\ll - 1$, as discussed in the main text, the ground-state manifold of the unperturbed Hamiltonian $H^0$ depends on the parity of $L$. Here, we consider $L$ to be even. In this case, $H^0$ has a ground state degeneracy of $\text{G.S.D.} = 3 + 2\cdot2^{L/2}$. There are two types of ground states: three ferromagnetic states
\begin{align*}
    \ket{\cdots\mathbb{1}\mathbb{1}\mathbb{1}\cdots},\, \ket{\cdots\psi\psi\psi\cdots},\, \ket{\cdots\sigma\sigma\sigma\cdots}
\end{align*}
and $2^{1+L/2}$ anti-ferromagnetic states
\begin{align*}
     \ket{\cdots\mu_i\sigma\mu_{i+1}\sigma\cdots}, \ket{\cdots\sigma\mu_i\sigma\mu_{i+1}\cdots}
\end{align*}
where $\mu_i = \mathbb{1}, \psi$. The two sectors of anti-ferromagnetic states are related by shifting a lattice constant. In low-order perturbation (of order much smaller than $L/2$), only the states within one of the anti-ferromagnetic states mix with each other. Hence, our discussion below focuses on a particular anti-ferromagnetic sector. The calculations for other cases are similar.

To set up our notation, let
\begin{align}
    H^0|n_\mu^{(0)}\rangle = E_n^{(0)} |n_\mu^{(0)}\rangle
\end{align}
where $\mu$ labels the degenerate states of the $n$th unperturbed energy level. Especially, $|0_\mu^{(0)}\rangle$ labels the ground states. We take the orthonormal condition $\langle m_\nu^{(0)}|n_\mu^{(0)}\rangle = \delta_{mn}\delta_{\mu\nu}$. Let $|\psi\rangle$ be an eigenstate of $H$, and $E$ be the corresponding energy. In the order of perturbation, they can be expanded as
\begin{align}
    |\psi\rangle & = |\psi^{(0)}\rangle + |\psi^{(1)}\rangle + |\psi^{(2)}\rangle + \cdots \nonumber \\
    E & =E^{(0)}+E^{(1)} + E^{(2)}+\cdots
\end{align}
where the superscript counts the order of the perturbative correction. We consider every state $|\psi\rangle $ that reduces to the unperturbed ground-state manifold when the perturbation $H^1$ is turned off, i.e., $|\psi^{(0)}\rangle$ is a linear superposition of $\{|0_\mu^{(0)}\rangle\}$. The unperturbed ground states have energy $E^{(0)}=E_0^{(0)}=0$.

To the first order in perturbation theory, Schrodinger's equation gives
\begin{equation}
    H^0\ket{\psi^{(1)}} + H^1\ket{\psi^{(0)}} = E^{(0)}\ket{\psi^{(1)}} + E^{(1)}\ket{\psi^{(0)}}
\end{equation}
One then obtains 
\begin{align}
    E^{(1)} \bra{0_\mu^{(0)}}\ket{\psi^{(0)}} & = \bra{0_\mu^{(0)}} H^1\ket{\psi^{(0)}} \label{eq:app_B_1}\\
    \bra{m^{(0)}}\ket{\psi^{(1)}} & = \frac{\bra{m^{(0)}} H^1 \ket{\psi^{(0)}}}{ \Delta_{0m}} \label{eq:app_B_2}
\end{align}
where $m\neq 0$ runs through all excited states (we have suppressed the degeneracy index $\mu$ for excited states), and $\Delta_{nm}=E_n^{(0)}-E_m^{(0)}$. Since $H^1$ flips an anyon type, it always maps a state within the ground state manifold to an excited state. Therefore, the right side of Eq.~\ref{eq:app_B_1} vanishes and $E^{(1)}=0$. Equation \ref{eq:app_B_2} gives the first-order correction to the state $|\psi\rangle$.

To the second order, we have
\begin{align}
        H^{0}&\ket{\psi^{(2)}} + H^1\ket{\psi^{(1)}} \nonumber\\
        & = E^{(0)}\ket{\psi^{(2)}} +E^{(1)}\ket{\psi^{(1)}}+ E^{(2)}\ket{\psi^{(0)}}
\end{align}
With some straightforward calculations, one can show that
\begin{equation}\label{2nd_order_pert}
    \sum_\nu \bra{0_\mu^{(0)}}\Lambda\ket{0_\nu^{(0)}}\bra{0_\nu^{(0)}}\ket{\psi^{(0)}} = E^{(2)}\bra{0_\mu^{(0)}}\ket{\psi^{(0)}}
\end{equation}
where $\mu,\nu$ are indices running through all unperturbed ground states. We have defined the operator
\begin{equation}
\label{2nd_order_pert2}
    \Lambda = \sum_{m\neq 0} \frac{H^1\ket{m^{(0)}}\bra{m^{(0)}}H^1}{\Delta_{0m}}
\end{equation}
Equation \ref{2nd_order_pert} is merely an eigenvalue problem. Therefore, we obtain the effective low energy Hamiltonian to the second order
\begin{equation}
\label{2nd_order_pert3}
    H_\text{eff}^{(2)} = P_0\Lambda P_0
\end{equation}
where $P_0 = \sum\limits_\nu\ket{0_\nu^{(0)}}\bra{0_\nu^{(0)}}$ is the projector onto the ground-state subspace. The second-order energy correction $E^{(2)}$ is an eigen-energy of the effective Hamiltonian $H_{\rm eff}^{(2)}$. 

It remains to calculate $\Lambda$. Note that $H^1=-\sum_i H_i^{\rm flip}$, where $H_i^{\rm flip}$ flips an anyon type at site $i$. Then, the only second-order process that brings a ground state back to the ground state manifold is the one with $H_i^{\rm flip}$ acting consecutively twice. Accordingly, we find that
\begin{equation}
\begin{split}
H_\text{eff}^{(2)}\ket{\cdots\mu_j\sigma\mu_{j+1}\cdots} = &\sum_{j}\frac{\cos^2\theta}{2r}\delta_{\mu_j,\mu_{j+1}}\ket{\cdots\mu_j\sigma\mu_{j+1}\cdots}\\ 
     + \sum_{j}\frac{\cos^2\theta}{4r}&(\ket{\cdots\mu_j\sigma\mu_{j+1}\cdots} +\ \ket{\cdots\bar{\mu}_j\sigma\mu_{j+1}\cdots})
\end{split}
\end{equation}
where $\bar{\mu}$ represents the $\mathbb{Z}_2$ flip $\mathbb{1}\leftrightarrow \psi$. The first sum on the right side comes from an $H_i^{\rm flip}$ that acts on a ``$\sigma$'' site, while the second sum comes from an $H_i^{\rm flip}$ that acts on a ``$\mu$'' site. 

Consider the short-hand notation for the anti-ferromagnetic states:
\begin{align}
    \ket{\cdots\mu_j\sigma\mu_{j+1}\sigma\cdots}\ \Rightarrow \ \ket{\cdots\mu_j\mu_{j+1}\cdots}
\end{align}
Then, the system can be thought of as a spin-$1/2$ chain, with the length being $L/2$. The effective Hamiltonian $H^{(2)}_{\rm eff}$ can be expressed by Pauli matrices. It is given by
\begin{equation}
    \begin{split}
        H_\text{eff}^{(2)} = &\frac{\cos^2\theta}{2r}\sum_j\left(\frac{1+\sigma^z_j\sigma^z_{j+1}}{2} + \frac{1+\sigma_j^x}{2}\right)\\
        = & \frac{\cos^2\theta}{4r}\left[L + \sum_j\sigma_j^z\sigma_{j+1}^z + \sum_j\sigma_j^x\right]
    \end{split}
\end{equation}
This is just the critical Ising model. The same calculation holds for the other anti-ferromagnetic sector. Therefore, to the second order, the anti-ferromagnetic sectors are deformed into two copies of Ising CFTs.

Finally, we need to compare the energy of the ferromagnetic states and anti-ferromagnetic states to the second order. The three ferromagnetic states remain degenerate eigenstates at second order, but they receive an energy correction. The correction can be calculated in a similar way as above, and we find 
\begin{equation}
    E_{\rm FM}^{(2)} = \frac{L\cos^2\theta}{2r}
\end{equation}
The ground state energy of the Ising model is known exactly \cite{pfeuty1970one}. Then, we find that the lowest energy in the anti-ferromagnetic sectors is
\begin{equation}
    E_{\rm AFM}^{(2)} = \frac{L\cos^2\theta}{4r}\left(1 + \frac{2}{\pi}\right)
\end{equation}
Therefore, the energy difference  between the FM and AFM sectors is
\begin{align}
    \Delta &=E_{\rm AFM}^{(2)}-E_{\rm FM}^{(2)} = \frac{L\cos^2\theta}{4r}\left(\frac{2}{\pi} - 1\right)
\end{align}
Since $\Delta > 0$, the actual ground states of the second order are the three FM states, the same as the odd-$L$ case.

\subsection{CatAFM phase: $r\gg 1$}
\label{app:pert2}

In this case, the model favors anti-ferromagnetic ground states. The ground state manifold of $H^0$ depends on $L$ modulo 4. For simplicity, we take $L$ to be a multiple of four (the other cases can be thought of as translational defects inserted into the ground states).   The ground state degeneracy is $4\cdot 2^{L/4}$. The unperturbed ground states are of the form
\begin{align*}
    \ket{\cdots\sigma\sigma\mu_j\mu_j\sigma\sigma\mu_{j+1}\mu_{j+1}\cdots}
\end{align*}
where $\mu_j=\mathbb{1}$ or $\psi$. The unperturbed energy is $E^{(0)}=-rL$. There are four sectors related by translating one, two, or three lattice sites. Similar to the CatFM case, different sectors do not mix under low order perturbations and we will focus on one of the sectors. 

We find that a fourth-order perturbation calculation is needed to split the ground state degeneracy. First, the correction to the energy vanishes at odd orders. This is because $H^1$ flips anyon types such that acting $H^1$ an odd number of times always takes a state out of the degenerate ground-state space. Second, at the second order,  the correction is a constant. To see that,  we make use of Eqs.~\ref{2nd_order_pert}, \ref{2nd_order_pert2} and \ref{2nd_order_pert3}. After a straightforward calculation, we have
\begin{align}
    H_{\rm eff}^{(2)} = E^{(2)} P_0
\end{align}
where $P_0$ is the projector onto the ground state subspace, and the second-order energy correction
\begin{equation}
    E^{(2)} = -\frac{L}{2r}\sin^2\theta
\end{equation}
It is a constant energy shift for all the states.

Accordingly, we must proceed to the fourth order perturbation to see a non-trivial energy dispersion. With a similar analysis as in Sec.~\ref{app:pert1},  we obtain the fourth-order energy correction, given by 
\begin{equation}
\label{4th-order-pert}
    \sum_\nu \bra{0_\mu^{(0)}}\Omega_1 + \Omega_2\ket{0_\nu^{(0)}}\bra{0_\nu^{(0)}}\ket{\psi^{(0)}} = E^{(4)}\bra{0_\mu^{(0)}}\ket{\psi^{(0)}}
\end{equation}
where 
\begin{align}
    &\Omega_1 = \sum_{l\neq 0} \sum_{m\neq 0} \sum_{k\neq 0} \frac{H^1\ket{l^{(0)}}H^1_{lm}H^1_{mk}\bra{k^{(0)}}H^1}{\Delta_{0l}\Delta_{0m}\Delta_{0k}}\nonumber\\
    &\Omega_2 = -E^{(2)}\sum_{m\neq 0} \frac{H^1 \ket{m^{(0)}}\bra{m^{(0)}}H^1}{\Delta_{0m}^2}
\end{align}
The matrix element $H_{lm}^1=\bra{l^0}H^1\ket{m^0}$. When deriving the above equations, we have used the fact that $E^{(1)}=E^{(3)}=0$. Equation \ref{4th-order-pert} is an eigenvalue problem; thus, we define the fourth-order effective Hamiltonian
\begin{equation}
    H_\text{eff}^{(4)} = P_0 (\Omega_1 + \Omega_2)P_0
\end{equation}
Diagonalizing $H_{\rm eff}^{(4)}$ in the ground state subspace, we will obtain a dispersive energy spectrum.

We still need to find the explicit form of $H_{\rm eff}^{(4)}$. Similar to $H_{\rm eff}^{(2)}$,  $\Omega_2$ only leads to a constant energy shift in the ground state subspace. More explicitly, we find
\begin{equation}
    P_0\Omega_2P_0 = \frac{L^2\sin^4\theta}{8r^3} P_0
    \label{eq:Omega2-shift}
\end{equation}
The $\Omega_1$ part is more complicated. A large number of cases should be considered to obtain the complete action of $\Omega_1$. However, in the majority of cases, the energy correction is again constant. In fact, we only have two non-trivial cases, which can be represented in the following diagram:
\begin{equation*}
    \begin{tikzpicture}[xscale=.7, yscale=.5, nodes={execute at begin node=$, execute at end node=$}]
        \node at (0,0) {\ket{\cdots\sigma\sigma\mu\mu\sigma\sigma\mu'\mu'\cdots}};
        \draw [->] (-0.5,-0.5)--(-2.5,-2) node[midway, left, scale=0.8]{\sin\theta};
        \node at (-2.5,-2.7) {\ket{\cdots\sigma\sigma\mu{\color{red}\sigma}\sigma\sigma\mu'\mu'\cdots}};
        \draw [->] (-2.5,-3.1) -- (-2.5,-5)node[midway, left, scale=0.8]{\displaystyle\frac{\cos\theta}{\sqrt{2}}};
        \node at (-2.5, -5.4) {\ket{\cdots\sigma\sigma{\color{red}\sigma}\sigma\sigma\sigma\mu'\mu'\cdots}};
        \draw [->] (-2.5, -5.8) -- (-2.5 , -7.7)node[midway, left, scale=0.8]{\displaystyle\frac{\cos\theta}{\sqrt{2}}};
        \node at (-2.5, -8.1) {\ket{\cdots\sigma\sigma\sigma{\color{red}\mathbb{\nu}}\sigma\sigma\mu'\mu'\cdots}};
        \draw[->] (-2.5, -8.5) -- (-2.5, -10.4)node[midway, left, scale=0.8]{\sin\theta};
        \node at (-2.5, -10.8){\ket{\cdots\sigma\sigma{\color{red}\nu}\nu\sigma\sigma\mu'\mu'\cdots}};
        \draw [->] (0.5,-0.5) -- (2.5, -2)node[midway, right, scale=0.8]{\,\,\sin\theta};
        \node at (2.5, -2.7) {\ket{\cdots\sigma\sigma\mu\mu{\color{red}\mu}\sigma\mu'\mu'\cdots}};
        \draw [->] (2.5, -3.1) -- (2.5, -5)node[midway, right, scale=0.8]{\delta_{\mu\mu'}\cos\theta};
        \node at (2.5, -5.4) {\ket{\cdots\sigma\sigma\mu\mu\mu{\color{red}\mu}\mu\mu\cdots}};
        \draw [->] (2.5, -5.8) -- (2.5, -7.7)node[midway, right, scale=0.8]{\cos\theta};
        \node at (2.5, -8.1) {\ket{\cdots\sigma\sigma\mu\mu{\color{red}\sigma}\mu\mu\mu\cdots}};
        \draw [->] (2.5, -8.5) -- (2.5, -10.4)node[midway, right, scale=0.8]{\sin\theta};
        \node at (2.5,-10.8) {\ket{\cdots\sigma\sigma\mu\mu\sigma{\color{red}\sigma}\mu\mu\cdots}};
    \end{tikzpicture}
\end{equation*}
where $\mu,\mu', \nu=\mathbb{1}$ or $\psi$, each arrow represents a matrix element of $H^1$, and the red symbols denote those flipped in the preceding step. The four arrows on each side of the diagram correspond to the sequence of $H^1$ matrix elements in $\langle 0_\mu^{(0)}|\Omega_1|0_\nu^{(0)}\rangle$.

The full fourth-order effective Hamiltonian should be $H_{\rm eff}^{(2)}+H_{\rm eff}^{(4)}$. Notice that the constant energy shift \ref{eq:Omega2-shift} from $\Omega_2$ is of the order of $ 1/r^3$, much smaller than the second order correction $E^{(2)}$ in the limit $r\gg 1$. The constant energy shifts from $\Omega_1$ are also of the order of $ 1/r^3$, much smaller than $E^{(2)}$. Hence, evaluating the nontrivial matrix elements of $\Omega_1$ and neglecting all constants of the order of $1/r^3$, we obtain an effective low-energy Hamiltonian
\begin{equation}
       H_\text{eff} = -\frac{L\sin^2\theta}{2r}+ \frac{\sin^2\theta\cos^2\theta}{8r^3}\left(\sum_i\sigma_i^z\sigma_{i+1}^z + \sum_i\sigma_i^x\right)
\label{eq:catAFM_Heff}
\end{equation} 
We have again introduced the short-hand notation for the states:
\begin{align*}
    |\cdots\sigma\sigma\mu_j\mu_j\sigma\sigma\mu_{j+1}\mu_{j+1}\cdots\rangle \ \Rightarrow |\cdots\mu_j\mu_{j+1}\cdots\rangle
\end{align*}
so that it becomes an effective spin-1/2 chain. Note that the effective model has a length $L/4$. By diagonalizing $H_{\rm eff}$, we obtain the fourth-order energy correction to the unperturbed ground states of $H^0$.

The effective model \eqref{eq:catAFM_Heff} is again the critical Ising model. The same calculation holds for the other three sectors (note that mixing between the sectors does not occur in any perturbation of an order smaller than $L/2$). Therefore, the IR theory in the CatAFM phase is described by four copies of Ising CFTs, i.e.,
\[H_\text{eff} \sim \bigoplus_4\text{Ising CFT}\]
It is a gapless phase with a fourfold ground state degeneracy.

\begin{widetext}

\section{Entanglement entropy in the CatAFM phase}
\label{app:entropy_gap2}

In this appendix, we discuss the gap in the entanglement entropy plot (Fig.~\ref{fig:entropy}). The ground state of our model deep in the CatAFM phase is given by Eq.~\ref{eq:catAFM_gs}. Since the $\sigma$ labels are irrelevant for the following discussions, we will omit them for simplicity and write the ground state as
\begin{align}
    |\Psi_{\rm GS}\rangle = \sum_{\{\mu_i\}} \Psi(\mu_1\cdots\mu_{2l})|\{\mu_i\}\rangle
\end{align}
where $l=L/4$ and $\mu_i=\pm 1$ label the two states at site $i$. The wave function is related to that of the transverse-field Ising model by
\begin{align}
    \Psi(\mu_1\cdots\mu_{2l}) = \delta^{\mu_1}_{\mu_2}\delta^{\mu_3}_{\mu_4}\cdots\delta^{\mu_{2l-1}}_{\mu_{2l}}\psi_{\rm Ising}(\mu_1\mu_3\cdots\mu_{2l-1})
\end{align}
where $\psi_{\rm Ising}(\nu_1\nu_2\cdots\nu_{l})$ is the ground-state wave function of the critical Ising chain of length $l$ with free boundary conditions. The density matrix is 
\begin{align}
    {\rho} = |\Psi_{\rm GS}\rangle\langle \Psi_{\rm GS}| = \sum_{\{\mu_i\},\{\mu_i'\}} \rho^{\mu_1\cdots\mu_{2l}}_{\mu_1'\cdots\mu_{2l}'} |\{\mu_i\}\rangle \langle \{\mu_i'\}|
\end{align}
with matrix elements
\begin{align}
    \rho^{\mu_1\cdots\mu_{2l}}_{\mu_1'\cdots\mu_{2l}'} & = \Psi(\mu_1\cdots\mu_{2l}) \Psi^*(\mu_1'\cdots\mu_{2l}') \nonumber\\
    & = (\delta^{\mu_1}_{\mu_2}\delta^{\mu'_1}_{\mu'_2})(\delta^{\mu_3}_{\mu_4}\delta^{\mu'_3}_{\mu'_4})\cdots(\delta^{\mu_{2l-1}}_{\mu_{2l}}\delta^{\mu'_{2l-1}}_{\mu'_{2l}}) \psi_{\rm Ising}(\mu_1\mu_3\cdots\mu_{2l-1})\psi^*_{\rm Ising}(\mu_1'\mu_3'\cdots\mu_{2l-1}')
\label{eq:rho}
\end{align}
Below, we distinguish two types of entanglement cuts: (i) those between sites $2k$ and $2k+1$, which correspond to type-B, C and D cuts in \eqref{eq:20}, and (ii) those between sites $2k-1$ and $2k$, which correspond to type-A cuts in \eqref{eq:20}. Let us discuss the two types separately.

\subsection{$2k|2k+1$ cut}

For an entanglement cut between $\mu_{2k}$ and $\mu_{2k+1}$, the reduced density matrix is obtained by tracing out the degrees of freedom $\mu_{2k+1},\cdots, \mu_{2l}$, i.e., 
\begin{align}
    {\rho}_r = \tr_{i\ge 2k+1}({\rho})
\end{align}
Combining it with Eq.~\ref{eq:rho}, we obtain the matrix element of the reduced density matrix
\begin{align}
    (\rho_r)^{\mu_1\cdots \mu_{2k}}_{\mu_1' \cdots \mu_{2k}'} = (\delta^{\mu_1}_{\mu_2}\delta^{\mu'_1}_{\mu'_2})\cdots(\delta^{\mu_{2k-1}}_{\mu_{2k}}\delta^{\mu'_{2k-1}}_{\mu'_{2k}}) \sum_{\{\mu_{2j+1}|j\ge k\}} 
    \psi_{\rm Ising}(\mu_1\dots\mu_{2k-1}\mu_{2k+1}\dots\mu_{2l-1})\psi^*_{\rm Ising}(\mu'_1\cdots\mu'_{2k-1}\mu_{2k+1}\cdots\mu_{2l-1})
    \label{eq:rdm-even}
\end{align}
where the sums over $\mu_{2j}$ with $j\ge k+1$ have been done. Next, we define new indices $\nu_{j}\equiv \mu_{2j-1}$ and $\eta_{j} \equiv \mu_{2j-1}\mu_{2j}$ (i.e., we use the eigenvalues of $\sigma_{2j-1}^z$ and $\sigma_{2j-1}^z\sigma_{2j}^z$ as new labels for the basis states, yet the basis remains unchanged). Then, the reduced density matrix is expressed as
\begin{align}
    (\rho_r)^{\nu_1\cdots \nu_{k};\eta_1\cdots\eta_k}_{\nu_1' \cdots \nu_{k}';\eta_1'\cdots\eta_k'} =(\delta^{\eta_1}_{1}\delta^{\eta'_1}_{1})\cdots(\delta^{\eta_k}_{1}\delta^{\eta'_k}_{1}) \sum_{\{\nu_j|j\ge k+1\}} \psi_{\rm Ising}(\nu_1\nu_2\cdots\nu_{k}\nu_{k+1}\cdots\nu_{l})\psi^*_{\rm Ising}(\nu'_1\nu'_2\cdots\nu'_{k}\nu_{k+1}\cdots\nu_{l})
\end{align}
It is easy to see that the reduced density matrix is non-zero only when $\eta_j =\eta_j'= 1$ for all $j\le k$. This gives a nonzero block of size $2^k\times 2^k$
\begin{align}
    (\rho_r)^{\nu_1\cdots \nu_{k};1\cdots1}_{\nu_1' \cdots \nu_{k}';1\cdots1} =\sum_{\{\nu_j|j\ge k+1\}}\psi_{\rm Ising}(\nu_1\nu_2\cdots\nu_{k}\nu_{k+1}\cdots\nu_{l})\psi^*_{\rm Ising}(\nu'_1\nu'_2\cdots\nu'_{k}\nu_{k+1}\cdots\nu_{l})
    \label{eq:rdm-ising}
\end{align}
Obviously, the right-hand side is the reduced density matrix of the ground state of the transverse-field Ising model, with the entanglement cut between sites $k$ and $k+1$. Therefore, the entanglement entropy $S(x)$ at $x=2k$ is equal to 
\begin{align}
    S(2k) = S[\rho_{\rm Ising}]
    \label{eq:entropy-2k}
\end{align}
where $\rho_{\rm Ising}$ is the reduced density matrix on the right-hand side of \ref{eq:rdm-ising}.

\subsection{$2k-1|2k$ cut}
For an entanglement cut between $\mu_{2k-1}$ and $\mu_{2k}$, the reduced density matrix is obtained by tracing out $\mu_{2k},\mu_{2k+1},\cdots, \mu_{2l}$, i.e.,
\begin{align}
    \rho_r = \tr_{i\ge 2k}(\rho)
\end{align}
Combining it with Eq.~\ref{eq:rho}, we obtain the matrix element of the reduced density matrix
\begin{align}
    (\rho_r)^{\mu_1\cdots \mu_{2k-1}}_{\mu_1' \cdots \mu_{2k-1}'} = & (\delta^{\mu_1}_{\mu_2}\delta^{\mu'_1}_{\mu'_2})\cdots(\delta^{\mu_{2k-3}}_{\mu_{2k-2}}\delta^{\mu'_{2k-3}}_{\mu'_{2k-2}})(\delta^{\mu_{2k-1}}_{\mu'_{2k-1}})\nonumber\\
    &  \sum_{\{\mu_{2j+1}|j\ge k\}} \psi_{\rm Ising}(\mu_1\dots\mu_{2k-1}\mu_{2k+1}\dots\mu_{2l-1})\psi^*_{\rm Ising}(\mu'_1\cdots\mu'_{2k-1}\mu_{2k+1}\cdots\mu_{2l-1})
\end{align}
where the sums over $\mu_{2j}$ with $j\ge k$ have been done, and the last $\delta$ function results from the sum over $\mu_{2k}$. We again define indices $\nu_{j}\equiv \mu_{2j-1}$ and $\eta_{j} \equiv \mu_{2j-1}\mu_{2j}$. The matrix element of the reduced density matrix is then written as
\begin{align}
    (\rho_r)^{\nu_1\cdots \nu_{k};\eta_1\cdots\eta_{k-1}}_{\nu_1' \cdots \nu_{k}';\eta_1'\cdots\eta_{k-1}'} =(\delta^{\eta_1}_{1}\delta^{\eta'_1}_{1})\cdots(\delta^{\eta_{k-1}}_{1}\delta^{\eta'_{k-1}}_{1})\delta^{\nu_k}_{\nu_k'} \sum_{\{\nu_j|j\ge k+1\}} \psi_{\rm Ising}(\nu_1\nu_2\cdots\nu_{k}\nu_{k+1}\cdots\nu_{l})\psi^*_{\rm Ising}(\nu'_1\nu'_2\cdots\nu'_{k}\nu_{k+1}\cdots\nu_{l})
\end{align}
The matrix element is non-zero only when $\eta_j=\eta_j'=1$ for all $j\le k-1$. This leads to a non-zero block of size $2^k\times 2^k$,
\begin{align}
    (\rho_r)^{\nu_1\cdots \nu_{k};1\cdots1}_{\nu_1' \cdots \nu_{k}';1\cdots1} =\delta^{\nu_k}_{\nu_k'}\sum_{\{\nu_j|j\ge k+1\}}\psi_{\rm Ising}(\nu_1\nu_2\cdots\nu_{k}\nu_{k+1}\cdots\nu_{l})\psi^*_{\rm Ising}(\nu'_1\nu'_2\cdots\nu'_{k}\nu_{k+1}\cdots\nu_{l})
    \label{eq:rdm-nonzero}
\end{align}
The right-hand side is not quite the reduced density matrix of the Ising ground state. Instead, it is block diagonal in $k$-th Ising spin indices $\nu_k$ and $\nu_k'$.  

\begin{figure}
    \centering
    \includegraphics[width=0.5\linewidth]{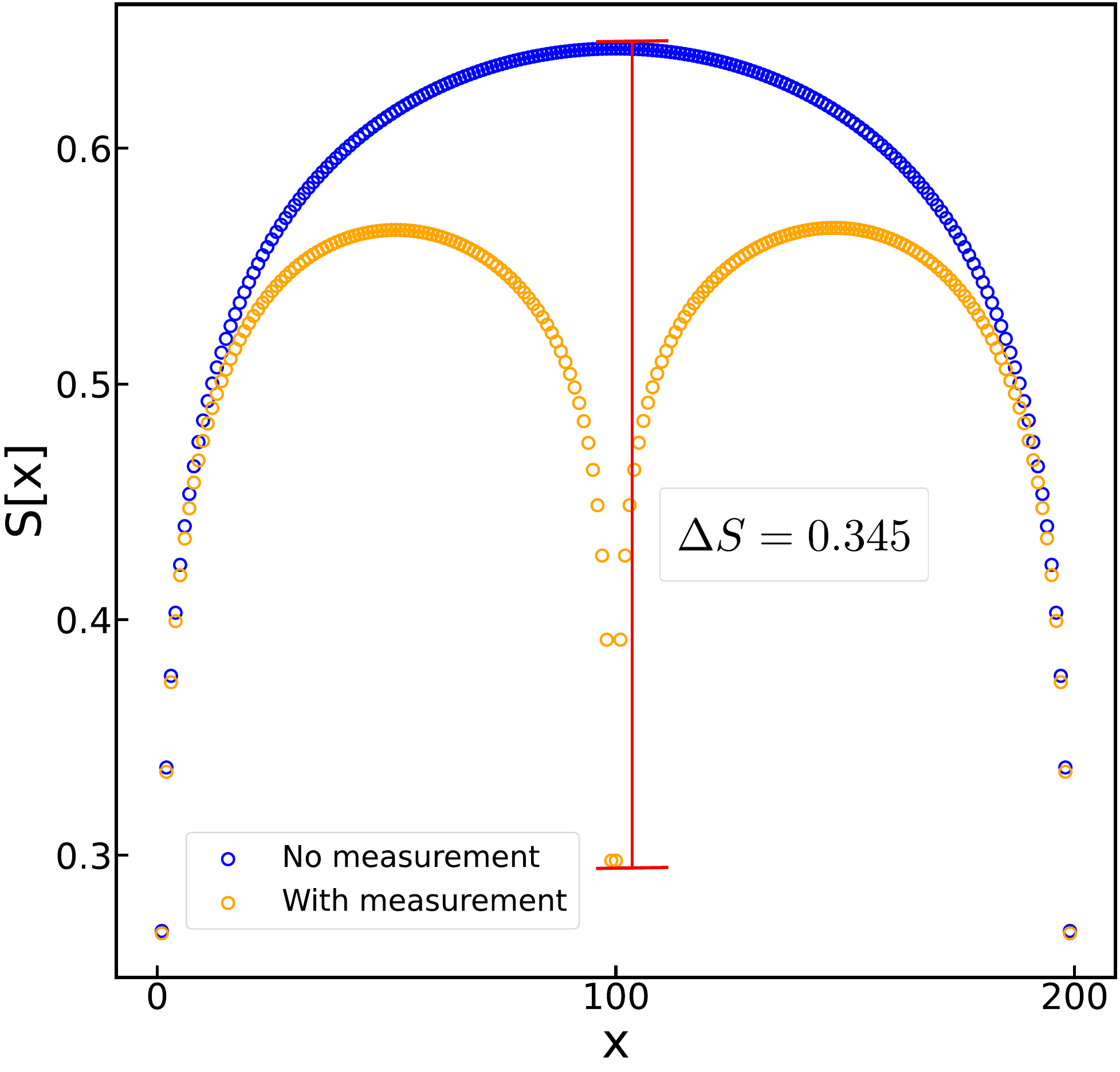}
    \caption{Entanglement entropy of the critical Ising ground state (obtained from DMRG) before (blue) and after (orange) a single-site projective measurement into the $+1$-eigenstate in the $\sigma^z$ basis. The length of the chain is $L = 200$, and the measurement has been performed at site $k = 100$. A drop of entanglement entropy $\Delta S \approx 0.345 \approx \frac{1}{2}\log 2$ is observed at the site of measurement.}
    \label{fig:measurement}
\end{figure}

To proceed, we define two states
\begin{align}
    |\psi^\pm_{\rm Ising}\rangle = \frac{1\pm\sigma_k^z}{\sqrt{2}}|\psi_{\rm Ising}\rangle
\end{align}
They are the states obtained after a projective measurement of $\sigma^z_k$ on the Ising ground state, with eigenvalues being $\pm1$, respectively. The coefficient has been set such that $|\psi^\pm_{\rm Ising}\rangle$ is normalized. Define the reduced density matrix
\begin{align}
    \rho^\pm_{\rm Ising} = \tr_{j\ge k} \left(|\psi^\pm_{\rm Ising}\rangle \langle\psi^\pm_{\rm Ising}|\right)
\end{align}
which are $2^{k-1}\times 2^{k-1}$ matrices. By explicitly comparing the matrix elements, one may easily verify that
\begin{align}
    \rho_r^{\rm block} =\frac{1}{2}
    \begin{pmatrix}
    \rho^+_{\rm Ising} & 0\\
    0 & \rho^-_{\rm Ising}
    \end{pmatrix}
\end{align}
where $\rho_r^{\rm block}$ is the nonzero block given by \ref{eq:rdm-nonzero}.
Then, according to the definition of the von Neumann entanglement entropy, we have 
\begin{equation}
    S(2k-1) = \log 2 + \frac{1}{2}(S[\rho^+_{\rm Ising}] + S[\rho^-_{\rm Ising}])
\end{equation}
where $S(2k-1)$ stands for the entanglement entropy of the original chain with the entanglement cut $x=2k-1$, i.e., between sites $2k-1$ and $2k$. Since $|\psi_{\rm Ising}\rangle$ is symmetric under the $\mathbb{Z}_2$ spin flip symmetry, $\rho^+_{\rm Ising}$ and $ \rho^-_{\rm Ising}$ are related by the  $\mathbb{Z}_2$ unitary transformation, thereby giving rise to equal entanglement entropy. We obtain
\begin{equation}
    S(2k-1) = \log 2+ S[\rho^+_{\rm Ising}].
    \label{eq:entropy-2k-1}
\end{equation}

With \eqref{eq:entropy-2k} and \eqref{eq:entropy-2k-1}, we obtain the difference in entanglement entropy
\begin{align}
    \Delta S &= S(2k-1)-S(2k) = \ln 2+ S[\rho^+_{\rm Ising}] - S[\rho_{\rm Ising}]
\end{align}
It remains to compute the entropy difference between the measured reduced density matrix $\rho_{\rm Ising}^+$ and the unmeasured reduced density matrix $\rho_{\rm Ising}$ of the transverse-field Ising model.  We have numerically computed this difference with DMRG, which gives
\begin{align}
   S[\rho^+_{\rm Ising}] - S[\rho_{\rm Ising}] \approx  - 0.345 \approx -\frac{1}{2}\log 2 
\end{align}
(see Fig.\ref{fig:measurement}). Note that this difference does not change much as $k$ varies, as long as it is not too close to the chain boundary. In general, it is expected that measurement will (though not always) disentangle a quantum state and hence lead to a drop in entropy (see Refs.~\cite{rajabpour2016entanglement, hoshino2025entanglement} for discussions on measurement-induced corrections to entanglement entropy in quantum  critical states.) It would be interesting in the future to analytically compute the difference $S[\rho^+_{\rm Ising}] - S[\rho_{\rm Ising}]$ since the ground state of the Ising model is known, and we conjecture $\frac{1}{2}\ln2$ is the exact value. Therefore, we have
\begin{align}
    \Delta S \approx \ln 2 - \frac{1}{2}\ln 2 = \frac{1}{2}\ln 2
\end{align}
This concludes our explanation of the entanglement gap observed in Fig.~\ref{fig:entropy}.

\end{widetext}

\bibliography{references}

\begin{thebibliography}{78}%
\makeatletter
\providecommand \@ifxundefined [1]{%
 \@ifx{#1\undefined}
}%
\providecommand \@ifnum [1]{%
 \ifnum #1\expandafter \@firstoftwo
 \else \expandafter \@secondoftwo
 \fi
}%
\providecommand \@ifx [1]{%
 \ifx #1\expandafter \@firstoftwo
 \else \expandafter \@secondoftwo
 \fi
}%
\providecommand \natexlab [1]{#1}%
\providecommand \enquote  [1]{``#1''}%
\providecommand \bibnamefont  [1]{#1}%
\providecommand \bibfnamefont [1]{#1}%
\providecommand \citenamefont [1]{#1}%
\providecommand \href@noop [0]{\@secondoftwo}%
\providecommand \href [0]{\begingroup \@sanitize@url \@href}%
\providecommand \@href[1]{\@@startlink{#1}\@@href}%
\providecommand \@@href[1]{\endgroup#1\@@endlink}%
\providecommand \@sanitize@url [0]{\catcode `\\12\catcode `\$12\catcode
  `\&12\catcode `\#12\catcode `\^12\catcode `\_12\catcode `\%12\relax}%
\providecommand \@@startlink[1]{}%
\providecommand \@@endlink[0]{}%
\providecommand \url  [0]{\begingroup\@sanitize@url \@url }%
\providecommand \@url [1]{\endgroup\@href {#1}{\urlprefix }}%
\providecommand \urlprefix  [0]{URL }%
\providecommand \Eprint [0]{\href }%
\providecommand \doibase [0]{https://doi.org/}%
\providecommand \selectlanguage [0]{\@gobble}%
\providecommand \bibinfo  [0]{\@secondoftwo}%
\providecommand \bibfield  [0]{\@secondoftwo}%
\providecommand \translation [1]{[#1]}%
\providecommand \BibitemOpen [0]{}%
\providecommand \bibitemStop [0]{}%
\providecommand \bibitemNoStop [0]{.\EOS\space}%
\providecommand \EOS [0]{\spacefactor3000\relax}%
\providecommand \BibitemShut  [1]{\csname bibitem#1\endcsname}%
\let\auto@bib@innerbib\@empty
\bibitem [{\citenamefont {Verlinde}(1988{\natexlab{a}})}]{Verlinde1988}%
  \BibitemOpen
  \bibfield  {author} {\bibinfo {author} {\bibfnamefont {E.}~\bibnamefont
  {Verlinde}},\ }\bibfield  {title} {\bibinfo {title} {Fusion rules and modular
  transformations in 2d conformal field theory},\ }\href
  {https://doi.org/https://doi.org/10.1016/0550-3213(88)90603-7} {\bibfield
  {journal} {\bibinfo  {journal} {Nuclear Physics B}\ }\textbf {\bibinfo
  {volume} {300}},\ \bibinfo {pages} {360} (\bibinfo {year}
  {1988}{\natexlab{a}})}\BibitemShut {NoStop}%
\bibitem [{\citenamefont {{McGreevy}}(2023)}]{McGreevyRev2023}%
  \BibitemOpen
  \bibfield  {author} {\bibinfo {author} {\bibfnamefont {J.}~\bibnamefont
  {{McGreevy}}},\ }\bibfield  {title} {\bibinfo {title} {{Generalized
  Symmetries in Condensed Matter}},\ }\href
  {https://doi.org/10.1146/annurev-conmatphys-040721-021029} {\bibfield
  {journal} {\bibinfo  {journal} {Annual Review of Condensed Matter Physics}\
  }\textbf {\bibinfo {volume} {14}},\ \bibinfo {pages} {57} (\bibinfo {year}
  {2023})},\ \Eprint {https://arxiv.org/abs/2204.03045} {arXiv:2204.03045}
  \BibitemShut {NoStop}%
\bibitem [{\citenamefont {{Shao}}(2023)}]{shao2023s}%
  \BibitemOpen
  \bibfield  {author} {\bibinfo {author} {\bibfnamefont {S.-H.}\ \bibnamefont
  {{Shao}}},\ }\bibfield  {title} {\bibinfo {title} {{What's Done Cannot Be
  Undone: TASI Lectures on Non-Invertible Symmetries}},\ }\href@noop {}
  {\bibfield  {journal} {\bibinfo  {journal} {arXiv e-prints}\ } (\bibinfo
  {year} {2023})},\ \Eprint {https://arxiv.org/abs/2308.00747}
  {arXiv:2308.00747} \BibitemShut {NoStop}%
\bibitem [{\citenamefont {Schäfer-Nameki}(2024)}]{schafer-nameki_ictp_2023}%
  \BibitemOpen
  \bibfield  {author} {\bibinfo {author} {\bibfnamefont {S.}~\bibnamefont
  {Schäfer-Nameki}},\ }\bibfield  {title} {\bibinfo {title} {{ICTP} lectures
  on (non-)invertible generalized symmetries},\ }\href
  {https://doi.org/https://doi.org/10.1016/j.physrep.2024.01.007} {\bibfield
  {journal} {\bibinfo  {journal} {Physics Reports}\ }\textbf {\bibinfo {volume}
  {1063}},\ \bibinfo {pages} {1} (\bibinfo {year} {2024})}\BibitemShut
  {NoStop}%
\bibitem [{\citenamefont {Bhardwaj}\ \emph
  {et~al.}(2024{\natexlab{a}})\citenamefont {Bhardwaj}, \citenamefont
  {Bottini}, \citenamefont {Fraser-Taliente}, \citenamefont {Gladden},
  \citenamefont {Gould}, \citenamefont {Platschorre},\ and\ \citenamefont
  {Tillim}}]{bhardwaj_lectures_2023}%
  \BibitemOpen
  \bibfield  {author} {\bibinfo {author} {\bibfnamefont {L.}~\bibnamefont
  {Bhardwaj}}, \bibinfo {author} {\bibfnamefont {L.~E.}\ \bibnamefont
  {Bottini}}, \bibinfo {author} {\bibfnamefont {L.}~\bibnamefont
  {Fraser-Taliente}}, \bibinfo {author} {\bibfnamefont {L.}~\bibnamefont
  {Gladden}}, \bibinfo {author} {\bibfnamefont {D.~S.}\ \bibnamefont {Gould}},
  \bibinfo {author} {\bibfnamefont {A.}~\bibnamefont {Platschorre}},\ and\
  \bibinfo {author} {\bibfnamefont {H.}~\bibnamefont {Tillim}},\ }\bibfield
  {title} {\bibinfo {title} {Lectures on generalized symmetries},\ }\href
  {https://doi.org/https://doi.org/10.1016/j.physrep.2023.11.002} {\bibfield
  {journal} {\bibinfo  {journal} {Physics Reports}\ }\textbf {\bibinfo {volume}
  {1051}},\ \bibinfo {pages} {1} (\bibinfo {year}
  {2024}{\natexlab{a}})}\BibitemShut {NoStop}%
\bibitem [{\citenamefont {Luo}\ \emph {et~al.}(2024)\citenamefont {Luo},
  \citenamefont {Wang},\ and\ \citenamefont {Wang}}]{LuoWangWangRev2024}%
  \BibitemOpen
  \bibfield  {author} {\bibinfo {author} {\bibfnamefont {R.}~\bibnamefont
  {Luo}}, \bibinfo {author} {\bibfnamefont {Q.-R.}\ \bibnamefont {Wang}},\ and\
  \bibinfo {author} {\bibfnamefont {Y.-N.}\ \bibnamefont {Wang}},\ }\bibfield
  {title} {\bibinfo {title} {Lecture notes on generalized symmetries and
  applications},\ }\href
  {https://doi.org/https://doi.org/10.1016/j.physrep.2024.02.002} {\bibfield
  {journal} {\bibinfo  {journal} {Physics Reports}\ }\textbf {\bibinfo {volume}
  {1065}},\ \bibinfo {pages} {1} (\bibinfo {year} {2024})}\BibitemShut
  {NoStop}%
\bibitem [{\citenamefont {{Feiguin}}\ \emph {et~al.}(2007)\citenamefont
  {{Feiguin}}, \citenamefont {{Trebst}}, \citenamefont {{Ludwig}},
  \citenamefont {{Troyer}}, \citenamefont {{Kitaev}}, \citenamefont {{Wang}},\
  and\ \citenamefont {{Freedman}}}]{feiguin_interacting_2007}%
  \BibitemOpen
  \bibfield  {author} {\bibinfo {author} {\bibfnamefont {A.}~\bibnamefont
  {{Feiguin}}}, \bibinfo {author} {\bibfnamefont {S.}~\bibnamefont {{Trebst}}},
  \bibinfo {author} {\bibfnamefont {A.~W.~W.}\ \bibnamefont {{Ludwig}}},
  \bibinfo {author} {\bibfnamefont {M.}~\bibnamefont {{Troyer}}}, \bibinfo
  {author} {\bibfnamefont {A.}~\bibnamefont {{Kitaev}}}, \bibinfo {author}
  {\bibfnamefont {Z.}~\bibnamefont {{Wang}}},\ and\ \bibinfo {author}
  {\bibfnamefont {M.~H.}\ \bibnamefont {{Freedman}}},\ }\bibfield  {title}
  {\bibinfo {title} {{Interacting Anyons in Topological Quantum Liquids: The
  Golden Chain}},\ }\href {https://doi.org/10.1103/PhysRevLett.98.160409}
  {\bibfield  {journal} {\bibinfo  {journal} {\prl}\ }\textbf {\bibinfo
  {volume} {98}},\ \bibinfo {eid} {160409} (\bibinfo {year} {2007})},\ \Eprint
  {https://arxiv.org/abs/cond-mat/0612341} {arXiv:cond-mat/0612341}
  \BibitemShut {NoStop}%
\bibitem [{\citenamefont {{Bhardwaj}}\ and\ \citenamefont
  {{Tachikawa}}(2018)}]{BhardwajTachikawa2018JHEP}%
  \BibitemOpen
  \bibfield  {author} {\bibinfo {author} {\bibfnamefont {L.}~\bibnamefont
  {{Bhardwaj}}}\ and\ \bibinfo {author} {\bibfnamefont {Y.}~\bibnamefont
  {{Tachikawa}}},\ }\bibfield  {title} {\bibinfo {title} {{On finite symmetries
  and their gauging in two dimensions}},\ }\href
  {https://doi.org/10.1007/JHEP03(2018)189} {\bibfield  {journal} {\bibinfo
  {journal} {Journal of High Energy Physics}\ }\textbf {\bibinfo {volume}
  {2018}},\ \bibinfo {eid} {189} (\bibinfo {year} {2018})},\ \Eprint
  {https://arxiv.org/abs/1704.02330} {arXiv:1704.02330} \BibitemShut {NoStop}%
\bibitem [{\citenamefont {{Chang}}\ \emph {et~al.}(2019)\citenamefont
  {{Chang}}, \citenamefont {{Lin}}, \citenamefont {{Shao}}, \citenamefont
  {{Wang}},\ and\ \citenamefont {{Yin}}}]{Chang_etal_2019JHEP}%
  \BibitemOpen
  \bibfield  {author} {\bibinfo {author} {\bibfnamefont {C.-M.}\ \bibnamefont
  {{Chang}}}, \bibinfo {author} {\bibfnamefont {Y.-H.}\ \bibnamefont {{Lin}}},
  \bibinfo {author} {\bibfnamefont {S.-H.}\ \bibnamefont {{Shao}}}, \bibinfo
  {author} {\bibfnamefont {Y.}~\bibnamefont {{Wang}}},\ and\ \bibinfo {author}
  {\bibfnamefont {X.}~\bibnamefont {{Yin}}},\ }\bibfield  {title} {\bibinfo
  {title} {{Topological defect lines and renormalization group flows in two
  dimensions}},\ }\href {https://doi.org/10.1007/JHEP01(2019)026} {\bibfield
  {journal} {\bibinfo  {journal} {Journal of High Energy Physics}\ }\textbf
  {\bibinfo {volume} {2019}},\ \bibinfo {eid} {26} (\bibinfo {year} {2019})},\
  \Eprint {https://arxiv.org/abs/1802.04445} {arXiv:1802.04445} \BibitemShut
  {NoStop}%
\bibitem [{\citenamefont {Ji}\ and\ \citenamefont {Wen}(2020)}]{JiWen2020PRR}%
  \BibitemOpen
  \bibfield  {author} {\bibinfo {author} {\bibfnamefont {W.}~\bibnamefont
  {Ji}}\ and\ \bibinfo {author} {\bibfnamefont {X.-G.}\ \bibnamefont {Wen}},\
  }\bibfield  {title} {\bibinfo {title} {Categorical symmetry and noninvertible
  anomaly in symmetry-breaking and topological phase transitions},\ }\href
  {https://doi.org/10.1103/PhysRevResearch.2.033417} {\bibfield  {journal}
  {\bibinfo  {journal} {Phys. Rev. Res.}\ }\textbf {\bibinfo {volume} {2}},\
  \bibinfo {pages} {033417} (\bibinfo {year} {2020})}\BibitemShut {NoStop}%
\bibitem [{\citenamefont {{Jia}}\ \emph {et~al.}(2025)\citenamefont {{Jia}},
  \citenamefont {{Luo}}, \citenamefont {{Tian}}, \citenamefont {{Wang}},\ and\
  \citenamefont {{Zhang}}}]{YiNanWangarXiv2025}%
  \BibitemOpen
  \bibfield  {author} {\bibinfo {author} {\bibfnamefont {Q.}~\bibnamefont
  {{Jia}}}, \bibinfo {author} {\bibfnamefont {R.}~\bibnamefont {{Luo}}},
  \bibinfo {author} {\bibfnamefont {J.}~\bibnamefont {{Tian}}}, \bibinfo
  {author} {\bibfnamefont {Y.-N.}\ \bibnamefont {{Wang}}},\ and\ \bibinfo
  {author} {\bibfnamefont {Y.}~\bibnamefont {{Zhang}}},\ }\bibfield  {title}
  {\bibinfo {title} {{Categorical Continuous Symmetry}},\ }\href@noop {}
  {\bibfield  {journal} {\bibinfo  {journal} {arXiv e-prints}\ } (\bibinfo
  {year} {2025})},\ \Eprint {https://arxiv.org/abs/2509.13170}
  {arXiv:2509.13170} \BibitemShut {NoStop}%
\bibitem [{\citenamefont {Etingof}\ \emph {et~al.}(2015)\citenamefont
  {Etingof}, \citenamefont {Gelaki}, \citenamefont {Nikshych},\ and\
  \citenamefont {Ostrik}}]{Etingof2015tensor}%
  \BibitemOpen
  \bibfield  {author} {\bibinfo {author} {\bibfnamefont {P.}~\bibnamefont
  {Etingof}}, \bibinfo {author} {\bibfnamefont {S.}~\bibnamefont {Gelaki}},
  \bibinfo {author} {\bibfnamefont {D.}~\bibnamefont {Nikshych}},\ and\
  \bibinfo {author} {\bibfnamefont {V.}~\bibnamefont {Ostrik}},\ }\href@noop {}
  {\emph {\bibinfo {title} {Tensor Categories}}}\ (\bibinfo  {publisher}
  {American Mathematical Society},\ \bibinfo {year} {2015})\BibitemShut
  {NoStop}%
\bibitem [{\citenamefont {{Gaiotto}}\ \emph {et~al.}(2015)\citenamefont
  {{Gaiotto}}, \citenamefont {{Kapustin}}, \citenamefont {{Seiberg}},\ and\
  \citenamefont {{Willett}}}]{gaiotto_generalized_2015}%
  \BibitemOpen
  \bibfield  {author} {\bibinfo {author} {\bibfnamefont {D.}~\bibnamefont
  {{Gaiotto}}}, \bibinfo {author} {\bibfnamefont {A.}~\bibnamefont
  {{Kapustin}}}, \bibinfo {author} {\bibfnamefont {N.}~\bibnamefont
  {{Seiberg}}},\ and\ \bibinfo {author} {\bibfnamefont {B.}~\bibnamefont
  {{Willett}}},\ }\bibfield  {title} {\bibinfo {title} {{Generalized global
  symmetries}},\ }\href {https://doi.org/10.1007/JHEP02(2015)172} {\bibfield
  {journal} {\bibinfo  {journal} {Journal of High Energy Physics}\ }\textbf
  {\bibinfo {volume} {2015}},\ \bibinfo {eid} {172} (\bibinfo {year} {2015})},\
  \Eprint {https://arxiv.org/abs/1412.5148} {arXiv:1412.5148} \BibitemShut
  {NoStop}%
\bibitem [{\citenamefont {'t~Hooft}(1980)}]{tHooft1979}%
  \BibitemOpen
  \bibfield  {author} {\bibinfo {author} {\bibfnamefont {G.}~\bibnamefont
  {'t~Hooft}},\ }\bibfield  {title} {\bibinfo {title} {{Naturalness, chiral
  symmetry, and spontaneous chiral symmetry breaking}},\ }\href
  {https://doi.org/10.1007/978-1-4684-7571-5_9} {\bibfield  {journal} {\bibinfo
   {journal} {NATO Sci. Ser. B}\ }\textbf {\bibinfo {volume} {59}},\ \bibinfo
  {pages} {135} (\bibinfo {year} {1980})}\BibitemShut {NoStop}%
\bibitem [{\citenamefont {{Douglas}}\ and\ \citenamefont
  {{Reutter}}(2018)}]{Douglas2018}%
  \BibitemOpen
  \bibfield  {author} {\bibinfo {author} {\bibfnamefont {C.~L.}\ \bibnamefont
  {{Douglas}}}\ and\ \bibinfo {author} {\bibfnamefont {D.~J.}\ \bibnamefont
  {{Reutter}}},\ }\bibfield  {title} {\bibinfo {title} {{Fusion 2-categories
  and a state-sum invariant for 4-manifolds}},\ }\href@noop {} {\bibfield
  {journal} {\bibinfo  {journal} {arXiv e-prints}\ } (\bibinfo {year}
  {2018})},\ \Eprint {https://arxiv.org/abs/1812.11933} {arXiv:1812.11933}
  \BibitemShut {NoStop}%
\bibitem [{\citenamefont {{Kong}}\ and\ \citenamefont
  {{Wen}}(2014)}]{KongWen2014}%
  \BibitemOpen
  \bibfield  {author} {\bibinfo {author} {\bibfnamefont {L.}~\bibnamefont
  {{Kong}}}\ and\ \bibinfo {author} {\bibfnamefont {X.-G.}\ \bibnamefont
  {{Wen}}},\ }\bibfield  {title} {\bibinfo {title} {{Braided fusion categories,
  gravitational anomalies, and the mathematical framework for topological
  orders in any dimensions}},\ }\href@noop {} {\bibfield  {journal} {\bibinfo
  {journal} {arXiv e-prints}\ } (\bibinfo {year} {2014})},\ \Eprint
  {https://arxiv.org/abs/1405.5858} {arXiv:1405.5858} \BibitemShut {NoStop}%
\bibitem [{\citenamefont {{Kong}}\ \emph {et~al.}(2020)\citenamefont {{Kong}},
  \citenamefont {{Lan}}, \citenamefont {{Wen}}, \citenamefont {{Zhang}},\ and\
  \citenamefont {{Zheng}}}]{KLWZZ_higherAlg_2020}%
  \BibitemOpen
  \bibfield  {author} {\bibinfo {author} {\bibfnamefont {L.}~\bibnamefont
  {{Kong}}}, \bibinfo {author} {\bibfnamefont {T.}~\bibnamefont {{Lan}}},
  \bibinfo {author} {\bibfnamefont {X.-G.}\ \bibnamefont {{Wen}}}, \bibinfo
  {author} {\bibfnamefont {Z.-H.}\ \bibnamefont {{Zhang}}},\ and\ \bibinfo
  {author} {\bibfnamefont {H.}~\bibnamefont {{Zheng}}},\ }\bibfield  {title}
  {\bibinfo {title} {{Algebraic higher symmetry and categorical symmetry: A
  holographic and entanglement view of symmetry}},\ }\href
  {https://doi.org/10.1103/PhysRevResearch.2.043086} {\bibfield  {journal}
  {\bibinfo  {journal} {Physical Review Research}\ }\textbf {\bibinfo {volume}
  {2}},\ \bibinfo {eid} {043086} (\bibinfo {year} {2020})},\ \Eprint
  {https://arxiv.org/abs/2005.14178} {arXiv:2005.14178} \BibitemShut {NoStop}%
\bibitem [{\citenamefont {{Freed}}\ \emph {et~al.}(2022)\citenamefont
  {{Freed}}, \citenamefont {{Moore}},\ and\ \citenamefont
  {{Teleman}}}]{Freed2022}%
  \BibitemOpen
  \bibfield  {author} {\bibinfo {author} {\bibfnamefont {D.~S.}\ \bibnamefont
  {{Freed}}}, \bibinfo {author} {\bibfnamefont {G.~W.}\ \bibnamefont
  {{Moore}}},\ and\ \bibinfo {author} {\bibfnamefont {C.}~\bibnamefont
  {{Teleman}}},\ }\bibfield  {title} {\bibinfo {title} {{Topological symmetry
  in quantum field theory}},\ }\href@noop {} {\bibfield  {journal} {\bibinfo
  {journal} {arXiv e-prints}\ } (\bibinfo {year} {2022})},\ \Eprint
  {https://arxiv.org/abs/2209.07471} {arXiv:2209.07471} \BibitemShut {NoStop}%
\bibitem [{\citenamefont {Blume}\ \emph {et~al.}(1971)\citenamefont {Blume},
  \citenamefont {Emery},\ and\ \citenamefont {Griffiths}}]{BlumeEmeryPRB1971}%
  \BibitemOpen
  \bibfield  {author} {\bibinfo {author} {\bibfnamefont {M.}~\bibnamefont
  {Blume}}, \bibinfo {author} {\bibfnamefont {V.~J.}\ \bibnamefont {Emery}},\
  and\ \bibinfo {author} {\bibfnamefont {R.~B.}\ \bibnamefont {Griffiths}},\
  }\bibfield  {title} {\bibinfo {title} {Ising model for the
  $\ensuremath{\lambda}$ transition and phase separation in
  {${\mathrm{He}}^{3}$-${\mathrm{He}}^{4}$} mixtures},\ }\href
  {https://doi.org/10.1103/PhysRevA.4.1071} {\bibfield  {journal} {\bibinfo
  {journal} {Phys. Rev. A}\ }\textbf {\bibinfo {volume} {4}},\ \bibinfo {pages}
  {1071} (\bibinfo {year} {1971})}\BibitemShut {NoStop}%
\bibitem [{\citenamefont {O'Brien}\ and\ \citenamefont
  {Fendley}(2018)}]{OBrienFendleyPRL2018}%
  \BibitemOpen
  \bibfield  {author} {\bibinfo {author} {\bibfnamefont {E.}~\bibnamefont
  {O'Brien}}\ and\ \bibinfo {author} {\bibfnamefont {P.}~\bibnamefont
  {Fendley}},\ }\bibfield  {title} {\bibinfo {title} {{Lattice Supersymmetry
  and Order-Disorder Coexistence in the Tricritical Ising Model}},\ }\href
  {https://doi.org/10.1103/PhysRevLett.120.206403} {\bibfield  {journal}
  {\bibinfo  {journal} {Phys. Rev. Lett.}\ }\textbf {\bibinfo {volume} {120}},\
  \bibinfo {pages} {206403} (\bibinfo {year} {2018})}\BibitemShut {NoStop}%
\bibitem [{\citenamefont {{Seiberg}}\ \emph {et~al.}(2024)\citenamefont
  {{Seiberg}}, \citenamefont {{Seifnashri}},\ and\ \citenamefont
  {{Shao}}}]{seiberg_non-invertible_2024}%
  \BibitemOpen
  \bibfield  {author} {\bibinfo {author} {\bibfnamefont {N.}~\bibnamefont
  {{Seiberg}}}, \bibinfo {author} {\bibfnamefont {S.}~\bibnamefont
  {{Seifnashri}}},\ and\ \bibinfo {author} {\bibfnamefont {S.-H.}\ \bibnamefont
  {{Shao}}},\ }\bibfield  {title} {\bibinfo {title} {{Non-invertible symmetries
  and LSM-type constraints on a tensor product Hilbert space}},\ }\href
  {https://doi.org/10.21468/SciPostPhys.16.6.154} {\bibfield  {journal}
  {\bibinfo  {journal} {SciPost Physics}\ }\textbf {\bibinfo {volume} {16}},\
  \bibinfo {eid} {154} (\bibinfo {year} {2024})},\ \Eprint
  {https://arxiv.org/abs/2401.12281} {arXiv:2401.12281} \BibitemShut {NoStop}%
\bibitem [{\citenamefont {Wen}(2017)}]{Wee_Zoo_paper_2017}%
  \BibitemOpen
  \bibfield  {author} {\bibinfo {author} {\bibfnamefont {X.-G.}\ \bibnamefont
  {Wen}},\ }\bibfield  {title} {\bibinfo {title} {Colloquium: Zoo of
  quantum-topological phases of matter},\ }\href
  {https://doi.org/10.1103/RevModPhys.89.041004} {\bibfield  {journal}
  {\bibinfo  {journal} {Rev. Mod. Phys.}\ }\textbf {\bibinfo {volume} {89}},\
  \bibinfo {pages} {041004} (\bibinfo {year} {2017})}\BibitemShut {NoStop}%
\bibitem [{\citenamefont {Wen}(2019)}]{Wen2019PRB}%
  \BibitemOpen
  \bibfield  {author} {\bibinfo {author} {\bibfnamefont {X.-G.}\ \bibnamefont
  {Wen}},\ }\bibfield  {title} {\bibinfo {title} {Emergent anomalous higher
  symmetries from topological order and from dynamical electromagnetic field in
  condensed matter systems},\ }\href
  {https://doi.org/10.1103/PhysRevB.99.205139} {\bibfield  {journal} {\bibinfo
  {journal} {Phys. Rev. B}\ }\textbf {\bibinfo {volume} {99}},\ \bibinfo
  {pages} {205139} (\bibinfo {year} {2019})}\BibitemShut {NoStop}%
\bibitem [{\citenamefont {{Lake}}(2018)}]{Lake2018}%
  \BibitemOpen
  \bibfield  {author} {\bibinfo {author} {\bibfnamefont {E.}~\bibnamefont
  {{Lake}}},\ }\bibfield  {title} {\bibinfo {title} {{Higher-form symmetries
  and spontaneous symmetry breaking}},\ }\href@noop {} {\bibfield  {journal}
  {\bibinfo  {journal} {arXiv e-prints}\ } (\bibinfo {year} {2018})},\ \Eprint
  {https://arxiv.org/abs/1802.07747} {arXiv:1802.07747} \BibitemShut {NoStop}%
\bibitem [{\citenamefont {Thorngren}\ and\ \citenamefont
  {Wang}(2024)}]{ThorngrenWang1}%
  \BibitemOpen
  \bibfield  {author} {\bibinfo {author} {\bibfnamefont {R.}~\bibnamefont
  {Thorngren}}\ and\ \bibinfo {author} {\bibfnamefont {Y.}~\bibnamefont
  {Wang}},\ }\bibfield  {title} {\bibinfo {title} {{Fusion category symmetry.
  Part I. Anomaly in-flow and gapped phases}},\ }\href
  {https://doi.org/10.1007/JHEP04(2024)132} {\bibfield  {journal} {\bibinfo
  {journal} {JHEP}\ }\textbf {\bibinfo {volume} {04}},\ \bibinfo {pages}
  {132}},\ \Eprint {https://arxiv.org/abs/1912.02817} {arXiv:1912.02817}
  \BibitemShut {NoStop}%
\bibitem [{\citenamefont {Choi}\ \emph {et~al.}(2022)\citenamefont {Choi},
  \citenamefont {Lam},\ and\ \citenamefont {Shao}}]{choi2022noninvertible}%
  \BibitemOpen
  \bibfield  {author} {\bibinfo {author} {\bibfnamefont {Y.}~\bibnamefont
  {Choi}}, \bibinfo {author} {\bibfnamefont {H.~T.}\ \bibnamefont {Lam}},\ and\
  \bibinfo {author} {\bibfnamefont {S.-H.}\ \bibnamefont {Shao}},\ }\bibfield
  {title} {\bibinfo {title} {Noninvertible global symmetries in the standard
  model},\ }\href {https://doi.org/10.1103/PhysRevLett.129.161601} {\bibfield
  {journal} {\bibinfo  {journal} {Phys. Rev. Lett.}\ }\textbf {\bibinfo
  {volume} {129}},\ \bibinfo {pages} {161601} (\bibinfo {year}
  {2022})}\BibitemShut {NoStop}%
\bibitem [{\citenamefont {Bhardwaj}\ \emph
  {et~al.}(2024{\natexlab{b}})\citenamefont {Bhardwaj}, \citenamefont
  {Bottini}, \citenamefont {Pajer},\ and\ \citenamefont
  {Sch\"afer-Nameki}}]{bhardwaj2024categorical}%
  \BibitemOpen
  \bibfield  {author} {\bibinfo {author} {\bibfnamefont {L.}~\bibnamefont
  {Bhardwaj}}, \bibinfo {author} {\bibfnamefont {L.~E.}\ \bibnamefont
  {Bottini}}, \bibinfo {author} {\bibfnamefont {D.}~\bibnamefont {Pajer}},\
  and\ \bibinfo {author} {\bibfnamefont {S.}~\bibnamefont {Sch\"afer-Nameki}},\
  }\bibfield  {title} {\bibinfo {title} {{Categorical Landau Paradigm for
  Gapped Phases}},\ }\href {https://doi.org/10.1103/PhysRevLett.133.161601}
  {\bibfield  {journal} {\bibinfo  {journal} {Phys. Rev. Lett.}\ }\textbf
  {\bibinfo {volume} {133}},\ \bibinfo {pages} {161601} (\bibinfo {year}
  {2024}{\natexlab{b}})}\BibitemShut {NoStop}%
\bibitem [{\citenamefont {{Bhardwaj}}\ \emph
  {et~al.}(2025{\natexlab{a}})\citenamefont {{Bhardwaj}}, \citenamefont
  {{Bottini}}, \citenamefont {{Pajer}},\ and\ \citenamefont
  {{Sch{\"a}fer-Nameki}}}]{Bhardwaj_generalizedLandau2}%
  \BibitemOpen
  \bibfield  {author} {\bibinfo {author} {\bibfnamefont {L.}~\bibnamefont
  {{Bhardwaj}}}, \bibinfo {author} {\bibfnamefont {L.~E.}\ \bibnamefont
  {{Bottini}}}, \bibinfo {author} {\bibfnamefont {D.}~\bibnamefont {{Pajer}}},\
  and\ \bibinfo {author} {\bibfnamefont {S.}~\bibnamefont
  {{Sch{\"a}fer-Nameki}}},\ }\bibfield  {title} {\bibinfo {title} {{Gapped
  phases with non-invertible symmetries: (1+1)d}},\ }\href
  {https://doi.org/10.21468/SciPostPhys.18.1.032} {\bibfield  {journal}
  {\bibinfo  {journal} {SciPost Physics}\ }\textbf {\bibinfo {volume} {18}},\
  \bibinfo {eid} {032} (\bibinfo {year} {2025}{\natexlab{a}})},\ \Eprint
  {https://arxiv.org/abs/2310.03784} {arXiv:2310.03784} \BibitemShut {NoStop}%
\bibitem [{\citenamefont {{Bhardwaj}}\ \emph
  {et~al.}(2025{\natexlab{b}})\citenamefont {{Bhardwaj}}, \citenamefont
  {{Bottini}}, \citenamefont {{Sch{\"a}fer-Nameki}},\ and\ \citenamefont
  {{Tiwari}}}]{bhardwaj_illustrating_2024}%
  \BibitemOpen
  \bibfield  {author} {\bibinfo {author} {\bibfnamefont {L.}~\bibnamefont
  {{Bhardwaj}}}, \bibinfo {author} {\bibfnamefont {L.~E.}\ \bibnamefont
  {{Bottini}}}, \bibinfo {author} {\bibfnamefont {S.}~\bibnamefont
  {{Sch{\"a}fer-Nameki}}},\ and\ \bibinfo {author} {\bibfnamefont
  {A.}~\bibnamefont {{Tiwari}}},\ }\bibfield  {title} {\bibinfo {title}
  {{Illustrating the categorical Landau paradigm in lattice models}},\ }\href
  {https://doi.org/10.1103/PhysRevB.111.054432} {\bibfield  {journal} {\bibinfo
   {journal} {\prb}\ }\textbf {\bibinfo {volume} {111}},\ \bibinfo {eid}
  {054432} (\bibinfo {year} {2025}{\natexlab{b}})},\ \Eprint
  {https://arxiv.org/abs/2405.05302} {arXiv:2405.05302} \BibitemShut {NoStop}%
\bibitem [{\citenamefont {{Moradi}}\ \emph {et~al.}(2023)\citenamefont
  {{Moradi}}, \citenamefont {{Faroogh Moosavian}},\ and\ \citenamefont
  {{Tiwari}}}]{Apoorv2023SciPost}%
  \BibitemOpen
  \bibfield  {author} {\bibinfo {author} {\bibfnamefont {H.}~\bibnamefont
  {{Moradi}}}, \bibinfo {author} {\bibfnamefont {S.}~\bibnamefont {{Faroogh
  Moosavian}}},\ and\ \bibinfo {author} {\bibfnamefont {A.}~\bibnamefont
  {{Tiwari}}},\ }\bibfield  {title} {\bibinfo {title} {{Topological Holography:
  Towards a Unification of Landau and Beyond-Landau Physics}},\ }\href
  {https://doi.org/10.21468/SciPostPhysCore.6.4.066} {\bibfield  {journal}
  {\bibinfo  {journal} {SciPost Physics Core}\ }\textbf {\bibinfo {volume}
  {6}},\ \bibinfo {eid} {066} (\bibinfo {year} {2023})},\ \Eprint
  {https://arxiv.org/abs/2207.10712} {arXiv:2207.10712} \BibitemShut {NoStop}%
\bibitem [{\citenamefont {{Chatterjee}}\ \emph {et~al.}(2024)\citenamefont
  {{Chatterjee}}, \citenamefont {{Aksoy}},\ and\ \citenamefont
  {{Wen}}}]{chatterjee2024quantum}%
  \BibitemOpen
  \bibfield  {author} {\bibinfo {author} {\bibfnamefont {A.}~\bibnamefont
  {{Chatterjee}}}, \bibinfo {author} {\bibfnamefont {{\"O}.~M.}\ \bibnamefont
  {{Aksoy}}},\ and\ \bibinfo {author} {\bibfnamefont {X.-G.}\ \bibnamefont
  {{Wen}}},\ }\bibfield  {title} {\bibinfo {title} {{Quantum phases and
  transitions in spin chains with non-invertible symmetries}},\ }\href
  {https://doi.org/10.21468/SciPostPhys.17.4.115} {\bibfield  {journal}
  {\bibinfo  {journal} {SciPost Physics}\ }\textbf {\bibinfo {volume} {17}},\
  \bibinfo {eid} {115} (\bibinfo {year} {2024})},\ \Eprint
  {https://arxiv.org/abs/2405.05331} {arXiv:2405.05331} \BibitemShut {NoStop}%
\bibitem [{\citenamefont {{C{\'o}rdova}}\ \emph {et~al.}(2024)\citenamefont
  {{C{\'o}rdova}}, \citenamefont {{Hong}}, \citenamefont {{Koren}},\ and\
  \citenamefont {{Ohmori}}}]{Cordova2024PRX}%
  \BibitemOpen
  \bibfield  {author} {\bibinfo {author} {\bibfnamefont {C.}~\bibnamefont
  {{C{\'o}rdova}}}, \bibinfo {author} {\bibfnamefont {S.}~\bibnamefont
  {{Hong}}}, \bibinfo {author} {\bibfnamefont {S.}~\bibnamefont {{Koren}}},\
  and\ \bibinfo {author} {\bibfnamefont {K.}~\bibnamefont {{Ohmori}}},\
  }\bibfield  {title} {\bibinfo {title} {{Neutrino Masses from Generalized
  Symmetry Breaking}},\ }\href {https://doi.org/10.1103/PhysRevX.14.031033}
  {\bibfield  {journal} {\bibinfo  {journal} {Physical Review X}\ }\textbf
  {\bibinfo {volume} {14}},\ \bibinfo {eid} {031033} (\bibinfo {year}
  {2024})}\BibitemShut {NoStop}%
\bibitem [{\citenamefont {{Zhao}}\ and\ \citenamefont
  {{Wan}}(2025)}]{ZhaoWan2025arXiv}%
  \BibitemOpen
  \bibfield  {author} {\bibinfo {author} {\bibfnamefont {Y.}~\bibnamefont
  {{Zhao}}}\ and\ \bibinfo {author} {\bibfnamefont {Y.}~\bibnamefont {{Wan}}},\
  }\bibfield  {title} {\bibinfo {title} {{Landau-Ginzburg Paradigm of
  Topological Phases}},\ }\href@noop {} {\bibfield  {journal} {\bibinfo
  {journal} {arXiv e-prints}\ } (\bibinfo {year} {2025})},\ \Eprint
  {https://arxiv.org/abs/2506.05319} {arXiv:2506.05319} \BibitemShut {NoStop}%
\bibitem [{\citenamefont {Gils}\ \emph {et~al.}(2013)\citenamefont {Gils},
  \citenamefont {Ardonne}, \citenamefont {Trebst}, \citenamefont {Huse},
  \citenamefont {Ludwig}, \citenamefont {Troyer},\ and\ \citenamefont
  {Wang}}]{Gils2013PRB}%
  \BibitemOpen
  \bibfield  {author} {\bibinfo {author} {\bibfnamefont {C.}~\bibnamefont
  {Gils}}, \bibinfo {author} {\bibfnamefont {E.}~\bibnamefont {Ardonne}},
  \bibinfo {author} {\bibfnamefont {S.}~\bibnamefont {Trebst}}, \bibinfo
  {author} {\bibfnamefont {D.~A.}\ \bibnamefont {Huse}}, \bibinfo {author}
  {\bibfnamefont {A.~W.~W.}\ \bibnamefont {Ludwig}}, \bibinfo {author}
  {\bibfnamefont {M.}~\bibnamefont {Troyer}},\ and\ \bibinfo {author}
  {\bibfnamefont {Z.}~\bibnamefont {Wang}},\ }\bibfield  {title} {\bibinfo
  {title} {Anyonic quantum spin chains: Spin-1 generalizations and topological
  stability},\ }\href {https://doi.org/10.1103/PhysRevB.87.235120} {\bibfield
  {journal} {\bibinfo  {journal} {Phys. Rev. B}\ }\textbf {\bibinfo {volume}
  {87}},\ \bibinfo {pages} {235120} (\bibinfo {year} {2013})}\BibitemShut
  {NoStop}%
\bibitem [{\citenamefont {{Trebst}}\ \emph {et~al.}(2009)\citenamefont
  {{Trebst}}, \citenamefont {{Troyer}}, \citenamefont {{Wang}},\ and\
  \citenamefont {{Ludwig}}}]{trebst_short_2008}%
  \BibitemOpen
  \bibfield  {author} {\bibinfo {author} {\bibfnamefont {S.}~\bibnamefont
  {{Trebst}}}, \bibinfo {author} {\bibfnamefont {M.}~\bibnamefont {{Troyer}}},
  \bibinfo {author} {\bibfnamefont {Z.}~\bibnamefont {{Wang}}},\ and\ \bibinfo
  {author} {\bibfnamefont {A.~W.~W.}\ \bibnamefont {{Ludwig}}},\ }\bibfield
  {title} {\bibinfo {title} {{A short introduction to Fibonacci anyon
  models}},\ }\href {https://doi.org/10.48550/arXiv.0902.3275} {\bibfield
  {journal} {\bibinfo  {journal} {arXiv e-prints}\ ,\ \bibinfo {eid}
  {arXiv:0902.3275}} (\bibinfo {year} {2009})}\BibitemShut {NoStop}%
\bibitem [{\citenamefont {{Aasen}}\ \emph {et~al.}(2016)\citenamefont
  {{Aasen}}, \citenamefont {{Mong}},\ and\ \citenamefont {{Fendley}}}]{AMF1}%
  \BibitemOpen
  \bibfield  {author} {\bibinfo {author} {\bibfnamefont {D.}~\bibnamefont
  {{Aasen}}}, \bibinfo {author} {\bibfnamefont {R.~S.~K.}\ \bibnamefont
  {{Mong}}},\ and\ \bibinfo {author} {\bibfnamefont {P.}~\bibnamefont
  {{Fendley}}},\ }\bibfield  {title} {\bibinfo {title} {{Topological defects on
  the lattice: I. The Ising model}},\ }\href
  {https://doi.org/10.1088/1751-8113/49/35/354001} {\bibfield  {journal}
  {\bibinfo  {journal} {Journal of Physics A Mathematical General}\ }\textbf
  {\bibinfo {volume} {49}},\ \bibinfo {eid} {354001} (\bibinfo {year}
  {2016})},\ \Eprint {https://arxiv.org/abs/1601.07185} {arXiv:1601.07185}
  \BibitemShut {NoStop}%
\bibitem [{\citenamefont {{Aasen}}\ \emph {et~al.}(2020)\citenamefont
  {{Aasen}}, \citenamefont {{Fendley}},\ and\ \citenamefont {{Mong}}}]{AMF2}%
  \BibitemOpen
  \bibfield  {author} {\bibinfo {author} {\bibfnamefont {D.}~\bibnamefont
  {{Aasen}}}, \bibinfo {author} {\bibfnamefont {P.}~\bibnamefont {{Fendley}}},\
  and\ \bibinfo {author} {\bibfnamefont {R.~S.~K.}\ \bibnamefont {{Mong}}},\
  }\bibfield  {title} {\bibinfo {title} {{Topological Defects on the Lattice:
  Dualities and Degeneracies}},\ }\href@noop {} {\bibfield  {journal} {\bibinfo
   {journal} {arXiv e-prints}\ } (\bibinfo {year} {2020})},\ \Eprint
  {https://arxiv.org/abs/2008.08598} {arXiv:2008.08598} \BibitemShut {NoStop}%
\bibitem [{\citenamefont {Vanhove}\ \emph {et~al.}(2018)\citenamefont
  {Vanhove}, \citenamefont {Bal}, \citenamefont {Williamson}, \citenamefont
  {Bultinck}, \citenamefont {Haegeman},\ and\ \citenamefont
  {Verstraete}}]{Vanhove2018PRL}%
  \BibitemOpen
  \bibfield  {author} {\bibinfo {author} {\bibfnamefont {R.}~\bibnamefont
  {Vanhove}}, \bibinfo {author} {\bibfnamefont {M.}~\bibnamefont {Bal}},
  \bibinfo {author} {\bibfnamefont {D.~J.}\ \bibnamefont {Williamson}},
  \bibinfo {author} {\bibfnamefont {N.}~\bibnamefont {Bultinck}}, \bibinfo
  {author} {\bibfnamefont {J.}~\bibnamefont {Haegeman}},\ and\ \bibinfo
  {author} {\bibfnamefont {F.}~\bibnamefont {Verstraete}},\ }\bibfield  {title}
  {\bibinfo {title} {Mapping topological to conformal field theories through
  strange correlators},\ }\href
  {https://doi.org/10.1103/PhysRevLett.121.177203} {\bibfield  {journal}
  {\bibinfo  {journal} {Phys. Rev. Lett.}\ }\textbf {\bibinfo {volume} {121}},\
  \bibinfo {pages} {177203} (\bibinfo {year} {2018})}\BibitemShut {NoStop}%
\bibitem [{\citenamefont {{Senthil}}(2015)}]{Senthil_sptReview_2015}%
  \BibitemOpen
  \bibfield  {author} {\bibinfo {author} {\bibfnamefont {T.}~\bibnamefont
  {{Senthil}}},\ }\bibfield  {title} {\bibinfo {title} {{Symmetry-Protected
  Topological Phases of Quantum Matter}},\ }\href
  {https://doi.org/10.1146/annurev-conmatphys-031214-014740} {\bibfield
  {journal} {\bibinfo  {journal} {Annual Review of Condensed Matter Physics}\
  }\textbf {\bibinfo {volume} {6}},\ \bibinfo {pages} {299} (\bibinfo {year}
  {2015})},\ \Eprint {https://arxiv.org/abs/1405.4015} {arXiv:1405.4015}
  \BibitemShut {NoStop}%
\bibitem [{\citenamefont {{Zou}}\ and\ \citenamefont
  {{Cheng}}(2026)}]{LSM-review}%
  \BibitemOpen
  \bibfield  {author} {\bibinfo {author} {\bibfnamefont {L.}~\bibnamefont
  {{Zou}}}\ and\ \bibinfo {author} {\bibfnamefont {M.}~\bibnamefont
  {{Cheng}}},\ }\bibfield  {title} {\bibinfo {title} {{Lieb-Schultz-Mattis
  Anomalies and Anomaly Matching}},\ }\href@noop {} {\bibfield  {journal}
  {\bibinfo  {journal} {arXiv e-prints}\ } (\bibinfo {year} {2026})},\ \Eprint
  {https://arxiv.org/abs/2604.00347} {arXiv:2604.00347} \BibitemShut {NoStop}%
\bibitem [{\citenamefont {{Chatterjee}}\ and\ \citenamefont
  {{Wen}}(2023)}]{Chatterjee2023PRB}%
  \BibitemOpen
  \bibfield  {author} {\bibinfo {author} {\bibfnamefont {A.}~\bibnamefont
  {{Chatterjee}}}\ and\ \bibinfo {author} {\bibfnamefont {X.-G.}\ \bibnamefont
  {{Wen}}},\ }\bibfield  {title} {\bibinfo {title} {{Symmetry as a shadow of
  topological order and a derivation of topological holographic principle}},\
  }\href {https://doi.org/10.1103/PhysRevB.107.155136} {\bibfield  {journal}
  {\bibinfo  {journal} {\prb}\ }\textbf {\bibinfo {volume} {107}},\ \bibinfo
  {eid} {155136} (\bibinfo {year} {2023})},\ \Eprint
  {https://arxiv.org/abs/2203.03596} {arXiv:2203.03596} \BibitemShut {NoStop}%
\bibitem [{\citenamefont {{Ning}}\ \emph {et~al.}(2024)\citenamefont {{Ning}},
  \citenamefont {{Mao}},\ and\ \citenamefont {{Wang}}}]{ning_building_2023}%
  \BibitemOpen
  \bibfield  {author} {\bibinfo {author} {\bibfnamefont {S.-Q.}\ \bibnamefont
  {{Ning}}}, \bibinfo {author} {\bibfnamefont {B.-B.}\ \bibnamefont {{Mao}}},\
  and\ \bibinfo {author} {\bibfnamefont {C.}~\bibnamefont {{Wang}}},\
  }\bibfield  {title} {\bibinfo {title} {{Building 1D lattice models with
  G-graded fusion category}},\ }\href
  {https://doi.org/10.21468/SciPostPhys.17.5.125} {\bibfield  {journal}
  {\bibinfo  {journal} {SciPost Physics}\ }\textbf {\bibinfo {volume} {17}},\
  \bibinfo {eid} {125} (\bibinfo {year} {2024})},\ \Eprint
  {https://arxiv.org/abs/2301.06416} {arXiv:2301.06416} \BibitemShut {NoStop}%
\bibitem [{\citenamefont {{Hung}}\ \emph {et~al.}(2025)\citenamefont {{Hung}},
  \citenamefont {{Ji}}, \citenamefont {{Shen}}, \citenamefont {{Wan}},\ and\
  \citenamefont {{Zhao}}}]{Hung2025arXiv}%
  \BibitemOpen
  \bibfield  {author} {\bibinfo {author} {\bibfnamefont {L.-Y.}\ \bibnamefont
  {{Hung}}}, \bibinfo {author} {\bibfnamefont {K.}~\bibnamefont {{Ji}}},
  \bibinfo {author} {\bibfnamefont {C.}~\bibnamefont {{Shen}}}, \bibinfo
  {author} {\bibfnamefont {Y.}~\bibnamefont {{Wan}}},\ and\ \bibinfo {author}
  {\bibfnamefont {Y.}~\bibnamefont {{Zhao}}},\ }\bibfield  {title} {\bibinfo
  {title} {{A 2D-CFT Factory: Critical Lattice Models from Competing Anyon
  Condensation Processes in SymTO/SymTFT}},\ }\href
  {https://doi.org/10.48550/arXiv.2506.05324} {\bibfield  {journal} {\bibinfo
  {journal} {arXiv e-prints}\ ,\ \bibinfo {eid} {arXiv:2506.05324}} (\bibinfo
  {year} {2025})},\ \Eprint {https://arxiv.org/abs/2506.05324}
  {arXiv:2506.05324} \BibitemShut {NoStop}%
\bibitem [{\citenamefont {{Vanhove}}\ \emph {et~al.}(2022)\citenamefont
  {{Vanhove}}, \citenamefont {{Lootens}}, \citenamefont {{Van Damme}},
  \citenamefont {{Wolf}}, \citenamefont {{Osborne}}, \citenamefont
  {{Haegeman}},\ and\ \citenamefont {{Verstraete}}}]{Haagerup1}%
  \BibitemOpen
  \bibfield  {author} {\bibinfo {author} {\bibfnamefont {R.}~\bibnamefont
  {{Vanhove}}}, \bibinfo {author} {\bibfnamefont {L.}~\bibnamefont
  {{Lootens}}}, \bibinfo {author} {\bibfnamefont {M.}~\bibnamefont {{Van
  Damme}}}, \bibinfo {author} {\bibfnamefont {R.}~\bibnamefont {{Wolf}}},
  \bibinfo {author} {\bibfnamefont {T.~J.}\ \bibnamefont {{Osborne}}}, \bibinfo
  {author} {\bibfnamefont {J.}~\bibnamefont {{Haegeman}}},\ and\ \bibinfo
  {author} {\bibfnamefont {F.}~\bibnamefont {{Verstraete}}},\ }\bibfield
  {title} {\bibinfo {title} {{Critical Lattice Model for a Haagerup Conformal
  Field Theory}},\ }\href {https://doi.org/10.1103/PhysRevLett.128.231602}
  {\bibfield  {journal} {\bibinfo  {journal} {\prl}\ }\textbf {\bibinfo
  {volume} {128}},\ \bibinfo {eid} {231602} (\bibinfo {year} {2022})},\ \Eprint
  {https://arxiv.org/abs/2110.03532} {arXiv:2110.03532} \BibitemShut {NoStop}%
\bibitem [{\citenamefont {{Huang}}\ \emph {et~al.}(2022)\citenamefont
  {{Huang}}, \citenamefont {{Lin}}, \citenamefont {{Ohmori}}, \citenamefont
  {{Tachikawa}},\ and\ \citenamefont {{Tezuka}}}]{Haagerup2}%
  \BibitemOpen
  \bibfield  {author} {\bibinfo {author} {\bibfnamefont {T.-C.}\ \bibnamefont
  {{Huang}}}, \bibinfo {author} {\bibfnamefont {Y.-H.}\ \bibnamefont {{Lin}}},
  \bibinfo {author} {\bibfnamefont {K.}~\bibnamefont {{Ohmori}}}, \bibinfo
  {author} {\bibfnamefont {Y.}~\bibnamefont {{Tachikawa}}},\ and\ \bibinfo
  {author} {\bibfnamefont {M.}~\bibnamefont {{Tezuka}}},\ }\bibfield  {title}
  {\bibinfo {title} {{Numerical Evidence for a Haagerup Conformal Field
  Theory}},\ }\href {https://doi.org/10.1103/PhysRevLett.128.231603} {\bibfield
   {journal} {\bibinfo  {journal} {\prl}\ }\textbf {\bibinfo {volume} {128}},\
  \bibinfo {eid} {231603} (\bibinfo {year} {2022})},\ \Eprint
  {https://arxiv.org/abs/2110.03008} {arXiv:2110.03008} \BibitemShut {NoStop}%
\bibitem [{\citenamefont {{Fuchs}}\ \emph {et~al.}(2002)\citenamefont
  {{Fuchs}}, \citenamefont {{Runkel}},\ and\ \citenamefont
  {{Schweigert}}}]{Fuchs1}%
  \BibitemOpen
  \bibfield  {author} {\bibinfo {author} {\bibfnamefont {J.}~\bibnamefont
  {{Fuchs}}}, \bibinfo {author} {\bibfnamefont {I.}~\bibnamefont {{Runkel}}},\
  and\ \bibinfo {author} {\bibfnamefont {C.}~\bibnamefont {{Schweigert}}},\
  }\bibfield  {title} {\bibinfo {title} {{TFT construction of RCFT correlators
  I: partition functions}},\ }\href
  {https://doi.org/10.1016/S0550-3213(02)00744-7} {\bibfield  {journal}
  {\bibinfo  {journal} {Nuclear Physics B}\ }\textbf {\bibinfo {volume}
  {646}},\ \bibinfo {pages} {353} (\bibinfo {year} {2002})},\ \Eprint
  {https://arxiv.org/abs/hep-th/0204148} {arXiv:hep-th/0204148} \BibitemShut
  {NoStop}%
\bibitem [{\citenamefont {{Fuchs}}\ \emph {et~al.}(2004)\citenamefont
  {{Fuchs}}, \citenamefont {{Runkel}},\ and\ \citenamefont
  {{Schweigert}}}]{Fuchs3}%
  \BibitemOpen
  \bibfield  {author} {\bibinfo {author} {\bibfnamefont {J.}~\bibnamefont
  {{Fuchs}}}, \bibinfo {author} {\bibfnamefont {I.}~\bibnamefont {{Runkel}}},\
  and\ \bibinfo {author} {\bibfnamefont {C.}~\bibnamefont {{Schweigert}}},\
  }\bibfield  {title} {\bibinfo {title} {{TFT construction of RCFT correlators.
  III: simple currents}},\ }\href
  {https://doi.org/10.1016/j.nuclphysb.2004.05.014} {\bibfield  {journal}
  {\bibinfo  {journal} {Nuclear Physics B}\ }\textbf {\bibinfo {volume}
  {694}},\ \bibinfo {pages} {277} (\bibinfo {year} {2004})},\ \Eprint
  {https://arxiv.org/abs/hep-th/0403157} {arXiv:hep-th/0403157 [hep-th]}
  \BibitemShut {NoStop}%
\bibitem [{\citenamefont {{Fr{\"o}hlich}}\ \emph {et~al.}(2004)\citenamefont
  {{Fr{\"o}hlich}}, \citenamefont {{Fuchs}}, \citenamefont {{Runkel}},\ and\
  \citenamefont {{Schweigert}}}]{Fuchs2}%
  \BibitemOpen
  \bibfield  {author} {\bibinfo {author} {\bibfnamefont {J.}~\bibnamefont
  {{Fr{\"o}hlich}}}, \bibinfo {author} {\bibfnamefont {J.}~\bibnamefont
  {{Fuchs}}}, \bibinfo {author} {\bibfnamefont {I.}~\bibnamefont {{Runkel}}},\
  and\ \bibinfo {author} {\bibfnamefont {C.}~\bibnamefont {{Schweigert}}},\
  }\bibfield  {title} {\bibinfo {title} {{Kramers-Wannier Duality from
  Conformal Defects}},\ }\href {https://doi.org/10.1103/PhysRevLett.93.070601}
  {\bibfield  {journal} {\bibinfo  {journal} {\prl}\ }\textbf {\bibinfo
  {volume} {93}},\ \bibinfo {eid} {070601} (\bibinfo {year} {2004})},\ \Eprint
  {https://arxiv.org/abs/cond-mat/0404051} {arXiv:cond-mat/0404051}
  \BibitemShut {NoStop}%
\bibitem [{\citenamefont {{Inamura}}\ and\ \citenamefont
  {{Ohmori}}(2024)}]{Inamura2024arXiv}%
  \BibitemOpen
  \bibfield  {author} {\bibinfo {author} {\bibfnamefont {K.}~\bibnamefont
  {{Inamura}}}\ and\ \bibinfo {author} {\bibfnamefont {K.}~\bibnamefont
  {{Ohmori}}},\ }\bibfield  {title} {\bibinfo {title} {{Fusion surface models:
  2+1d lattice models from fusion 2-categories}},\ }\href
  {https://doi.org/10.21468/SciPostPhys.16.6.143} {\bibfield  {journal}
  {\bibinfo  {journal} {SciPost Physics}\ }\textbf {\bibinfo {volume} {16}},\
  \bibinfo {eid} {143} (\bibinfo {year} {2024})},\ \Eprint
  {https://arxiv.org/abs/2305.05774} {arXiv:2305.05774} \BibitemShut {NoStop}%
\bibitem [{\citenamefont {Chen}\ \emph {et~al.}(2013)\citenamefont {Chen},
  \citenamefont {Gu}, \citenamefont {Liu},\ and\ \citenamefont
  {Wen}}]{ChengGuLiuWen-SPT}%
  \BibitemOpen
  \bibfield  {author} {\bibinfo {author} {\bibfnamefont {X.}~\bibnamefont
  {Chen}}, \bibinfo {author} {\bibfnamefont {Z.-C.}\ \bibnamefont {Gu}},
  \bibinfo {author} {\bibfnamefont {Z.-X.}\ \bibnamefont {Liu}},\ and\ \bibinfo
  {author} {\bibfnamefont {X.-G.}\ \bibnamefont {Wen}},\ }\bibfield  {title}
  {\bibinfo {title} {Symmetry protected topological orders and the group
  cohomology of their symmetry group},\ }\href
  {https://doi.org/10.1103/PhysRevB.87.155114} {\bibfield  {journal} {\bibinfo
  {journal} {Phys. Rev. B}\ }\textbf {\bibinfo {volume} {87}},\ \bibinfo
  {pages} {155114} (\bibinfo {year} {2013})}\BibitemShut {NoStop}%
\bibitem [{\citenamefont {Barkeshli}\ \emph {et~al.}(2019)\citenamefont
  {Barkeshli}, \citenamefont {Bonderson}, \citenamefont {Cheng},\ and\
  \citenamefont {Wang}}]{SET-StationQ}%
  \BibitemOpen
  \bibfield  {author} {\bibinfo {author} {\bibfnamefont {M.}~\bibnamefont
  {Barkeshli}}, \bibinfo {author} {\bibfnamefont {P.}~\bibnamefont
  {Bonderson}}, \bibinfo {author} {\bibfnamefont {M.}~\bibnamefont {Cheng}},\
  and\ \bibinfo {author} {\bibfnamefont {Z.}~\bibnamefont {Wang}},\ }\bibfield
  {title} {\bibinfo {title} {Symmetry fractionalization, defects, and gauging
  of topological phases},\ }\href {https://doi.org/10.1103/PhysRevB.100.115147}
  {\bibfield  {journal} {\bibinfo  {journal} {Phys. Rev. B}\ }\textbf {\bibinfo
  {volume} {100}},\ \bibinfo {pages} {115147} (\bibinfo {year}
  {2019})}\BibitemShut {NoStop}%
\bibitem [{\citenamefont {{Yang}}\ \emph {et~al.}(2015)\citenamefont {{Yang}},
  \citenamefont {{Gu}},\ and\ \citenamefont {{Wen}}}]{yang_loop_2017}%
  \BibitemOpen
  \bibfield  {author} {\bibinfo {author} {\bibfnamefont {S.}~\bibnamefont
  {{Yang}}}, \bibinfo {author} {\bibfnamefont {Z.-C.}\ \bibnamefont {{Gu}}},\
  and\ \bibinfo {author} {\bibfnamefont {X.-G.}\ \bibnamefont {{Wen}}},\
  }\bibfield  {title} {\bibinfo {title} {{Loop optimization for tensor network
  renormalization}},\ }\href {https://doi.org/10.48550/arXiv.1512.04938}
  {\bibfield  {journal} {\bibinfo  {journal} {arXiv e-prints}\ ,\ \bibinfo
  {eid} {arXiv:1512.04938}} (\bibinfo {year} {2015})},\ \Eprint
  {https://arxiv.org/abs/1512.04938} {arXiv:1512.04938} \BibitemShut {NoStop}%
\bibitem [{\citenamefont {Fishman}\ \emph {et~al.}(2022)\citenamefont
  {Fishman}, \citenamefont {White},\ and\ \citenamefont
  {Stoudenmire}}]{fishman_itensor_2022}%
  \BibitemOpen
  \bibfield  {author} {\bibinfo {author} {\bibfnamefont {M.}~\bibnamefont
  {Fishman}}, \bibinfo {author} {\bibfnamefont {S.~R.}\ \bibnamefont {White}},\
  and\ \bibinfo {author} {\bibfnamefont {E.~M.}\ \bibnamefont {Stoudenmire}},\
  }\bibfield  {title} {\bibinfo {title} {{The ITensor Software Library for
  Tensor Network Calculations}},\ }\href
  {https://doi.org/10.21468/SciPostPhysCodeb.4} {\bibfield  {journal} {\bibinfo
   {journal} {SciPost Phys. Codebases}\ ,\ \bibinfo {pages} {4}} (\bibinfo
  {year} {2022})}\BibitemShut {NoStop}%
\bibitem [{\citenamefont {Jones}\ and\ \citenamefont
  {Metlitski}(2021)}]{jones2021majorana-spt}%
  \BibitemOpen
  \bibfield  {author} {\bibinfo {author} {\bibfnamefont {R.~A.}\ \bibnamefont
  {Jones}}\ and\ \bibinfo {author} {\bibfnamefont {M.~A.}\ \bibnamefont
  {Metlitski}},\ }\bibfield  {title} {\bibinfo {title} {One-dimensional lattice
  models for the boundary of two-dimensional majorana fermion
  symmetry-protected topological phases: Kramers-wannier duality as an exact
  ${Z}_{2}$ symmetry},\ }\href {https://doi.org/10.1103/PhysRevB.104.245130}
  {\bibfield  {journal} {\bibinfo  {journal} {Phys. Rev. B}\ }\textbf {\bibinfo
  {volume} {104}},\ \bibinfo {pages} {245130} (\bibinfo {year}
  {2021})}\BibitemShut {NoStop}%
\bibitem [{\citenamefont {Simon}(2023)}]{simon2023topological}%
  \BibitemOpen
  \bibfield  {author} {\bibinfo {author} {\bibfnamefont {S.~H.}\ \bibnamefont
  {Simon}},\ }\href@noop {} {\emph {\bibinfo {title} {Topological quantum}}}\
  (\bibinfo  {publisher} {Oxford University Press},\ \bibinfo {year}
  {2023})\BibitemShut {NoStop}%
\bibitem [{\citenamefont {{Buican}}\ and\ \citenamefont
  {{Gromov}}(2017)}]{buican_anyonic_2017}%
  \BibitemOpen
  \bibfield  {author} {\bibinfo {author} {\bibfnamefont {M.}~\bibnamefont
  {{Buican}}}\ and\ \bibinfo {author} {\bibfnamefont {A.}~\bibnamefont
  {{Gromov}}},\ }\bibfield  {title} {\bibinfo {title} {{Anyonic Chains,
  Topological Defects, and Conformal Field Theory}},\ }\href
  {https://doi.org/10.1007/s00220-017-2995-6} {\bibfield  {journal} {\bibinfo
  {journal} {Communications in Mathematical Physics}\ }\textbf {\bibinfo
  {volume} {356}},\ \bibinfo {pages} {1017} (\bibinfo {year} {2017})},\ \Eprint
  {https://arxiv.org/abs/1701.02800} {arXiv:1701.02800} \BibitemShut {NoStop}%
\bibitem [{\citenamefont {Verlinde}(1988{\natexlab{b}})}]{VERLINDE1988360}%
  \BibitemOpen
  \bibfield  {author} {\bibinfo {author} {\bibfnamefont {E.}~\bibnamefont
  {Verlinde}},\ }\bibfield  {title} {\bibinfo {title} {Fusion rules and modular
  transformations in 2d conformal field theory},\ }\href
  {https://doi.org/https://doi.org/10.1016/0550-3213(88)90603-7} {\bibfield
  {journal} {\bibinfo  {journal} {Nuclear Physics B}\ }\textbf {\bibinfo
  {volume} {300}},\ \bibinfo {pages} {360} (\bibinfo {year}
  {1988}{\natexlab{b}})}\BibitemShut {NoStop}%
\bibitem [{\citenamefont {{Kitaev}}(2006)}]{kitaev_anyons_2006}%
  \BibitemOpen
  \bibfield  {author} {\bibinfo {author} {\bibfnamefont {A.}~\bibnamefont
  {{Kitaev}}},\ }\bibfield  {title} {\bibinfo {title} {{Anyons in an exactly
  solved model and beyond}},\ }\href
  {https://doi.org/10.1016/j.aop.2005.10.005} {\bibfield  {journal} {\bibinfo
  {journal} {Annals of Physics}\ }\textbf {\bibinfo {volume} {321}},\ \bibinfo
  {pages} {2} (\bibinfo {year} {2006})},\ \Eprint
  {https://arxiv.org/abs/cond-mat/0506438} {arXiv:cond-mat/0506438}
  \BibitemShut {NoStop}%
\bibitem [{\citenamefont {White}(1992)}]{White1992PRL}%
  \BibitemOpen
  \bibfield  {author} {\bibinfo {author} {\bibfnamefont {S.~R.}\ \bibnamefont
  {White}},\ }\bibfield  {title} {\bibinfo {title} {Density matrix formulation
  for quantum renormalization groups},\ }\href
  {https://doi.org/10.1103/PhysRevLett.69.2863} {\bibfield  {journal} {\bibinfo
   {journal} {Phys. Rev. Lett.}\ }\textbf {\bibinfo {volume} {69}},\ \bibinfo
  {pages} {2863} (\bibinfo {year} {1992})}\BibitemShut {NoStop}%
\bibitem [{\citenamefont {Xiang}(2023)}]{xiang2023density}%
  \BibitemOpen
  \bibfield  {author} {\bibinfo {author} {\bibfnamefont {T.}~\bibnamefont
  {Xiang}},\ }\href@noop {} {\emph {\bibinfo {title} {Density Matrix and Tensor
  Network Renormalization}}}\ (\bibinfo  {publisher} {Cambridge University
  Press},\ \bibinfo {year} {2023})\BibitemShut {NoStop}%
\bibitem [{\citenamefont
  {Schollw\"ock}(2005)}]{schollwock_density-matrix_2005}%
  \BibitemOpen
  \bibfield  {author} {\bibinfo {author} {\bibfnamefont {U.}~\bibnamefont
  {Schollw\"ock}},\ }\bibfield  {title} {\bibinfo {title} {The density-matrix
  renormalization group},\ }\href {https://doi.org/10.1103/RevModPhys.77.259}
  {\bibfield  {journal} {\bibinfo  {journal} {Rev. Mod. Phys.}\ }\textbf
  {\bibinfo {volume} {77}},\ \bibinfo {pages} {259} (\bibinfo {year}
  {2005})}\BibitemShut {NoStop}%
\bibitem [{\citenamefont
  {{Schollw{\"o}ck}}(2011)}]{schollwock_density-matrix_2011}%
  \BibitemOpen
  \bibfield  {author} {\bibinfo {author} {\bibfnamefont {U.}~\bibnamefont
  {{Schollw{\"o}ck}}},\ }\bibfield  {title} {\bibinfo {title} {{The
  density-matrix renormalization group in the age of matrix product states}},\
  }\href {https://doi.org/10.1016/j.aop.2010.09.012} {\bibfield  {journal}
  {\bibinfo  {journal} {Annals of Physics}\ }\textbf {\bibinfo {volume}
  {326}},\ \bibinfo {pages} {96} (\bibinfo {year} {2011})},\ \Eprint
  {https://arxiv.org/abs/1008.3477} {arXiv:1008.3477} \BibitemShut {NoStop}%
\bibitem [{\citenamefont {Cardy}(1986)}]{cardy1986operator}%
  \BibitemOpen
  \bibfield  {author} {\bibinfo {author} {\bibfnamefont {J.~L.}\ \bibnamefont
  {Cardy}},\ }\bibfield  {title} {\bibinfo {title} {Operator content of
  two-dimensional conformally invariant theories},\ }\href
  {https://doi.org/https://doi.org/10.1016/0550-3213(86)90552-3} {\bibfield
  {journal} {\bibinfo  {journal} {Nuclear Physics B}\ }\textbf {\bibinfo
  {volume} {270}},\ \bibinfo {pages} {186} (\bibinfo {year}
  {1986})}\BibitemShut {NoStop}%
\bibitem [{\citenamefont {{Calabrese}}\ and\ \citenamefont
  {{Cardy}}(2004)}]{calabrese2004entanglement}%
  \BibitemOpen
  \bibfield  {author} {\bibinfo {author} {\bibfnamefont {P.}~\bibnamefont
  {{Calabrese}}}\ and\ \bibinfo {author} {\bibfnamefont {J.}~\bibnamefont
  {{Cardy}}},\ }\bibfield  {title} {\bibinfo {title} {{Entanglement entropy and
  quantum field theory}},\ }\href
  {https://doi.org/10.1088/1742-5468/2004/06/P06002} {\bibfield  {journal}
  {\bibinfo  {journal} {Journal of Statistical Mechanics: Theory and
  Experiment}\ }\textbf {\bibinfo {volume} {2004}},\ \bibinfo {pages} {06002}
  (\bibinfo {year} {2004})},\ \Eprint {https://arxiv.org/abs/hep-th/0405152}
  {arXiv:hep-th/0405152} \BibitemShut {NoStop}%
\bibitem [{\citenamefont {Levin}\ and\ \citenamefont
  {Nave}(2007)}]{LevinNavePRL2007}%
  \BibitemOpen
  \bibfield  {author} {\bibinfo {author} {\bibfnamefont {M.}~\bibnamefont
  {Levin}}\ and\ \bibinfo {author} {\bibfnamefont {C.~P.}\ \bibnamefont
  {Nave}},\ }\bibfield  {title} {\bibinfo {title} {Tensor renormalization group
  approach to two-dimensional classical lattice models},\ }\href
  {https://doi.org/10.1103/PhysRevLett.99.120601} {\bibfield  {journal}
  {\bibinfo  {journal} {Phys. Rev. Lett.}\ }\textbf {\bibinfo {volume} {99}},\
  \bibinfo {pages} {120601} (\bibinfo {year} {2007})}\BibitemShut {NoStop}%
\bibitem [{\citenamefont {Evenbly}\ and\ \citenamefont
  {Vidal}(2015)}]{evenbly2015tensor}%
  \BibitemOpen
  \bibfield  {author} {\bibinfo {author} {\bibfnamefont {G.}~\bibnamefont
  {Evenbly}}\ and\ \bibinfo {author} {\bibfnamefont {G.}~\bibnamefont
  {Vidal}},\ }\bibfield  {title} {\bibinfo {title} {Tensor network
  renormalization},\ }\href {https://doi.org/10.1103/PhysRevLett.115.180405}
  {\bibfield  {journal} {\bibinfo  {journal} {Phys. Rev. Lett.}\ }\textbf
  {\bibinfo {volume} {115}},\ \bibinfo {pages} {180405} (\bibinfo {year}
  {2015})}\BibitemShut {NoStop}%
\bibitem [{\citenamefont {Bao}(2019)}]{bao2019loop}%
  \BibitemOpen
  \bibfield  {author} {\bibinfo {author} {\bibfnamefont {C.}~\bibnamefont
  {Bao}},\ }\emph {\bibinfo {title} {Loop optimization of tensor network
  renormalization: algorithms and applications}},\ \href@noop {} {Ph.D.
  thesis},\ \bibinfo  {school} {University of Waterloo} (\bibinfo {year}
  {2019})\BibitemShut {NoStop}%
\bibitem [{\citenamefont {Li}\ \emph {et~al.}(2022)\citenamefont {Li},
  \citenamefont {Pai},\ and\ \citenamefont {Gu}}]{li_tensor-network_2022}%
  \BibitemOpen
  \bibfield  {author} {\bibinfo {author} {\bibfnamefont {G.}~\bibnamefont
  {Li}}, \bibinfo {author} {\bibfnamefont {K.~H.}\ \bibnamefont {Pai}},\ and\
  \bibinfo {author} {\bibfnamefont {Z.-C.}\ \bibnamefont {Gu}},\ }\bibfield
  {title} {\bibinfo {title} {Tensor-network renormalization approach to the
  $q$-state clock model},\ }\href
  {https://doi.org/10.1103/PhysRevResearch.4.023159} {\bibfield  {journal}
  {\bibinfo  {journal} {Phys. Rev. Res.}\ }\textbf {\bibinfo {volume} {4}},\
  \bibinfo {pages} {023159} (\bibinfo {year} {2022})}\BibitemShut {NoStop}%
\bibitem [{\citenamefont {{Wei}}\ and\ \citenamefont
  {{Gu}}(2023)}]{wei_tensor_2023}%
  \BibitemOpen
  \bibfield  {author} {\bibinfo {author} {\bibfnamefont {Y.-J.}\ \bibnamefont
  {{Wei}}}\ and\ \bibinfo {author} {\bibfnamefont {Z.-C.}\ \bibnamefont
  {{Gu}}},\ }\bibfield  {title} {\bibinfo {title} {{Tensor network
  renormalization: application to dynamic correlation functions and
  non-hermitian systems}},\ }\href@noop {} {\bibfield  {journal} {\bibinfo
  {journal} {arXiv e-prints}\ } (\bibinfo {year} {2023})},\ \Eprint
  {https://arxiv.org/abs/2311.18785} {arXiv:2311.18785} \BibitemShut {NoStop}%
\bibitem [{\citenamefont {Di~Francesco}\ \emph {et~al.}(1997)\citenamefont
  {Di~Francesco}, \citenamefont {Mathieu},\ and\ \citenamefont
  {Sénéchal}}]{di_francesco_conformal_1997}%
  \BibitemOpen
  \bibfield  {author} {\bibinfo {author} {\bibfnamefont {P.}~\bibnamefont
  {Di~Francesco}}, \bibinfo {author} {\bibfnamefont {P.}~\bibnamefont
  {Mathieu}},\ and\ \bibinfo {author} {\bibfnamefont {D.}~\bibnamefont
  {Sénéchal}},\ }\href {https://doi.org/10.1007/978-1-4612-2256-9} {\emph
  {\bibinfo {title} {Conformal {Field} {Theory}}}},\ Graduate {Texts} in
  {Contemporary} {Physics}\ (\bibinfo  {publisher} {Springer New York},\
  \bibinfo {address} {New York, NY},\ \bibinfo {year} {1997})\BibitemShut
  {NoStop}%
\bibitem [{\citenamefont {Rahmani}\ \emph
  {et~al.}(2015{\natexlab{a}})\citenamefont {Rahmani}, \citenamefont {Zhu},
  \citenamefont {Franz},\ and\ \citenamefont {Affleck}}]{RahmaniPRL2015}%
  \BibitemOpen
  \bibfield  {author} {\bibinfo {author} {\bibfnamefont {A.}~\bibnamefont
  {Rahmani}}, \bibinfo {author} {\bibfnamefont {X.}~\bibnamefont {Zhu}},
  \bibinfo {author} {\bibfnamefont {M.}~\bibnamefont {Franz}},\ and\ \bibinfo
  {author} {\bibfnamefont {I.}~\bibnamefont {Affleck}},\ }\bibfield  {title}
  {\bibinfo {title} {Emergent supersymmetry from strongly interacting majorana
  zero modes},\ }\href {https://doi.org/10.1103/PhysRevLett.115.166401}
  {\bibfield  {journal} {\bibinfo  {journal} {Phys. Rev. Lett.}\ }\textbf
  {\bibinfo {volume} {115}},\ \bibinfo {pages} {166401} (\bibinfo {year}
  {2015}{\natexlab{a}})}\BibitemShut {NoStop}%
\bibitem [{\citenamefont {Rahmani}\ \emph
  {et~al.}(2015{\natexlab{b}})\citenamefont {Rahmani}, \citenamefont {Zhu},
  \citenamefont {Franz},\ and\ \citenamefont {Affleck}}]{RahmaniPRB2015}%
  \BibitemOpen
  \bibfield  {author} {\bibinfo {author} {\bibfnamefont {A.}~\bibnamefont
  {Rahmani}}, \bibinfo {author} {\bibfnamefont {X.}~\bibnamefont {Zhu}},
  \bibinfo {author} {\bibfnamefont {M.}~\bibnamefont {Franz}},\ and\ \bibinfo
  {author} {\bibfnamefont {I.}~\bibnamefont {Affleck}},\ }\bibfield  {title}
  {\bibinfo {title} {Phase diagram of the interacting majorana chain model},\
  }\href {https://doi.org/10.1103/PhysRevB.92.235123} {\bibfield  {journal}
  {\bibinfo  {journal} {Phys. Rev. B}\ }\textbf {\bibinfo {volume} {92}},\
  \bibinfo {pages} {235123} (\bibinfo {year} {2015}{\natexlab{b}})}\BibitemShut
  {NoStop}%
\bibitem [{\citenamefont {Cogburn}\ \emph {et~al.}(2024)\citenamefont
  {Cogburn}, \citenamefont {Fitzpatrick},\ and\ \citenamefont
  {Geng}}]{CameronSciPost2024}%
  \BibitemOpen
  \bibfield  {author} {\bibinfo {author} {\bibfnamefont {C.~V.}\ \bibnamefont
  {Cogburn}}, \bibinfo {author} {\bibfnamefont {A.~L.}\ \bibnamefont
  {Fitzpatrick}},\ and\ \bibinfo {author} {\bibfnamefont {H.}~\bibnamefont
  {Geng}},\ }\bibfield  {title} {\bibinfo {title} {{CFT and lattice correlators
  near an RG domain wall between minimal models}},\ }\href
  {https://doi.org/10.21468/SciPostPhysCore.7.2.021} {\bibfield  {journal}
  {\bibinfo  {journal} {SciPost Phys. Core}\ }\textbf {\bibinfo {volume} {7}},\
  \bibinfo {pages} {021} (\bibinfo {year} {2024})}\BibitemShut {NoStop}%
\bibitem [{\citenamefont {{Cortes Cubero}}\ \emph {et~al.}(2022)\citenamefont
  {{Cortes Cubero}}, \citenamefont {{Konik}}, \citenamefont {{Lencs{\'e}s}},
  \citenamefont {{Mussardo}},\ and\ \citenamefont
  {{Takacs}}}]{cortes2022duality}%
  \BibitemOpen
  \bibfield  {author} {\bibinfo {author} {\bibfnamefont {A.}~\bibnamefont
  {{Cortes Cubero}}}, \bibinfo {author} {\bibfnamefont {R.}~\bibnamefont
  {{Konik}}}, \bibinfo {author} {\bibfnamefont {M.}~\bibnamefont
  {{Lencs{\'e}s}}}, \bibinfo {author} {\bibfnamefont {G.}~\bibnamefont
  {{Mussardo}}},\ and\ \bibinfo {author} {\bibfnamefont {G.}~\bibnamefont
  {{Takacs}}},\ }\bibfield  {title} {\bibinfo {title} {{Duality and form
  factors in the thermally deformed two-dimensional tricritical Ising model}},\
  }\href {https://doi.org/10.21468/SciPostPhys.12.5.162} {\bibfield  {journal}
  {\bibinfo  {journal} {SciPost Physics}\ }\textbf {\bibinfo {volume} {12}},\
  \bibinfo {eid} {162} (\bibinfo {year} {2022})},\ \Eprint
  {https://arxiv.org/abs/2109.09767} {arXiv:2109.09767} \BibitemShut {NoStop}%
\bibitem [{\citenamefont {Sakurai}\ and\ \citenamefont
  {Napolitano}(2020)}]{sakurai2020modern}%
  \BibitemOpen
  \bibfield  {author} {\bibinfo {author} {\bibfnamefont {J.~J.}\ \bibnamefont
  {Sakurai}}\ and\ \bibinfo {author} {\bibfnamefont {J.}~\bibnamefont
  {Napolitano}},\ }\href@noop {} {\emph {\bibinfo {title} {Modern quantum
  mechanics}}}\ (\bibinfo  {publisher} {Cambridge University Press},\ \bibinfo
  {year} {2020})\BibitemShut {NoStop}%
\bibitem [{\citenamefont {Pfeuty}(1970)}]{pfeuty1970one}%
  \BibitemOpen
  \bibfield  {author} {\bibinfo {author} {\bibfnamefont {P.}~\bibnamefont
  {Pfeuty}},\ }\bibfield  {title} {\bibinfo {title} {The one-dimensional ising
  model with a transverse field},\ }\href
  {https://doi.org/https://doi.org/10.1016/0003-4916(70)90270-8} {\bibfield
  {journal} {\bibinfo  {journal} {Annals of Physics}\ }\textbf {\bibinfo
  {volume} {57}},\ \bibinfo {pages} {79} (\bibinfo {year} {1970})}\BibitemShut
  {NoStop}%
\bibitem [{\citenamefont {{Rajabpour}}(2016)}]{rajabpour2016entanglement}%
  \BibitemOpen
  \bibfield  {author} {\bibinfo {author} {\bibfnamefont {M.~A.}\ \bibnamefont
  {{Rajabpour}}},\ }\bibfield  {title} {\bibinfo {title} {{Entanglement entropy
  after a partial projective measurement in 1{\,}{\,}+{\,}{\,}1 dimensional
  conformal field theories: exact results}},\ }\href
  {https://doi.org/10.1088/1742-5468/2016/06/063109} {\bibfield  {journal}
  {\bibinfo  {journal} {Journal of Statistical Mechanics: Theory and
  Experiment}\ }\textbf {\bibinfo {volume} {6}},\ \bibinfo {pages} {063109}
  (\bibinfo {year} {2016})},\ \Eprint {https://arxiv.org/abs/1512.03940}
  {arXiv:1512.03940} \BibitemShut {NoStop}%
\bibitem [{\citenamefont {{Hoshino}}\ \emph {et~al.}(2025)\citenamefont
  {{Hoshino}}, \citenamefont {{Oshikawa}},\ and\ \citenamefont
  {{Ashida}}}]{hoshino2025entanglement}%
  \BibitemOpen
  \bibfield  {author} {\bibinfo {author} {\bibfnamefont {M.}~\bibnamefont
  {{Hoshino}}}, \bibinfo {author} {\bibfnamefont {M.}~\bibnamefont
  {{Oshikawa}}},\ and\ \bibinfo {author} {\bibfnamefont {Y.}~\bibnamefont
  {{Ashida}}},\ }\bibfield  {title} {\bibinfo {title} {{Entanglement swapping
  in critical quantum spin chains}},\ }\href
  {https://doi.org/10.1103/PhysRevB.111.155143} {\bibfield  {journal} {\bibinfo
   {journal} {\prb}\ }\textbf {\bibinfo {volume} {111}},\ \bibinfo {eid}
  {155143} (\bibinfo {year} {2025})},\ \Eprint
  {https://arxiv.org/abs/2406.12377} {arXiv:2406.12377} \BibitemShut {NoStop}%
\end{thebibliography}%
\end{document}